\documentclass[prx,showpacs,superscriptaddress,twocolumn]{revtex4-2} 
\usepackage{amsmath}
\usepackage{amssymb}
\usepackage{amsthm}
\usepackage{bm}
\usepackage{float}
\usepackage{hyperref}
\usepackage{titlesec}
\usepackage[T1]{fontenc}
\usepackage{mathtools} 
\usepackage[dvipsnames]{xcolor}
\usepackage{graphicx}
\newtheorem{lemma}{Lemma}
\newtheorem{proposition}{Proposition}

\hypersetup{
  colorlinks   = true,
  urlcolor     = blue,
  linkcolor    = black,
  citecolor   = black 
}

\begin{document}
\title{Collective is different: Information exchange and speed-accuracy trade-offs \\ in self-organized patterning}

\author{Ashutosh Tripathi}

\affiliation{Department of Mathematics, Massachusetts Institute of Technology,
77 Massachusetts Avenue, Cambridge, MA 02139, USA}

\affiliation{Center for Computational Biology, Flatiron Institute, 162 5th Ave, New York, NY 10010, USA}

\author{J\"orn Dunkel}
\affiliation{Department of Mathematics, Massachusetts Institute of Technology,
77 Massachusetts Avenue, Cambridge, MA 02139, USA}
\author{Dominic J. Skinner}
\thanks{dskinner@flatironinstitute.org}
\affiliation{Center for Computational Biology, Flatiron Institute, 162 5th Ave, New York, NY 10010, USA}

\begin{abstract}
    During development, highly ordered structures emerge as cells collectively coordinate with each other. While recent advances have clarified how individual cells process and respond to external signals, understanding collective cellular decision making remains a major challenge. Here, we introduce a minimal, analytically tractable, model of cell patterning via local cell-cell communication. Using this framework, we identify a trade-off between the speed and accuracy of collective pattern formation and, by adapting techniques from stochastic chemical kinetics, quantify how information flows between cells during patterning.  Our analysis reveals counterintuitive features of collective patterning: globally optimized solutions do not necessarily maximize intercellular information transfer and individual cells may appear suboptimal in isolation.  
    Moreover, the model predicts that instantaneous information shared between cells can be non-monotonic in time as patterning occurs. An analysis of recent experimental data from lateral inhibition in \emph{Drosophila} pupal abdomen finds a qualitatively similar effect.
\end{abstract}
\maketitle

\section{Introduction}
Creating a functional organism requires complex multicellular coordination~\cite{Bruckner2024,Corson2019}. To this end, eukaryotic cells have evolved a multitude of mechanisms to sense external cues~\cite{Bruckner2024, Corson2019, Kicheva2023} and respond to them by changing their internal state~\cite{PETKOVA2019, Hajji2025, Wang2024}, migrating~\cite{Leathers2022}, or signaling~\cite{Bray2025, Corson2019} to other cells. 
The underlying principles of multicellular self-organization remain elusive, despite numerous examples of developmental self-organization observed across biology~\cite{Phan2024, Corson2017, Bocci2020}. Does development optimize for particular objectives, such as robustness or speed? If so, what constraints shape the outcomes? In the case of a single cell processing an exogenous signal, theoretical progress has been made to determine how cells could optimally process a static~\cite{PETKOVA2019} or dynamic~\cite{Siggia2013,Kobayashi2010} signal, and how cells can sense their environment and optimally act upon it~\cite{Kobayashi2021, tottori2024}. Many developmental contexts fit this paradigm: one population emits a signal and another responds without providing feedback to the sender. For example, in \emph{C. elegans} vulval development an anchor cell secretes an EGF-like ligand which is received by vulval precursor cells but no reciprocal signal is sent~\cite{Corson2012, Sternberg2005}. Similarly, during ascidian development, vegetal cells secrete an FGF-like ligand, which induces a neural fate in specific animal cells and no reciprocal signal is sent~\cite{Roure2014, bettoni2024}.

Frequently, however, cells both send and receive signals, leading to complex non-linear feedback~\cite{Bruckner2024, Corson2019, kato2025}. Indeed, although the receiving cells in the previous examples do not feedback on the signal sender, they coordinate among themselves, through Notch signaling in \emph{C. elegans} vulva~\cite{Corson2012} and EphrinA signaling in ascidian neural induction~\cite{Ohta2013}.  A single cell making noisy measurements of its external environment and acting on them can be modeled as a partially observed Markov decision process, for which optimal control is governed by the Bellman equation~\cite{tottori2022,tottori2024}. In contrast, analyzing interacting cells engaged in a collective task, such as patterning, represents a decentralized partially observed Markov decision process, where computing an optimal strategy is generically NP-hard, though finding good solutions is becoming increasingly practical~\cite{Rashid2020, samvelyan2019}. Difficulties arising from decentralization are well known in computer science; for instance, the two generals' problem demonstrates that perfect coordination between decentralized actors is impossible when they communicate over noisy channels~\cite{Durfee2001, Gray1978}. Echoing ``more is different''~\cite{Strogatz2022, Anderson1972}, while multicellular systems are indeed made up of individual cells, the emergent principles that govern the self-organizing collective are not simple extensions of single-cell behavior. Development provides a proof-by-example that decentralized systems are capable of robust self-organization. Yet, despite numerous models demonstrating collective self-organization~\cite{Corson2017, COLLIER1996, Friston2015}, the core principles, such as optimality and information flow, remain largely unexplored.

\begin{figure}[t]%
\centering
\includegraphics{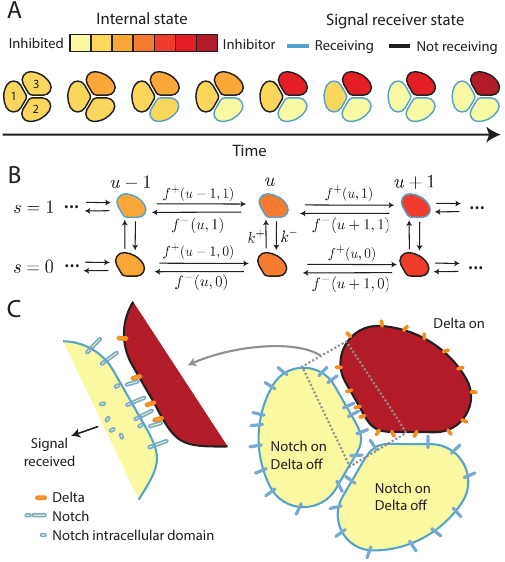}
\caption{Self-organization through lateral inhibition. (A)~In our model, cells have an internal state abstractly ranging from an inhibitor-like state to an inhibited-like state, along with a receiver state which can be activated by neighboring cells. This allows cells to self-organize into a target pattern, here 3 cells organize so that there is one inhibitor cell and two inhibited cells. Throughout, we number cells counterclockwise starting from the leftmost cell; here cell 3 is the inhibitor. (B) Possible transitions for a single cell in the model and corresponding transition rates. Only the rate of receiving the signal, $k^+$, is a function of the neighboring cell states.  (C)~Delta-Notch signaling as a motivating biological example of lateral inhibition. When Delta (orange) on one cell binds to Notch (blue) on a neighboring cell, Notch is cleaved, releasing its intracellular domain and activating downstream transcriptional responses. }
\label{fig1}
\end{figure}

To explore these fundamental questions on decentralized self-organization in an analytically and numerically tractable framework, we focus on a minimal model of self-organization through local cell-cell communication, Fig.~\ref{fig1}. In our model, motivated by  lateral inhibition~\cite{Bray2025, Corson2017, Kuintzle_2024}, cells communicate imperfectly with their neighbors and are tasked with forming a pattern in which exactly one cell adopts an ``inhibitor''-like state while its neighbors adopt an ``inhibited''-like state. With this simple task we find that the patterning strategies obey a speed-accuracy trade-off: arbitrarily high patterning accuracy is possible, but at the cost of increasing the time required to pattern. By reframing the problem as a stochastic reaction network~\cite{Moor2023}, we can precisely compute information theoretic quantities, such as the mutual information between cells, allowing us to track how information flows in our system. We contrast this dynamic information between trajectories with simpler instantaneous information quantities that can be computed from data, finding that instantaneous quantities can sometimes decrease as patterning occurs. Next, we show that after optimizing for a multicellular objective, individual cells do not appear to behave optimally when considered in isolation, and that the collectively optimal strategies neither maximize nor minimize the information transferred between cells. Finally, we connect these theoretical results to recent live-imaging experiments of collective patterning through Delta-Notch signaling~\cite{Phan2024}, by computing the instantaneous mutual information from experimental measurements. We find, similarly to the model, that the instantaneous measures are non-monotonic as patterning occurs.

\section{Collective patterning model}

To explore the principles of communication and self-organization in a multicellular system, we take a model that accounts for the following essential biological features: (i) cells have some internal state, which could be used to classify cells into cell types or to specify a target pattern, (ii) cells have a way to communicate with each other, (iii) cells control their internal state based on signals they have received, and control the signals they send based on their internal state, (iv)  both sending and receiving of signals, as well as control of their internal state, are imperfect and subject to stochastic fluctuations. These criteria are general enough to describe a host of complex developmental feats of self-organization, from neural induction to digit specification. While development regularly features multiple signaling pathways, long range morphogen diffusion, cell divisions, and physical rearrangements, for the sake of tractability we focus on the simpler example of a small fixed number of cells patterning through lateral inhibition with communication across physical cell-cell contacts. 

\subsection{Lateral inhibition as motivating example of collective cellular patterning}
Lateral inhibition is a way to sort a population of cells into two states in a controlled manner by having cells in one state, or advancing towards that state, inhibit their neighbors from similarly advancing towards that state, Fig.~\ref{fig1}. In development, this is often implemented through Notch signaling, although lateral inhibition appears more generally across biology, for instance, to increase sensory perception in neurons~\cite{Hartline1956}. Often, Delta-Notch signaling is combined with some initial pre-patterning to create highly ordered structures~\cite{Corson2017}. However, even from an initial population of identical cells, Delta-Notch signaling will sort cells into two states through lateral inhibition and stochasticity in intra-cellular dynamics. A 2D tissue of identical cells can then form a ``salt and pepper'' pattern of the two cell states under Delta-Notch dynamics.

Notch and Delta are transmembrane proteins located on the surface of cells. When a Delta protein in cell $A$ binds to a Notch receptor on cell $B$, cleavage of Notch is triggered, releasing the Notch intracellular domain. In turn, the Notch intracellular domain translocates to the nucleus where it leads to transcriptional changes within cell $B$. Having received this signal, cell $B$ suppresses the production of Delta~\cite{Bray2025}. In this way, a cell that is expressing Delta will inhibit its neighbors from similarly expressing Delta, thus forming a laterally inhibiting system, Fig.~\ref{fig1}C. The two final states being cells with high Notch and low Delta expression, and cells with high Delta expression. To model the full complexities of a specific Delta-Notch patterning system, one would have to account for both the Notch and Delta proteins, their mRNA transcripts, as well as a number of transcription factors and their mRNA transcripts, like Scute and E(spl)m3-HLH. Indeed, there are many existing models of Delta-Notch dynamics that capture varying levels of biological details~\cite{Bocci2020, COLLIER1996, Barad2010}. However, the key principles of lateral inhibition can be captured with a model where cells have a single internal variable~\cite{Corson2017}, analogous to a reaction coordinate between the two terminal cell states~\cite{Wang2024}.

\subsection{Minimal model of lateral inhibition}Specifically, we will study a modified version of the Delta-Notch model in Ref.~\cite{Corson2017}, where instead of a continuous internal variable, our cells have a discrete internal state $u\in \{0,1,\dots, N\}$. To model cell-cell communication, each cell has a signal receiver state $s$ where $s=0$ means no signal is being received and $s=1$ means that a signal is being received. In reality, the level of signal depends on how many Notch receptors have been recently cleaved and will not be binary. Cells can adjust their internal state stochastically, Fig~\ref{fig1}B, 
\begin{equation}\label{eq:intRate}
    u \xrightleftharpoons[f^-(u+1,s)]{f^+(u,s)} u + 1,
\end{equation}
where the transition rates $f^\pm(u,s)$ depends on the cell's internal state as well as its signal receiver state. The receiver state can also change stochastically,
\begin{equation}\label{eq:sigRate}
    s =0  \xrightleftharpoons[k^-]{k^+}  s =  1, 
\end{equation}
where $k^-$ is taken to be a fixed parameter, but $k^+$ is a function of the neighboring cells. Specifically, for cell $i$, we take
\begin{equation}\label{eq:adjRate}
    k^+ = \sum_{j} A_{ij} \, g(u_j),
\end{equation}
where $g(u)$ determines the strength at which a cell sends a signal to its neighbors, $A_{ij}$ is an adjacency matrix where $A_{ij}=1$ if cells $i$ and $j$ are neighbors at time $t$, and $A_{ij}=0$ otherwise. This form could be easily modified so that the value of the adjacency matrix depends on the area of contact between cells, or even an area of contact that changes with time, but for now we assume that all neighbors have equal and fixed contact areas. Finally, we assume that the states $u=0$ and $u=N$ are absorbing states, once a cell has entered those states it will not leave. Since cells can implement an effective threshold, these absorbing states model a level of Notch or Delta which triggers a fate commitment in that cell~\cite{Siggia2013, Bray2025}.

With the model specified, we want to know if there exists some set of rates $\{f^{\pm},k^-,g\}$ under which a system of $M$ cells will coordinate to make a target pattern. Eventually all cells will reach an absorbing state, either $0$ or $N$, after which there are no cell state transitions, and we refer to this as a terminal state. We focus on a system of $M=3$ cells where every cell is in contact with every other cell, and the target pattern is one cell in the $N$ state and two cells in the $0$ state, Fig.~\ref{fig1}. This is the smallest network that exhibits decentralized symmetry breaking with each cell coordinating with multiple neighbors.

Crucially, every cell has the exact same set of transition rates; if one cell had rates with $f^+ >0$, $f^-=0$ and the other two cells had different rates with $f^- >0$, $f^+=0$ then a correct patterning would always be achieved. Instead, we give every cell the same rates and identical initial conditions. If each cell were to stochastically choose an absorbing state without communication, our target pattern would be achieved with probability at most $4/9$ (Appendix~\ref{A:NoCom}). To reliably reach the target pattern, cells must coordinate with each other through signaling. In the next section, we will explore the optimal strategies for reaching the target state, before quantifying the information transferred between cells executing an optimal strategy in section~\ref{sec:InfoTrans}.

\section{\label{sec:OptimalPatterning} Speed-accuracy trade-off for optimal patterning}
\begin{figure*}[t]%
\centering
  \includegraphics[width=\linewidth]{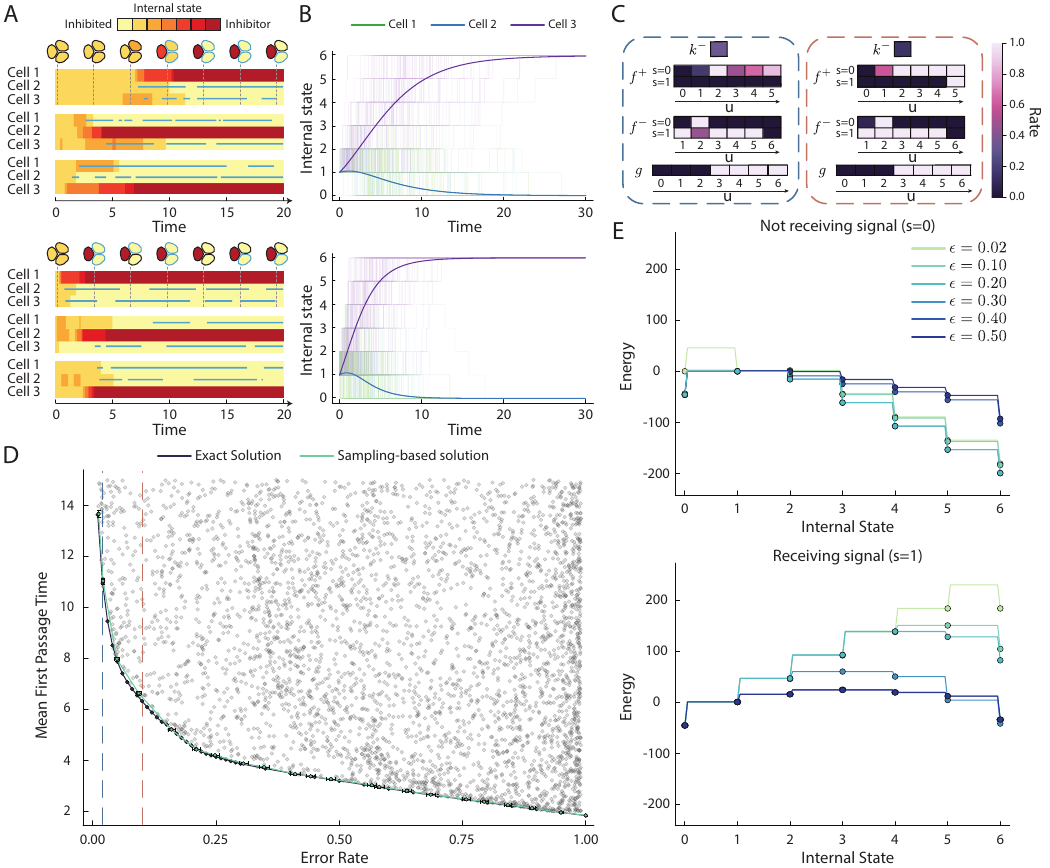}
\caption{Self-organizing model is constrained by a speed-accuracy trade-off. (A) Kymographs showing example trajectories from optimized system with error rate $\epsilon = 0.02$ (top) and $\epsilon = 0.1$ (bottom), illustrating how one cell advances stochastically towards the inhibitor state and signals to its neighbors. Color represents internal state, blue line represents a cell receiving a signal. The corresponding multicellular state for select time points is illustrated above for the first kymograph. Throughout this figure, a model with $M=3$ cells with $N=6$ internal states is shown. (B)  Average internal state for each cell conditioned on a successful trajectory where the inhibitor state was reached by cell 3, for $\epsilon=0.02$ (top) and $\epsilon = 0.1$ (bottom). Solid lines show the exact average, faint lines show $100$ stochastic realizations.  (C) Optimized model parameters for $\epsilon=0.02$ (left) and $\epsilon = 0.1$ (right). (D)  Optimizing the mean first passage time over the set of model parameters $p$, while constraining the error, finds a speed-accuracy trade-off curve that all models are bounded by (black points). The same trade-off curve obtained using our sampling-based approach (teal points) was computed using $10,000$ Gillespie simulations, with error bars showing $1.96 \times$ standard error. Additionally, $5000$ randomly sampled parameter values (gray points) are shown, demonstrating that the entire region above the trade-off curve is accessible. Blue and red dotted lines show the error rates $\epsilon=0.02$ and $\epsilon = 0.1$ respectively. (E) 
 Effective ``energy landscape'' for internal state dynamics are shown for a cell that is not receiving a signal ($s=0$, top) and a cell that is receiving a signal ($s=1$, bottom). Points represent the ``energy'' level of each state and lines represent the height of the effective energy barrier between states (SM Sec. II). }
\label{fig:tradeoff}
\end{figure*}

In our system, cells can receive a signal (through $s$), control their internal state (through $f^{\pm}$) and send a signal to neighboring cells (through $g$). With these essential ingredients for self-organization, our first question is whether regions of the parameter space, $p = \{ f^\pm,k^-,g\}$ can achieve patterning with a small error, $\epsilon$, which we define as the probability of our system ending up in a non-target terminal state. Scaling by the fastest time scale, so that $f^\pm, k^-, g \leq 1$, our model allows for an arbitrarily small error rate (Appendix~\ref{A:ExactAsymp}), although this comes at the price of taking infinitely long to reach the terminal states (Appendix~\ref{A:Bound}). This trade-off is generic, occurring for any value of $M\geq 3$ and any adjacency matrix (Appendix~\ref{A:Bound}). Therefore, a more biologically motivated optimization problem is to specify the degree of error that is permitted, $\epsilon_{tol}$, and minimize the average time to reach the terminal states, $\tau$, under the constraint that $\epsilon \leq \epsilon_{tol}$. We note the same region of the speed-error plane is accessible whether one constrains $\epsilon \leq \epsilon_{tol}$ and minimizes $\tau$ or if one constraints $\tau \leq \tau_{\max}$ and minimizes $\epsilon$.

The level of permitted error, $\epsilon_{tol}$ in a fate specification depends on its context within development, and whether any mistake can be corrected or compensated for later on. For instance, ABp specification via Notch-signaling is critical for normal development in \emph{C. elegans}~\cite{CElegansBook}, whereas when Notch-signaling specifies sensory organ precursor cells in \emph{Drosophila} pupal abdomen, errors can be compensated by later cell rearrangements and in any case a single misspecification is not critical to survival~\cite{Phan2024}. Ideally, we want to examine how the minimum average time to reach a terminal state changes as a function of the error constraint $\epsilon_{tol}$, a Pareto front that, barring additional considerations such as evolvability~\cite{Francois_2008} or thermodynamic cost of transcription, translation, and signaling~\cite{zoller2025}, evolutionarily optimized systems should be near. Similar trade-off structures appear for speed and dissipation in transcription~\cite{zoller2025}, for fluctuations in biochemical reaction networks~\cite{Hilfinger2016, Kell2023, Jiawei2019}, and for dissipation, speed, and accuracy for synchronized oscillators~\cite{Zhang2020, Zhang2025}.

Mathematically, our model is a continuous time Markov chain (CTMC) for the full state of the system $x_\alpha = ((u_1,s_1), \dots, (u_M, s_M))$, tracking both the internal and receiver states of every cell. A probability distribution over the states evolves according to the master equation,
\begin{equation}
    \frac{\mathrm{d}}{\mathrm{d}t} P_t(x_\alpha) = \sum_\beta Q_{\alpha \beta} P_t(x_\beta),
\end{equation}
where $P_t(x_\alpha)$ is the probability of the system being in state $x_\alpha$ at time $t$, $Q_{\alpha\beta}$ is the rate at which $x_\beta$ transitions to $x_\alpha$ for $\alpha\neq\beta$ and $Q_{\alpha\alpha} = -\sum_{\beta\neq \alpha} Q_{\beta\alpha}$. We can solve this master equation directly, or simulate realizations of the process with the Gillespie algorithm. Finding the error rate $\epsilon$ or the average time to terminal states $\tau$ can be done directly from the rate matrix $Q$, by posing them as first passage-like problems. To do so, we first define $\tau_\alpha$ to be the expected time to reach a terminal state starting from state $x_\alpha$ and $\epsilon_\alpha$ to be the probability that, starting from $x_\alpha$, the eventual terminal state does not achieve the target pattern. Calling $\mathcal{T}^G$ the set of terminal states which achieve the target pattern and $\mathcal{T}^B$ the set of terminal states that do not achieve the target pattern, then $\tau_\alpha = 0$ for $\alpha\in \mathcal{T}^G\cup \mathcal{T}^B$, $\epsilon_\alpha = 0$ for $\alpha\in\mathcal{T}^G$, and $\epsilon_\alpha = 1$ for $\alpha\in\mathcal{T}^B$. For the remaining $\alpha \notin \mathcal{T}^G\cup \mathcal{T}^B$, it follows that 
\begin{align}\label{eq:MFP}
    \tau_\alpha &= -\frac{1}{Q_{\alpha\alpha}} - \sum_{\beta\neq\alpha} \frac{Q_{\beta\alpha}}{Q_{\alpha\alpha}} \tau_\beta, \\
    \epsilon_\alpha &= - \sum_{\beta\neq\alpha} \frac{Q_{\beta\alpha}}{Q_{\alpha\alpha}} \epsilon_\beta,
\end{align}
and these linear equations can be solved to find $\tau_\alpha$, and $\epsilon_\alpha$. Since the rate matrix depends on the parameters $p = \{ f^{\pm},k^-,g\}$, and $\tau_\alpha$, $\epsilon_\alpha$ depend on the rate matrix, we have that  $\tau_\alpha = \tau_\alpha(p)$, $\epsilon_\alpha = \epsilon_\alpha(p)$. Supposing that we initialize the system in state $\alpha^*$, we can now pose the minimization problem as
\begin{align}\label{eq:optim}
    \min_{p} \tau_{\alpha^*}(p) \quad s.t. \quad \epsilon_{\alpha^*}(p) \leq \epsilon_{tol},
\end{align}
where, after minimization, we will find both the optimal $\tau$ along with a set of parameters $p$ that generate it. As the cells have a discrete internal state, there are only finitely many parameters that we are optimizing over, specifically $4(N-1)$ for the $f^\pm$, 1 for $k^-$ and $N+1$ for $g$. The optimization problem is equivalent to a quadratic problem with quadratic constraints (SM Sec. II), a generically hard problem~\cite{Luo2010}. In our numerical parameter scans however, we consistently find solutions converge to a small number of local minima from different initializations, whether solving as interior point optimization or through gradient descent (SM Sec. II).

Minimizing across many values of $\epsilon_{tol}$, with $N=6$ internal states and $\alpha^* = ((1,0),(1,0),(1,0))$, identifies the optimal trade-off curve between error and average time to reach the terminal state, Fig.~\ref{fig:tradeoff}D. Without penalizing error, by setting  $\epsilon_{tol} = 1$, all cells immediately head towards the closest absorbing state, $u=0$, and consequently never reach a target pattern. In the other limit of extreme precision, $\epsilon_{tol}\to 0$, it takes increasingly long to reach a terminal state. For intermediate values, the optimal strategy is able to consistently reach a terminal state in a relatively short time compared to the fastest timescale of the system, Fig.~\ref{fig:tradeoff}D. 
Above the Pareto front, any parameter combination is accessible, Fig.~\ref{fig:tradeoff}D. While our focus in the main text is on the tractable three-cell case, it is possible to explore the speed–accuracy trade-off in larger systems by using a sampling based gradient descent scheme (SM Sec. II). To illustrate this, we extended the model to seven cells, modeling an asymmetric two-dimensional epithelial region with non-trivial neighborhood structure, and obtained a  Pareto front comparable to the three-cell case (SM Fig. S2).

Generally, the optimal strategy appears to be one of cells stochastically increasing their internal state, the first cell to leave the initial state signals to its neighbors who decrease their internal state, Fig.~\ref{fig:tradeoff}(A-C). These strategies often implement a sharp change in $f^\pm$ as a function of internal state and receiver state and $g$ as a function of internal state, Fig.~\ref{fig:tradeoff}C. Such sharp responses are a common feature of gene regulatory networks and so this does not represent an unphysical aspect of the model. The non-monotonicity of the rates is also plausible given that we are not modeling a single gene but the output of a gene regulatory network which can be highly non-linear. In any case, constraining the rates to be monotonic has minimal impact on the Pareto front. These strategies can be mathematically interpreted as each cell, for some given receiver state, navigating an effective energy landscape~\cite{Owen2020}, analogous to a Waddington landscape~\cite{Corson2012} (SM Sec. II), as shown in Fig~\ref{fig:tradeoff}E. From the landscape interpretation, we see that for small error rate, such as $\epsilon=0.02$, cells that are not receiving a signal are prevented from reaching $u=0$ by an effective  energy barrier, and will slowly increase their internal state. Once a cell receives a signal, it rapidly descends to the inhibited state, Fig.~\ref{fig:tradeoff}E. For larger error rates, the landscape flattens, decreasing the energy barrier for fate commitment, Fig~\ref{fig:tradeoff}E. In this case, cells with a large internal state that are receiving a signal can still advance towards the inhibitor state. We observe qualitatively similar optimal strategies for systems with  a larger number of cells (SM Fig. S2).

This strategy, of a single cell stochastically increasing its internal state before signaling to its neighbors, Fig.~\ref{fig:tradeoff}A, differs qualitatively from the model in Ref.~\cite{Corson2017}, where all cells increase their internal state concomitantly before one cell eventually becomes dominant and inhibits its neighbors. Both concomitant~\cite{Corson2017} and non-concomitant~\cite{Phan2024} dynamics have been observed in recent live-imaging of Delta-Notch patterning. In particular, a recent experiment measured Scute expression, a transcription factor that up‑regulates Delta, while sensory organ precursor (SOP) specification was occurring in the \emph{Drosophila} dorsal histoblast and saw minimal concomitant increase in Scute~\cite{Phan2024}. Scute expression appeared to increase in future SOP cells while remaining low in non-SOP cells, which then began to increasingly express Notch~\cite{Phan2024}.

\section{\label{sec:InfoTrans}Dynamic information transfer}
\begin{figure*}[t]
\centering
\includegraphics[width=\linewidth]{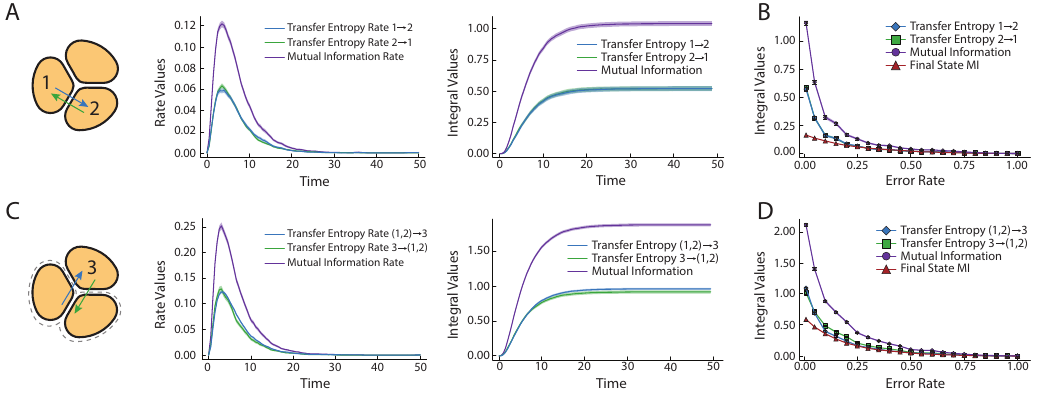}
\caption{Quantifying information shared between cell trajectories  finds a non-monotonic rate of information transfer as well as a greater amount of information shared between trajectories than final states alone. (A) For any pair of cells, we can compute the transfer entropy (TE) rates in both directions $\dot{\mathcal{T}}^{u_1\to u_2}$, $\dot{\mathcal{T}}^{u_2\to u_1}$, the mutual information (MI) rate $\dot{I}= \dot{\mathcal{T}}^{u_1\to u_2} + \dot{\mathcal{T}}^{u_2\to u_1}$ (left), as well as the corresponding integrals (right), shown here for the optimized model with $\epsilon = 0.02$.  (B) Total information transferred between any pair of cells for optimized models as the error rate is varied. Each value represents the limiting mutual information or transfer entropy for a given model at large time.  Instantaneous mutual information shared between the final states of both cells is also shown. (C) Transfer entropy rates between cell 3 and the remaining cells, $\dot{\mathcal{T}}^{u_3\to [u_1,u_2]}$, the converse $\dot{\mathcal{T}}^{[u_1,u_2]\to u_3}$, as well as the mutual information rate (left) along with the corresponding integrals (right), shown here for the optimized model with $\epsilon = 0.02$. (D) Total transfer entropy from one cell to the remaining cells and the converse, as well as the mutual information for optimized models as the allowed error rate is varied. Instantaneous mutual information shared between the final state of cell 3 and the remaining cells is also shown. Throughout, each computation is averaged over $n=10,000$ Monte-Carlo samples and shaded regions show $\pm 1.96\times$  standard error. Integral values are expressed in units of nats, while rates are given in nats per unit time.}
\label{fig:InfoQuant}
\end{figure*}

Successful collective self-organization requires cells to exchange information. In our system, in the absence of communication cells can only reach the target pattern with a probability of at most $4/9$.
(Appendix~\ref{A:NoCom}). With communication, however, they can find the correct pattern to arbitrary accuracy. In this section we quantify the information transferred between cells using techniques of information theory~\cite{Moor2023,Bruckner2024,Razo2020, Dubuis2013}. The central object will be the mutual information, 
\begin{equation}\label{eq:MI}
    I(X;Y) = \mathbb{E}\left[ \log \frac{ P(X,Y)}{ P(X) P(Y)} \right]
\end{equation}
where $X$ and $Y$ are random variables, $P(X)$, $P(Y)$ are their marginal distributions, and the expectation is taken over $P(X,Y)$, their joint distribution. The mutual information quantifies how much information one gains about $X$ upon seeing $Y$ or vice-versa, and is the natural measure of the information shared between $X$ and $Y$~\cite{Dubuis2013,Bialek2012pt1}. We could use the terminal state of the system to quantify the information shared between neighboring cells in the final pattern, as has been explored recently~\cite{Bruckner2024}. However, as we will see, the final state of the system does not quantify all the dynamic information shared between cells during patterning. Quantifying dynamic information requires computing the mutual information between trajectories $I(X_0^T;Y_0^T)$, where $X_0^T$ denotes the trajectory of a time dependent variable $X(t)$ for $0\leq t \leq T$. If $X(t)=u_1(t)$ and $Y(t)=u_2(t)$, then $I(X_0^T;Y_0^T)$ quantifies the total information that is shared between cell 1 and cell 2 up until time $T$. We will also make use of the transfer entropy rate, defined as
\begin{align}
\dot{\mathcal{T}}^{{Y}\to{X}}(t)
=&\lim_{dt\to 0}\frac{1}{dt}\mathbb{E}\left[\log{\frac{P_{t+dt}[X(t+dt)|X^{t}_{0},Y^{t}_{0}]}{P_{t+dt}[X(t+dt)  |X^{t}_{0}]}}\right],
\end{align}
which quantifies the directed information transfer rate from $Y$ to $X$, and provided $X$ and $Y$ cannot simultaneously change~\cite{Moor2023}, is related to the mutual information by
\begin{align}\label{eq:MI2}
    I(X_0^T;Y_0^T) = \int_0^T \left[ \dot{\mathcal{T}}^{{Y}\to{X}}(t) + \dot{\mathcal{T}}^{{X}\to{Y}}(t) \right]\, dt .
\end{align}
In this section, we measure the magnitude of the information flow, and by computing conditional transfer entropy rates, we explore how information is transferred from inhibitor to inhibited cells and vice-versa, as well as between inhibited cells.

\subsection{Feedback in lateral inhibition makes information calculations challenging}It is well established theoretically that gene regulatory networks can  process information, and that the mutual information between the network's inputs and its outputs can be quantified~\cite{Bialek2012pt1, Bialek2012pt2, Bialek2012pt3, Hahn2023, Wolde2025, zoller2025}. Information has been quantified in experimental systems, both directly from trajectories~\cite{Tang2021, witteveen2025, kramar2024}, and from building detailed models that are fit to experimental data~\cite{Razo2020, Cheong2011}. A handful of developmental problems have been studied from the perspective of information theory, for instance the information contained in the \emph{Drosophila} gap gene pattern was experimentally quantified~\cite{Dubuis2013, PETKOVA2019, Bauer2021}, and it has been proposed that the gene regulatory networks optimally create an information rich pattern~\cite{Sokolowski2025}. In this existing literature, the signal is taken as exogenous; a cell's response to a signal does not impact the signal's future values. While this is often an appropriate assumption, such as in ascidian neural induction~\cite{Roure2014, bettoni2024}, this need not hold in general, and indeed does not hold for lateral inhibition. In our model, cells are both sending and receiving signals and the signals they receive affect the future signals they will send. Any information theory analysis of our system must take this into account.

In addition to this feedback, other common approximations which simplify the computation of information theoretic quantities do not hold for a laterally inhibiting system. As we have seen, such systems are time dependent and cannot be approximated as a series of static input and output relationships. Moreover, the multi-modal nature of cell fate transitions means that we cannot approximate the dynamics as a uni-modal Gaussian process, for which computations are more tractable~\cite{moor2025, Tostevin2009}. It is, in theory, possible to simulate our system with the Gillespie algorithm and, treating the simulated data like it were experimental data, attempt to directly estimate the mutual information. However, direct estimation of mutual information remains challenging due to the high dimensionality of the space of trajectories and such estimators do not take advantage of any of the known structure of our model. 

\subsection{Reframing the model as a stochastic reaction network enables tractable information calculations }Computing the mutual information between trajectories is difficult even when the underlying stochastic model is known. To see this, suppose we wanted to compute a mutual information, $I(X_{0}^T;Y_{0}^T)$, where  $X(t)$ could be $X(t)=u_1(t)$ or $X(t) = [u_1(t),u_2(t)]$, similarly for $Y(t)$. While the path measure of a CTMC has a closed form expression, working out the mutual information requires computing marginal path measures, such as $P(X_0^T)$, which are intractable analytically. Recently,  Monte-Carlo sampling techniques that take advantage of the known model structure, have been used to estimate otherwise intractable terms in mutual-information-like computations~\cite{das2024,Reinhardt2023}. In these approaches an outer expectation is estimated by Monte-Carlo sampling but once a sample has been drawn, actually computing the corresponding value of the integrand requires another round of Monte-Carlo sampling.  While advanced sampling techniques like these may be required for many problems, here we can exactly compute the integrand in equation~\eqref{eq:MI2} following Ref.~\cite{Moor2023}. There, the authors consider a stochastic reaction network with $K$ reaction channels, $n$ chemical species $Z_1,\dots,Z_n$, with vector $Z(t)$ recording the copy number of each species at time $t$,  and the $k^{th}$ reaction occurring at a rate $\lambda_k(Z(t))$. If $X(t)$ and $Y(t)$ are variables, or disjoint subsets of variables, in $Z(t)$ that do not change simultaneously, then
\begin{align}\label{eq:SRNMI}
    I(X_0^T; Y_0^T) =& \mathbb{E}\left( \sum_{k\in R_{X}}\int_0^T \log \frac{\lambda_k^{XY}(s)}{\lambda^X_k(s)} dN_k(s)\right. \\ \notag
    &\quad \left. +\sum_{k\in R_{Y}} \int_0^T \log \frac{\lambda_k^{XY}(s)}{\lambda^Y_k(s)} dN_k(s) \right),
\end{align}
where $R_A$ denotes the set of reactions that involve a change in $A$, $dN_k(t)$ is the increment of $N_k(t)$ which counts the number of times reaction $k$ has occurred, and $\lambda_k^A = \mathbb{E}[\lambda_k(Z(t))| A_0^T]$ is the  expected rate of the $k^{th}$ reaction given $A_0^T$. The first and second terms in the expectation correspond to the transfer entropies $\mathcal{T}^{Y\to X}(T)$ and $\mathcal{T}^{X\to Y}(T)$ respectively~\cite{Moor2023}, where throughout we define the transfer entropy to be $\mathcal{T}^{A\to B}(T) = \int_0^T \dot{\mathcal{T}}^{A\to B}(t)dt $. Note that equation~\eqref{eq:SRNMI} appears as equation A6 in Ref.~\cite{Moor2023} although here we have neglected the integrals with expectation zero. 

To actually compute the integrand, we need to compute $\lambda^A_k$, which requires knowing the conditional distribution $ P(\bar{Z}(t)=\bar{z}|A_0^T)$, where $\bar{Z}(t)$ is a vector that tracks all molecular abundances except those in $A$. This conditional distribution obeys a stochastic differential equation known as the \textit{filtering equation}~\cite{Duso2018, Moor2023} (SM Sec. III). Keeping track of the probability for every possible  count of the latent species $\bar{z}$ is typically not practical in chemical reaction networks, and the filtering equation is instead used to construct moment closure approximations~\cite{Moor2023,Duso2018}.

We can interpret our model, equations~\eqref{eq:intRate}--\eqref{eq:adjRate}, as a stochastic reaction network, albeit an unusual one with non-linear reaction rates and with the copy number of each species bounded above by a finite number. This finiteness of copy number means that we need not approximate the filtering equation, and instead can solve it exactly. To compute the mutual information, we use Monte-Carlo sampling to approximate the outer expectation in eq.~\eqref{eq:SRNMI}, but for each sample we exactly compute the term inside the expectation by directly solving the filtering equation (SM Sec. III).

Similarly to the information $I(X_0^T;Y_0^T)$, we can compute an information rate $\frac{dI(X_0^t;Y_0^t)}{dt}$, as well as transfer entropy rates, $\dot{\mathcal{T}}^{X\to Y}(t)$ and $\dot{\mathcal{T}}^{Y\to X}(t)$, representing the directed rate of information transfer from $X$ to $Y$ and vice-versa. See SM for additional numerical details.

\subsection{Quantifying information transfer rates numerically }We are now in a position to quantify information flows in our system. For the optimized model with error $\epsilon = 0.02$, and taking  $X=u_1$, $Y=u_2$, we find that there is a sharp increase in the rate of mutual information which then sharply decreases, Fig.~\ref{fig:InfoQuant}A. Each cell will eventually reach an absorbing state after which no further ``reactions'' involving $X$ or $Y$ occur, and hence that trajectory makes no further contributions to the mutual information in equation~\eqref{eq:SRNMI}. The probability that a cell has not reached an absorbing state decays exponentially with time, and indeed, the total mutual information between the trajectories of $u_1$ and $u_2$ clearly asymptotes as $T\to\infty$, Fig.~\ref{fig:InfoQuant}A.  Due to the symmetry of the problem, any pair of internal variables will have the same mutual information between their trajectories as any other pair and the transfer entropy between them in both directions will be exactly half the value of the mutual information. 

For any given value of the error, we can find the optimal model, and compute the exact same information theoretic quantities as we have above for $\epsilon = 0.02$. We find that the total information exchanged between two trajectories decreases monotonically as the allowed error increases, reaching around zero at roughly the point when a zero-communication strategy is possible, Fig.~\ref{fig:InfoQuant}B. In the limit of $\epsilon \to 0$, it is possible for the trajectory of cells to share an arbitrarily large amount of information with each other (Appendix \ref{A:MI}).

\begin{figure}[t]%
\centering
\includegraphics[width=\linewidth]{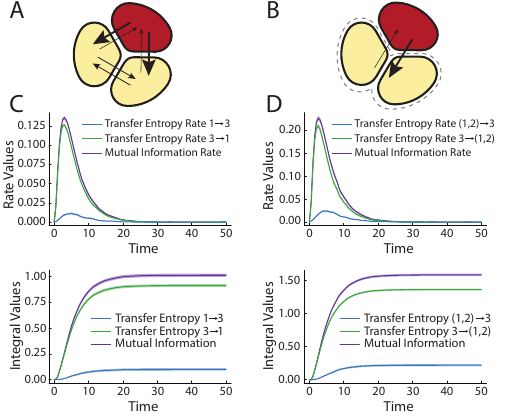}
\caption{Conditioning on final cell fates reveals persistent and asymmetric information transfer between cells.  Here, all information theoretic quantities are computed for a process conditioned on the terminal state of cell $3$ ending up as an inhibitor $u_3=N$, while cells 1 and 2 are inhibited, $u_1=u_2=0$.  (A-B) Schematic where arrow thickness is proportional to the long time conditional transfer entropy between pairs of cells (A) and the inhibitor cell and the remaining pair (B). 
(C) Conditioned mutual information and transfer entropy rates between cell $2$ and cell $3$ (top) as well as their corresponding integrals (bottom). Interestingly, the conditional information transferred from the inhibited cells is small but positive. 
(D) Conditional mutual information and transfer entropy rates between cell $3$ and and its neighbors (top), alongside their corresponding integrals (bottom). 
Throughout, each computation is averaged over $n=10,000$ Monte-Carlo samples, and the shaded regions show $1.96 \times$ standard error. Integral values are expressed in units of nats, while rates are given in nats per unit time.}
\label{fig:condInfo}
\end{figure}

In addition to looking at pairs of cells, we can take one cell and look at the information it has about the remaining cells, or $X = u_1$, $Y = [u_2,u_3]$. We similarly see a sharp increase in the information shared between them before this curve asymptotes,  at around $1.89$ nats for $\epsilon = 0.02$, Fig.~\ref{fig:InfoQuant}C. Interestingly, this quantity greatly exceeds the $0.693$ nats ($1$ bit) that can be shared between a cell and a fixed binary external signal. Additionally the transfer entropy also exceeds $1$ bit, and is asymmetric with each cell receiving more information from its neighbors than it sends, Fig.~\ref{fig:InfoQuant}C. This shows that, whether through signaling timing or repeated signal activation, the binarized receiver state is capable of transferring more than one bit of information. Similarly to the pair of cells, the mutual information between one cell and the remaining cells decreases monotonically as the allowed error increases, Fig.~\ref{fig:InfoQuant}D. 

\subsection{Successful patterning displays directed communication between all cells }Due to the inherent symmetry of our system, each cell is initially equally likely to be the inhibitor cell, and this symmetry obscures the way information is transferred. For instance, suppose that information only flowed from the inhibitor cell to the inhibited cells. The naive transfer entropy calculation would still find every pairwise transfer entropy rate to be equal, say $\dot{\mathcal{T}}^{1\to2}(t) = \dot{\mathcal{T}}^{2\to1}(t)$ since this quantity is computed over trajectories where cell 1 is the inhibitor and trajectories where cell 2 is the inhibitor. Instead, we would like to decompose our information flows by somehow removing the symmetry that any one of the cells is equally likely to be the inhibitor. In systems with feedback, such as this one, such decompositions can violate the data processing inequality or subtly introduce fictitious dependencies, and thus require careful interpretation. With those caveats in mind, we can define transfer entropy rates conditioned on some terminal event, for instance $\dot{\mathcal{T}}^{1\to2 | Z}(t)$ where $Z$ is the event that terminal state is $u_1 = N$, $u_2=u_3=0$ is reached. In general, for the event $Z$ that some final absorbing state is ultimately reached, conditioning the CTMC on $Z$ results in another CTMC. Specifically, using Doob's h-transform~\cite{Norris_1997}, the rate matrix of the conditioned system is 
\begin{equation}
    Q_{\alpha \beta | Z} = Q_{\alpha \beta} \frac{P(Z|\alpha)}{P(Z|\beta)},
\end{equation}
for $\alpha \neq \beta$ and $Q_{\beta\beta | Z} = -\sum_{\alpha\neq \beta} Q_{\alpha\beta|Z}$, where $P(Z|\alpha)$ is the probability of the event $Z$ given the system is in state $\alpha$. Quantities of the form $P(Z | \alpha)$ can be computed in the same way we compute $\epsilon_\alpha$ (equation~\eqref{eq:MFP}), after which we can use this new conditioned CTMC, to  compute information theoretic quantities exactly as before. We find that the conditioned transfer entropy rates are significantly higher from the inhibitor cell to the inhibited cells then in the reverse direction, Fig.~\ref{fig:condInfo}A,C. Nevertheless, the flow of information goes both ways with a non-zero conditioned transfer rates from the inhibited states to the inhibitor state, Fig.~\ref{fig:condInfo}C. Further, there is a non-zero conditioned transfer rate between the two inhibited cells (Appendix~\ref{app:condInfo_inhibited}). Exploring these  information theoretic quantities highlights the complex picture of information flows in a self-organizing system.

\section{Instantaneous versus dynamic information transfer}\label{sec:Instantaneous}
\begin{figure}[t]%
\centering
\includegraphics{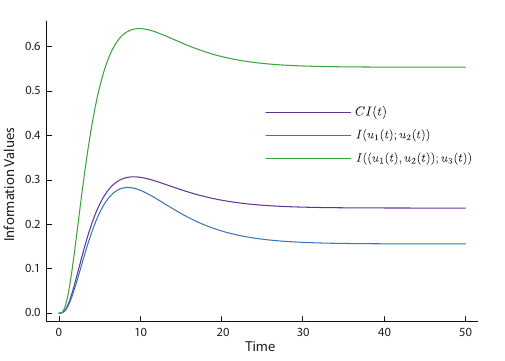}
\caption{Instantaneous information quantities can be non-monotonic in time during self-organization. The information quantities, $CI$, $I(u_1(t);u_2(t))$, and $I([u_1(t),u_2(t)];u_3(t))$, are computed from the instantaneous state of the system at time $t$, shown here for $\epsilon=0.02$.  All quantities are expressed in units of nats.}
\label{fig:StatvDyna}
\end{figure}

To completely quantify the information shared between cells and to account for situations where a change in one cell only affects another cell at a later time, information quantities should be computed between entire trajectories, as we have done so far. However, while it is possible to compute these quantities exactly in our model, computing the mutual information between trajectories directly from biological data remains impractical due to the high dimensional space which trajectories inhabit. A more practical approach is to compute information theoretic quantities using only the instantaneous state of the system, rather than full trajectories. In this section, we explore what can be learned from instantaneous information quantities alone.

Consider the mutual information between the internal state of two cells at some fixed time $t$, $I(u_1(t);u_2(t))$, which we will refer to as the instantaneous mutual information. Due to the data processing inequality the instantaneous mutual information is always less than the mutual information between the trajectories up until that point, or $I(X_0^t;Y_0^t) \geq I(X(t);Y(t))$. A related quantity called correlational information (CI) was recently proposed~\cite{Bruckner2024}, defined as
\begin{equation}
    CI(t) = \frac{1}{M}\mathbb{E}\left[\log \frac{P(u_1(t),\dots,u_M(t))}{\prod_{i=1}^M P_i(u_i(t))}\right],
\end{equation}
where in our system, all cells are equivalent and hence the marginal probability $P_i$ is the same for every cell. Note that $M\times CI$ is sometimes referred to as the multi-information or the total correlation of the random variable $(u_1(t),\dots,u_M(t))$~\cite{Watanabe1960}. Analogously to how we compute mutual information between trajectories, we can compute a trajectory version of correlational information (SM Sec. III).

Plotting the instantaneous mutual information for a particular choice of $\epsilon=0.02$, we find that the instantaneous information is indeed less than the full dynamic information by a factor of around $4-6$, Fig.~\ref{fig:StatvDyna}. In the long time limit, $t\to\infty$, the instantaneous mutual information between a pair of cells is around $0.156$ nats, or around $15.3\%$ of the total information shared between the full trajectories. The final state instantaneous mutual information is smaller than the dynamic mutual information for all values of $\epsilon$, with the largest difference occurring at small $\epsilon$, Fig.~\ref{fig:InfoQuant}B, C. 
In fact, since the final state can only take one of two values, $u_1(t\to\infty)\in\{0,N\}$, we have $I(u_1(t\to\infty);u_2(t\to\infty)) \leq H(u_1(t\to\infty)) \leq \log 2$.
In contrast, the dynamic information shared between cells can be unbounded in the $\epsilon\to0$ limit (Appendix~\ref{A:MI}).  

Intriguingly, the shape of the instantaneous mutual information curve, as well as the correlational information curve, is non-monotonic, exhibiting a local maxima at a finite time, Fig.~\ref{fig:StatvDyna}. Heuristically, we can understand this through the following argument. Suppose after some long time you learn that one cell is at state $N$. You can deduce that its neighbors are likely to be in state $0$. If you know one cell is at state $0$ you deduce that one of its neighbors is likely in state $N$ and the other one is likely in state $0$. Now suppose that we view the system at some intermediate, but later time. Upon learning that one cell is in state $N$ or $0$, we may make similar deductions as before. However, if the cell is in an intermediate state, we can deduce that the neighboring cells are also likely to be in an intermediate state. Therefore at the intermediate time, there are more possible states for the neighbors to be in, and hence one can learn more information about the neighbors by knowing the state of a cell than at the final time. This highlights how a careful interpretation of instantaneous mutual information is required, and that a decrease in instantaneous mutual information does not necessarily mean that the system is becoming less coupled or more disordered.

\section{Collective and individual optimality\label{sec:Collective}}
\begin{figure}[t]%
\centering
\includegraphics[width=\linewidth,keepaspectratio]{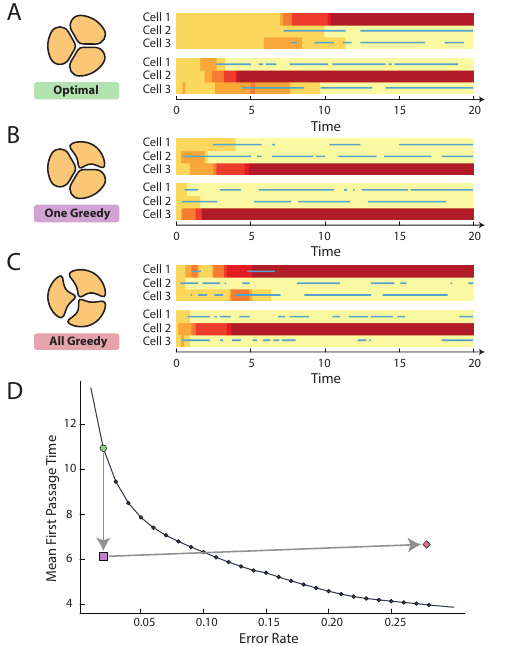}
\caption{Collective optimal strategy does not appear locally optimal. Example kymographs for (A) all cells collectively optimized ($\epsilon=0.02$, as in Fig.~\ref{fig:tradeoff}A), (B) optimizing a single cell with the remaining cells fixed, (C) three copies of the individually optimized cell. (D) Speed-error trade-off plot (from Fig.~\ref{fig:tradeoff}A), showing collective optimal strategy (green circle), a single optimized cell with remaining cells fixed (purple square) and  three individually optimized  cells combined (red diamond).}
\label{fig:Local}
\end{figure}
Through evolutionary selection, development is tuned to robustly generate viable, functional offspring. Many cellular processes, including Delta-Notch signaling, are highly conserved and have been tuned across millions of years of evolution. It is therefore natural to ask questions about optimality such as, how gene regulatory networks can optimally process intra-cellular information~\cite{Bialek2012pt1, Bialek2012pt2, Bialek2012pt3}, how cells can optimally infer an external signal~\cite{Siggia2013, Kobayashi2010, Mattingly2021} and how cells can act optimally to control their environment~\cite{tottori2024}. Similarly, development has been modeled by cells acting as optimal Bayesian agents, seeking to minimize uncertainty about their cellular identity~\cite{Friston2015}. However, fitness is defined at the level of the organism or even the level of the community, making it challenging to ascribe optimality to any particular component of the system. We find that by optimizing a collective objective, the actions of individual cells need not appear individually optimal. Additionally we find that the collectively optimal system does not maximize information flow. In this section, we briefly examine how  collective optimization results in individual cells appearing to act sub-optimally.

\subsection{Local sacrifices enable global gains }Previously, in equation~\eqref{eq:optim} we minimized over one set of parameters $p$, that every cell shared. Equivalently, we could have given each cell its own set of parameters $p_1$, $p_2$, and $p_3$, and optimized over all of these together with the constraint that $p_1=p_2=p_3$. Fix the parameters of two cells, say cells $1$ and $2$, to follow the collective optimum, $p_1^*=p_2^*$ and consider the remaining cell $3$. From the perspective of this cell alone, the strategy given by the collectively optimized parameters $p_3^*$ is not optimal. Cell $3$ can optimize its parameters to find a new strategy $\tilde{p}_3$ which achieves a faster time to the terminal state with the same error, Fig.~\ref{fig:Local}. However all cells adopting $p_3^*$ is superior to all cells adopting $\tilde{p}_3$ as (i) the resulting system is more error prone, and (ii) at the new error rate, the strategy is far from optimal, Fig.~\ref{fig:Local}. This reminds us that for self-organizational problems the objective is a collective one. The actions of a single cell, when treating the remaining cells and signaling environment as an exogenous mean field, may appear suboptimal. Only in the context of the collective problem are cells' actions optimal. Similar trade-offs where individual strategies appear suboptimal yet enable collective coordination have been recently observed at larger biological scales~\cite{Brandl2025}.

\begin{figure}[t]%
\centering
\includegraphics[width=\linewidth,keepaspectratio]{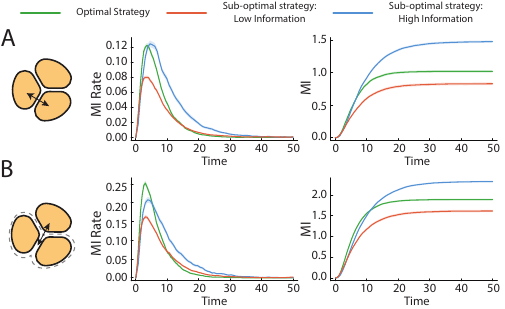}
\caption{Collective strategy does not optimize for information flow. For a fixed error rate of $\epsilon=0.02$, sub-optimal strategies that transfer more or less information between cells than the optimal strategy can be found. (A-B) Mutual information rate (middle) and total mutual information (right) transferred between (A) a pair of cells, and (B) a cell and its neighbors, in the collectively optimal system (green), a system that transfers more information for the same error rate (blue) and one that transfers less information (orange). Throughout, each computation is averaged over $n=10,000$ Monte-Carlo samples, and the shaded regions show $1.96 \times$ standard error. Integral values are expressed in units of nats, while rates are given in nats per unit time.}
\label{fig:SubOptimalInfo}
\end{figure}

\subsection{Total information flow is not the objective }A signaling pathway, taken in an isolated cell with an exogenous signal is often considered as optimally translating the information from the external signal into an internal state~\cite{zoller2025, Bialek2012pt1}. However for a collective, self-interacting system with feedback, where the signal is sent and received by cells, it becomes less clear what optimal signal processing should look like. For a given rate of error, we find suboptimal solutions that both transfer more information between cells than the optimal solution as well as less information, Fig.~\ref{fig:SubOptimalInfo}. When the objective is simply to reach the target state as quickly as possible with some allowed error rate, cells do not optimize the information transferred between them.

\section{Self-organized patterning in \textit{Drosophila} sensory organ formation}
\begin{figure*}[t]%
\centering
\includegraphics{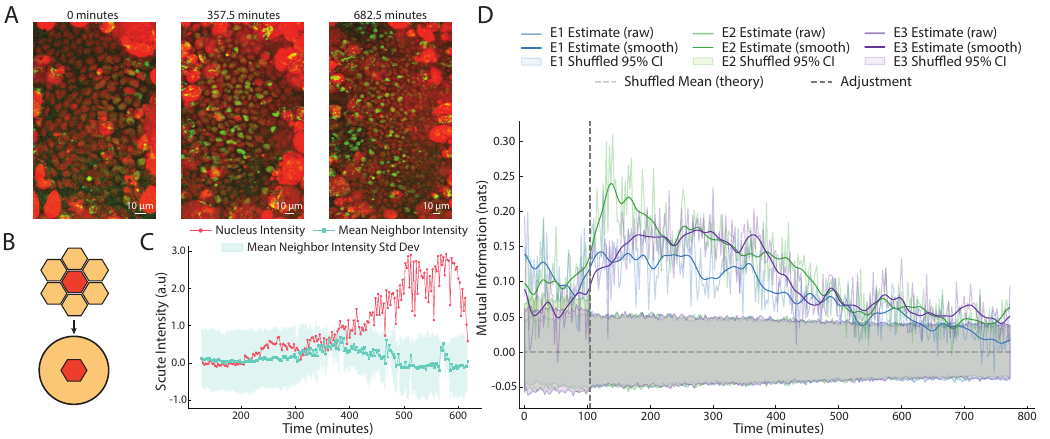}
\caption{Instantaneous mutual information between cells for experimental measurements of \emph{Drosophila} pupal abdomen patterning show non-monotonicity in time. (A) Snapshots of live imaging experiments from Ref.~\cite{Phan2024}, showing a nuclear marker (red) and Scute intensity (green). Maximum intensity projections of a three-dimensional raw image are shown. Throughout, time is measured from the start of the experiment, around 12 hours after puparium formation. (B) We compare the intensity of a cell to the average intensity of its spatial neighbors, which may change with time as cells rearrange and divide. (C) Example Scute intensity trajectory of a future SOP cell compared to its neighbors over the time window when patterning occurs (SM Sec. VII). Scute ultimately decreases once cell commits to SOP fate. (D) Instantaneous mutual information between cells and their neighbors across time and for 3 separate experiments (E1-E3). The raw mutual information time series is overlaid with a temporally smoothed trend (SM Sec. VII). Dotted line indicates discontinuity in the imaging region, and number of segmented cells, due to scope adjustment (SM Sec. VII).
Shuffling cell-neighbor intensity pairs gives a null data set with zero true mutual information, numerical confidence intervals ($95\%$ shown) show values of estimator applied to null data.}
\label{fig:ExpData}
\end{figure*}

To explore our theoretical predictions, we examine recent experimental data of Scute expression during sensory organ patterning in the \emph{Drosophila} pupal abdomen, an example of lateral inhibition through Delta-Notch signaling~\cite{Phan2024}. Directly computing the full mutual information between trajectories remains impractical from finite experimental data due to the high-dimensionality of trajectories~\cite{Reinhardt2023}. Instead, we compute the instantaneous mutual information directly from data without assuming any particular underlying model. We find that this measure varies non-monotonically over the course of cellular patterning, consistent with our theoretical analysis in Section~\ref{sec:Instantaneous}.

We analyze data from Ref.~\cite{Phan2024}, who performed live imaging of sensory organ formation in \emph{Drosophila}  pupal abdomen. In these experiments, the transcription factor Scute, along with a nuclear marker, were endogenously tagged with fluorescent reporters. 
Scute is part of the Delta-Notch feedback loop, it upregulates expression of Delta, and it is indirectly suppressed by the activation of Notch receptors. Thus, Scute serves as a proxy for the ``reaction coordinate'' between the inhibited state and the inhibitor state, with high Scute expression indicating the inhibitor state and low Scute expression indicating the inhibited state. Moreover, the fluorescence intensity of Scute within a nucleus provides a quantitative readout of its expression for each cell simultaneously. During the imaging period, cells coordinate through Delta-Notch signaling to create a somewhat regular pattern of Delta expressing cells, known as Sensory Organ Precursor (SOP) cells, Fig.~\ref{fig:ExpData}A. Errors where two neighboring cells are both SOPs occur around 10$\%$ of the time, but can be corrected through cell rearrangements~\cite{Phan2024}. In total, three live-imaging experiments of this form were performed. Each experiment was imaged for around 12 hours and has around 300 time points each of resolution $660\times900\times24$ pixels$^3$ corresponding to a physical region of around $257.4\times351\times31.9\,\mu m^3$.

A direct application of the mutual information formula is complicated by the fact that cells divide, die, and move within the tissue. Following Ref.~\cite{Phan2024}, we instead take the intensity of Scute in a cell, and compare this to the average intensity of its neighbors, Fig.~\ref{fig:ExpData}B-C. The neighborhood is determined by taking the three-dimensional Delaunay tessellation of the nuclei centroids and retaining only edges that are shorter than $12\mu m$ (SM Sec. VII). Note that this neighborhood can change in time.  
We then apply a difference-of-Gaussian filter to the intensities to remove regional variations in background fluorescence, which could erroneously correlate low Scute intensity cells and their neighbors (SM Sec. VII). Having processed the data, for each time point we have a series of pairs $(u_i(t),v_i(t))$ where $u_i(t)$ is the intensity of the $i^{th}$ cell, and $v_i(t)$ the mean intensity of its neighbors. To compute the mutual information between $u$ and $v$, we apply the  Kraskov–Stögbauer–Grassberger estimator~\cite{Kraskov2004} (SM Sec. IV). Estimating mutual information from finite data is challenging and data from uncorrelated variables can result a non-zero point estimate. To confirm that our estimate of mutual information is statistically significant, we create a null data set by randomly reassigning the identity of each neighbor, essentially creating a set of pairs $(u_i(t), v_{\sigma(i)}(t))$ where $\sigma$ is a random permutation. In this null data set the marginals are preserved but now $u$ and $v$ are approximately independent. For much of the time when patterning is occurring, the mutual information estimation exceeds $95\%$ of estimations computed from the null data set, demonstrating that there is statistically significant information shared between a cell and its neighbors, Fig.~\ref{fig:ExpData}D. We additionally see that the instantaneous information decreases as patterning occurs and this is observed across all experiments, Fig.~\ref{fig:ExpData}D, a potentially misleading feature of instantaneous information that we also observed in the model in Section.~\ref{sec:Instantaneous}.

\section{Discussion}

To explore the underlying principles of decentralized self-organization, we introduced a tractable model of a laterally inhibiting system and demonstrated that it is capable of reaching a target pattern starting from initially identical cells. Our analysis showed that a trade-off exists between patterning accuracy and time taken to pattern, resulting in a Pareto front of optimized solutions with varying error rates. By reframing the model as a stochastic reaction network, we were able to directly compute quantities such as the mutual information and transfer entropies between cell trajectories, revealing how information is transferred between cells. Having optimized for collective patterning speed and error rate, the solutions do not appear to optimize for the flow of information, nor do cells appear individually optimal when considered in isolation. Computing the mutual information between trajectories from data alone remains challenging, and so we explored more tractable information measures, such as the instantaneous mutual information, finding that these quantities may display counterintuitive behavior such as non-monotonicity in time. Finally, we computed such quantities in experimental measurements of Delta-Notch patterning and once again found that instantaneous quantities can be non-monotonic in time as patterning occurs, even as the total dynamic information shared between cells strictly increases in time. Our main focus was to extend the analysis of optimal behavior and information transfer beyond the single cell context to a system of decentralized interacting cells engaged in a self-organization problem.

In development, the starting point of patterning is often inhomogeneous. For example, in addition to Notch-signaling, there is also a Delta pre-pattern in SOP formation in \emph{Drosophila} dorsal thorax~\cite{Corson2017}, allowing for a more regular array of SOP cells. Rather than all cells being initially identical, we could consider them to have some initial pre-patterning containing some useful, but not perfect, information. This would still mean that cells need to coordinate to achieve a target pattern, but presumably could do so faster than without the pre-pattern for the same accuracy requirement. How optimal strategies and information flow change as the information contained in the pre-pattern change remains an open question for future study.

An assumption of our model is that every cell has exactly the same rates as every other cell. In reality, cells are heterogeneous and will differ somewhat in their initial condition as well as their parameters and hence response to a signal~\cite{Schwayer2025, kramar2024}. This initial heterogeneity need not represent a pre-pattern or contain useful information, it could be entirely stochastic. Additional sources of heterogeneity come from the non-regular initial packing of cells,  so that different cells experience different, and dynamic, neighborhoods. Even so, heterogeneity may well enhance the patterning ability of our system~\cite{Phan2024}, by breaking the initial symmetry between cells. It would be interesting to explore how varying amounts of heterogeneity change the optimal speed-accuracy trade-off curve. 

We have focused on a minimal patterning motif, but in principle our framework can be extended through the sampling-based gradient descent scheme (SM Sec. II). Such an extension is needed to investigate how the character of the optimized solutions depends on the specific choices of our modeling framework.  For instance, our framework has  discrete internal states and  our analysis is carried out at a fixed, moderate value of $N$, where stochastic fluctuations are significant. While we do not observe sensitivity to the particular value of $N$ in this regime, we have not systematically explored the $N\to\infty$ limit, or included additional states of signal receiving.
Another modification of the model would be to dynamically generate the cell-cell contact topology through a vertex model, allowing for cell movement, neighbor exchange and tissue deformation to feed back onto the signaling network~\cite{Okuda2018, Glen2019, Bajpai2022, Dullweber2023}. Similarly, one could incorporate cellular growth, division, and death, all of which are observed in SOP formation in \emph{Drosophila}  pupal abdomen~\cite{Phan2024}. When adding such complexity, however,  it becomes challenging if not impossible to numerically solve the master equation or directly compute hitting times and error rates. Nonetheless, the sampling-based approach (SM Sec. II) can, in principle, accommodate these additions as long as the likelihood of each simulation can be computed. The extent to which further physical constraints, such as energetic constraints, robustness, or evolvability, shape the character of the solutions remains an interesting open question.

Although our model reproduces behavior that is observed in real experimental systems, such as the scale of information transfer and non-monotonicity, it is not intended to be a detailed biological model. Relatedly, due to the complexities of gene regulatory networks, rather than modeling every molecular event conceptual progress has been made on understanding cell fate transitions by using  ``gene free'' phenomenological models~\cite{Corson2012, Corson2017, Hajji2025}. Similarly, simple discrete models have proved powerful for exploring the concepts underlying cell fate patterning without explicitly modeling the complex underlying gene regulatory networks~\cite{Smart2023, faldor2025cax, Richardson2024}. Here, we were able to gain insight into optimal patterning and information flow by studying a simplified phenomenological model. Supposing instead, that we wished to precisely calculate information theoretic quantities in a specific biological system. As we have seen, some simple information  measures, such as the instantaneous mutual information shared between cells, can be computed directly from data but they can be misleading. For instance, the instantaneous mutual information can decrease while the system patterns. While it is conceivable that alternative instantaneous information measures exist without these specific issues, any instantaneous quantity will necessarily be less insightful than the trajectory-based mutual information. Estimating full trajectory quantities directly from experimental data remains impractical with existing estimators due to the high dimensionality of trajectories, although this remains an active area of research~\cite{reinhardt2025}. Seemingly, the most promising approach is to build a detailed model, fit this to data, and then compute the information transfer within this model~\cite{Razo2020, zoller2025}. While such computations have not yet been attempted in a  detailed model with multiple cells and feedback, advances in computing transfer entropy rates~\cite{das2024} might render such computations  possible, albeit numerically expensive. With the ability to compute information transfer in multicellular systems, it will be possible to ask precise information theoretic questions, such as along which direction is information maximally transferred in, or how does information transfer depend on the cell contact topology? In this sense, an understanding of how information propagates in a self-organizing system requires a method to precisely compute information theoretic quantities.

Throughout, for computational simplicity, we bound all rates above by 1, which essentially constrains the fastest time scale in our system. Therefore, strategies which create a high accuracy pattern at the cost of taking a long time, have a time scale separation between the patterning time scale and the time scale of the internal dynamics or signaling. In other words, accurate self-organized patterning requires transcriptional and signaling dynamics to occur on a faster time scale than patterning. For many developmental processes, this may well be a limiting constraint. The time scale of transcriptional and translational dynamics is on the order of tens of minutes~\cite{LAMMERS2020}, which would constrain accurate self-organized patterning to be on the order of hundreds of minutes, which is indeed a typical patterning time scale. While we prove that such a trade-off exists in our model, we expect that this is a generic effect any cell faces in a system with a noisy mechanism of communication: spend longer acquiring information about the environment or act faster at the risk of making an incorrect patterning choice. Indeed, in a continuous model of lateral inhibition, a separation of time scales between the time scale over which cells commit to an SOP fate and the time scale at which a cell can inhibit its neighbors was identified as necessary for accurate patterning~\cite{Barad2010}. Beyond lateral inhibition, there are numerous contexts in developmental biology where the duration of a signal, and not just the strength, determines whether cells commit to a particular fate~\cite{Dessaud2007, MARSHALL1995}.
Perhaps the required time scale separation is one reason why pre-patterning and exogenous signals are often used in combination with self-organization: self-organization of identical cells can achieve accurate patterning, but such approaches take inordinately long compared to self-organization with pre-patterning.

\section*{Acknowledgments}
We are grateful to François Schweisguth for sharing experimental data with us and Minh Son Phan for image processing advice. We thank Henry Mattingly, Matthew Smart, and David Denberg for helpful discussions.
A.T. acknowledges the support of the Summer@Simons program.
The authors also acknowledge the MIT Office of Research Computing and Data for providing high performance computing resources that have contributed to the research results reported within this paper. This research received support through Schmidt Sciences, LLC (J.D.), the MathWorks Professorship
Fund (J.D.).
\appendix
\section*{Appendix}
\section{No communication strategy\label{A:NoCom}}
Without the ability to communicate, each cell can only adjust its internal state to reach either absorbing state, 0 or $N$. The specific choice of transition rates affects the time it takes to reach an absorbing state, but whatever the choice of rates there will be a probability $q$ of reaching $0$ first and a probability $1-q$ of reaching $N$ first. Since each cell has to have the same transition rates, and without communication the cells are independent, the probability that exactly one cell reaches $N$ and that two cells reach $0$ is $3q^2(1-q)$. Maximizing over $q$, we find $q=2/3$ and the probability is $4/9$.

\section{Error bounds the average time to pattern\label{A:Bound}}
Consider a system of $M\geq3$ cells and a set of good terminal states where for every good terminal state the number of cells in state $N$, or $n_d$, is some fixed number, $0 < n_d < M$. It need not be the case that any terminal configuration with $n_d$ cells in state $N$ is a good state, only that all good states have $n_d$ cells in state $N$. Throughout the text, $M=3$ and a good terminal state has exactly one cell in state $N$, or $n_d = 1$. For the more general system, suppose there exists some set of parameters $p$ such that the error rate is $\epsilon$. In this case we can show that the average time taken to reach a terminal state will be at least $M^{-2}(3M+3)^{-1}(1-\epsilon)\epsilon^{-(2N+2)^{-M}}$, demonstrating that precision comes at the cost of time. To prove this, we will need the following lemma.
\begin{lemma}\label{lemma1}
 If there is a non-zero probability of reaching a good terminal state, there is a non-zero probability of reaching a bad terminal state.
\end{lemma}
Let $\mathcal{X}$ be the set of \textit{self-sufficient} single cell states: those from which the cell has a non-zero probability of reaching an absorbing state ($0$ or $N$) without receiving any further signals. This could include states where the signal receiving state must turn off to reach an absorbing state. The absorbing states are trivially included in this set. We need not assume anything about the cell-cell adjacency matrix although it suffices to prove the lemma for a connected graph and apply the result to the disconnected components of a general adjacency matrix.

\paragraph*{Case $1$: $(1,0)\in \mathcal{X}$.} Then each cell can, with non-zero probability, reach an absorbing state without ever receiving any signal from its initial state. Hence there is a path with non-zero probability through which all cells can reach the same absorbing state, which would result in a bad terminal state.

\paragraph*{Case $2$: $(1,0) \notin \mathcal{X}$.}
This means that every cell must receive a signal to progress to an absorbing state. We can trace the steps a cell takes to go from $(1,0)$ to when it first enters $\mathcal{X}$. In doing so, it must be possible for a cell to reach a state $(w,0)$ where $g(w)>0$ before the cell receives a signal. If not, then no cell could ever receive a signal. Also, there must be a self-sufficient state that is reachable from $(1,0)$, $(v,s)\in \mathcal{X}$, for which $g(v)>0$. This is because a successful trajectory starts with no cells being in $\mathcal{X}$, ends with $M$ cells being in $\mathcal{X}$. Since only one cell state changes at a time, there must be a point at which $M-1$ cells occupy a state in $\mathcal{X}$ and one does not. To be successful, this final cell needs to receive a signal to enter $\mathcal{X}$ which requires a non-zero $g$ from one of the cells already in $\mathcal{X}$.

Now consider the following finite sequence of steps:
\begin{itemize}
\item Move cell $1$ into the state $(w,0)$ with $g(w)>0$.
\item Move the remaining cells, starting with the neighbors of cell 1, into the self-sufficient state $(v,s)$ with $g(v)>0$.
\item Move cell $1$ to $(N,s)$ (the exact receiver state does not matter). 
\item Now move the remaining cells into the absorbing state $N$ if it is accessible from $(v,s)$, else move them to $0$.
\end{itemize}
If state $N$ is accessible from $(v,s)$, or if $n_d\neq1$ then this produces a path  with non-zero probability that reaches a bad terminal state. If $N$ is not accessible from $(v,s)$ and $n_d=1$, then consider the modified sequence of steps.
\begin{itemize}
\item Move cell $1$ into the state $(w,0)$ with $g(w)>0$.
\item Move $M-2$ of the remaining $M-1$ cells, starting with the neighbors of cell 1, into the self-sufficient state $(v,s)$ with $g(v)>0$.
\item Move the final cell to $(N,s)$ (the exact receiver state does not matter). 
\item Move cell $1$ to $(N,s)$ (the exact receiver state does not matter). 
\item Now move the remaining cells into the absorbing state $0$.
\end{itemize}
The final state has 2 cells in state $N$ and hence this represents a path to a bad terminal state with non-zero probability.

\begin{proposition}
For an $M$ cell system with error rate $\epsilon < 1$, the average time to reach the terminal states $\tau$ satisfies $\tau \geq M^{-2}(3M+3)^{-1}(1-\epsilon)\epsilon^{-(2N+2)^{-M}}$
\end{proposition}
One can interpret a CTMC as a Markov chain for the sequence of states, along with a set of residence times drawn from an exponential distribution. Specifically, we can write the probability of making a particular transition from $j\to i$, given that we are in state $j$ as $\mathbb{P}(j\to i |j) = W_{ij}/\sum_{k\neq j} W_{kj}$. Since, in our system, only $3M$ transitions are possible from any given state, and each transition rate is bounded by 1, we can conclude that $\mathbb{P}(j\to i |j) \geq W_{ij} / 3M$. For a path $\mathcal{P} = \{\alpha_1 \to \alpha_2 \to \dots \to \alpha_{l+1}\}$, we have that 
\begin{align}
    \mathbb{P}(\mathcal{P}|\alpha_1) &\geq \prod_{k=1}^{l} \left( \frac{W_{\alpha_{k+1} \alpha_{k}}}{3M} \right) \\ \notag
    &\geq \left[\min_{1\leq k \leq l} W_{\alpha_{k+1} \alpha_{k}}/3M \right]^{l},
\end{align}
where by considering solely the Markov chain path, we have effectively marginalized over the possible waiting times. If we take $\alpha_1$ as the initial condition ($\alpha^*$), and $\alpha_{l+1}$ as a bad terminal state, then $\mathbb{P}(\mathcal{P}|\alpha_1) \leq \epsilon$, and hence
\begin{equation}
    3M \epsilon^{1/l} \geq \min_{1\leq k \leq l} W_{\alpha_{k+1}\alpha_k}.
\end{equation}

Since $\epsilon<1$, there is a non-zero probability of reaching a good terminal state and so Lemma~\ref{lemma1} tells us there is a non-zero probability of reaching a bad terminal state. Taking such a path to a bad terminal state, the  smallest rate along this path satisfies $ 0 <W_{\alpha_k,\alpha_{k+1}} \leq 3M\epsilon^{1/l}$. Without loss of generality, we can remove any loops in this path (places where $\alpha_r = \alpha_q$, $r\neq q$) leaving a loop free path with non-zero probability, and since there are at most $(2N+2)^M$ states along this path we have $ 0 <W_{\alpha_k,\alpha_{k+1}} \leq 3M\epsilon^{(2N+2)^{-M}}$. 

Now we progressively prune our rates. In particular, we formally set whatever parameter determines this smallest rate to zero. This parameter will either correspond to exactly one of the $f^\pm,k^-$. If it corresponds to a $k^+$, then all the $g$ values involved must be smaller than $3M\epsilon^{(2N+2)^{-M}}$, and so set them all to zero in the new system. If it is possible to reach a good terminal state in this new system, it is possible to reach a bad terminal state by Lemma~\ref{lemma1}, and hence there exists a path with non-zero probability which, as before. So, we prune again. We repeat this procedure until there are no possible paths to the good terminal state (which must exist as there are only finitely many parameters). At this point, we can conclude that to reach the good terminal state, the system must make a transition where the rate is ``slow''. Typically that means that the transition rate is at most $ 3M\epsilon^{(2N+2)^{-M}}$, although in the case of a $k^+$, it could be the combination of $M-1$ small $g$ values and is at most $ 3(M-1)M\epsilon^{(2N+2)^{-M}}$. The average time to reach a good terminal state, $\tau_g$ satisfies $\tau \geq (1-\epsilon)\tau_g$, and hence if we can bound $\tau_g$ we can bound $\tau$. To bound $\tau_g$ we can ask how long  it takes on average to make one of these slow transitions, given that at least one of these transitions must be made to reach the good state. Since at most $3M$ transitions are possible from any given state (of which, at most $M$ correspond to a $k^\pm$) even if the system was in a state where every possible transition was a slow transition, the rate at which a slow transition occurs would be at most $M^2(3M+3)\epsilon^{(2N+2)^{-M}}$. Thus, this quantity bounds the rate at which a slow transition occurs, whatever state the system is in. Hence, the average time for such a transition to occur is at least 
$M^{-2}(3M+3)^{-1}\epsilon^{-(2N+2)^{-M}}$, and hence in total,
$\tau \geq M^{-2}(3M+3)^{-1}(1-\epsilon)\epsilon^{-(2N+2)^{-M}}$. Although not the tightest bound possible, this still shows that there must exist a speed-accuracy tradeoff and $\tau\to\infty$ as $\epsilon\to0$.

\section{Exact asymptotic system\label{A:ExactAsymp}}
Here we explicitly construct a solution that can achieve an arbitrarily small error rate. Motivated by the appearance of numerically optimized solutions, let us take
$f^+(1,0) = \eta$, $f^+(i,s) = 1$ for $i>1$, $f^+(1,1) = 0$. $f^-(i,0) = 0, \ \forall i$, $f^-(1,1) = 1$, $f^-(i,1) = 0$  for $i>1$, $k^-=0$, $g(0)=g(1)=0$ $g(i)=1$ for $i>1$, $\eta \ll 1$. 

To compute the error rate, suppose that the first transition has occurred and hence a cell has transitioned from $(1,0)\to (2,0)$. At this point it will reach $N$, and contribute a constant signal $g=1$ to its neighbors. Precisely when it transitions or whether it receives a signal is of no relevance for computing the error rate. Each of the next cells has a choice, they could transition to $(2,0)$ with probability $\eta/(1+\eta)$ or transition to $(1,1)$ with probability $1/(1+\eta)$. A transition to $(2,0)$ guarantees the system will reach a bad terminal state, whereas a transition to $(1,1)$ guarantees that cell will eventually reach $(0,1)$ and will not signal to its neighbors. Thus if it does transition to $(0,1)$ in order for the system to reach the good terminal state the remaining cell has to also transition to $(0,1)$ which occurs again with probability $1/(1+\eta)$. Hence the probability of failure is $1 - 1/(1+\eta)^2 \approx 2\eta$. 

The expected time to the first transition is $1/3\eta$. After which time, each cell can only ever make at most $N-1$ (for $N>2$) transitions, each of which have a waiting time with mean at most $1$. For fixed $N$ this gives $\tau = 1/3\eta + O(1)$, or $\tau = 2/(3\epsilon) + O(1)$. This is not  asymptotically optimal, this strategy can be optimized by choosing $f^\pm(i,1)$ more carefully.

\section{Unbounded mutual information\label{A:MI}}
In this section, we will show that the full mutual information between a pair of trajectories in the system in Appendix~\ref{A:ExactAsymp} is unbounded. Let us consider $I(X_0^T;Y_0^T)$ with $X=u_1(t)$, and $Y=u_2(t)$. Using the chain rule for mutual information we have that 
\begin{align} 
    I(X_0^T;Y_0^T,Z) &= I(X_0^T;Y_0^T) + I(X_0^T;Z|Y_0^T), \\\notag
    &= I(X_0^T;Z) + I(X_0^T;Y_0^T|Z),
\end{align}
for any random variable $Z$. Choosing a $Z$ with a finite state space of size $N_Z$ (and the size of this state space is independent of $\eta$), means that the entropy of $Z$, or entropy of $Z$ conditioned on another variable is bounded above by $H(Z) \leq \log N_Z$, and hence any mutual information term between $Z$ and another variable is similarly bounded by $\log N_Z$. Hence, we can determine that as $\epsilon \to 0$,
\begin{align}\notag
    I(X_0^T;Y_0^T) \text{ bounded } \iff& I(X_0^T;Y_0^T,Z) \text{ bounded } \\\notag \iff& I(X_0^T,Z;Y_0^T,Z) \text{ bounded }
    \\ \iff& I(X_0^T;Y_0^T|Z) \text{ bounded. }
\end{align}
For the transition rates in Appendix~\ref{A:ExactAsymp}, the set of possible paths (sans waiting times) from the initial to condition to the final is finite, so set $Z$ to be the random variable representing which path is taken. Further, let us take the path where $u_1$ first transitions from $1\to2$, followed by $s_2$ transitioning from $0\to1$, and then $u_2$ transitioning from $1\to 0$. This particular path has a $O(1)$ probability of occurring, and so if the mutual information conditioned on this path diverges, then the overall mutual information diverges. Further, from the data processing inequality, we can apply any coarse graining function to the trajectories $X_0^T$ and $Y_0^T$ and only decrease the mutual information. Thus, if we reduce the trajectory $X_0^T$ to the first time at which $u_1$ changes, and similarly $Y_0^T$ to the first time at which $u_2$ changes, essentially we are left computing the mutual information, $I(\tau_1:\hat\tau)$, with $\hat\tau = \tau_1+\tau_2+\tau_3$, where all the $\tau_i$'s are drawn from exponential distributions, with rates $3\eta$ for $\tau_1$,  $(3+2\eta)$ for $\tau_2$ and $(3 + \eta)$ for $\tau_3$ (even after conditioning on a path the waiting time distribution is unchanged and reflects the sum of possible transition rates before conditioning). Decomposing the mutual information, 
\begin{align}
I(\tau_1;\hat\tau) &= H(\hat\tau) - H(\hat\tau|\tau_1) \\\notag 
&\geq \max\{H(\tau_1),H(\tau_2),H(\tau_3)\} - H(\tau_2+\tau_3) \\ \notag
&\geq H(\tau_1) - H(\tau_2) - H(\tau_3) 
\\ \notag
&= O(\log1/\eta)
\end{align}
using the fact that $H(X+Y)\geq \max \{H(X),H(Y)\}$, $H(X,Y) \geq H(X+Y)$, $H(X+Y) = H(X) + H(Y)$ if $X$ and $Y$ are independent, and if $X\sim \text{Exponential}(\lambda)$ then $H(X) = 1 - \log \lambda$. Thus the mutual information diverges at least as fast as $O(\log 1/\eta)$.

Intuitively, in this system cell 1 can transition at any point in an $O(1/\eta)$ time. If you observe a transition $1\to0$ in cell $2$ at time $T$, you know that with $O(1)$ probability, cell 1 transitioned at a time $T + O(1)$, narrowing down from an $O(1/\eta)$ uncertainty and thus gaining a significant amount of information. However, this intuition also suggests a strategy that would have a finite amount of mutual information, while still having an arbitrarily small error rate. If we keep $g(i)=1$ for $i>1$ but set $f^-(1,1) = \eta$, $f^+(i,s)=\eta$ for $i>1$, the error rate is unaffected. However, observing a transition $1\to0$ in cell $2$ doesn't narrow down the time at which another cell transitioned all that much, since the cell 2 transition occurs at a time $O(1/\eta)$ after the first transition. If we took the information, $I( (u_1,s_1) ; (u_2,s_2))$ this would again diverge, in some sense in this new system there is divergent mutual information it is just not stored in the internal state. We suspect that the mutual information  $I( (u_1,s_1) ; (u_2,s_2))$ always becomes divergent as the error goes to zero.
\section{\label{app:condInfo_inhibited}Conditional dynamical information between laterally inhibited states}
\begin{figure}[!h]%
\centering
\includegraphics{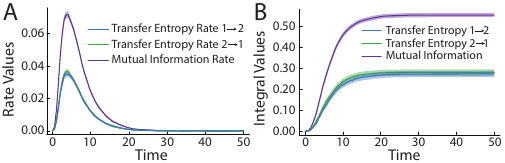}
\caption{Conditioning on final cell fates reveals non-trivial communication between the laterally inhibited states. 
Conditioned mutual information and transfer entropy rates between the inhibited cell $1$ and inhibited cell $2$ (A) as well as the corresponding integrals (B). Throughout, each computation is averaged over $n=10,000$ Monte-Carlo samples, and the shaded regions show $1.96 \times $standard error.}
\label{fig:condinfo_appendix}
\end{figure}

In Section~\ref{sec:InfoTrans}, we showed that successful patterning depends on asymmetric, non-trivial communication among cells. While Fig.~\ref{fig:condInfo} depicts transfer between inhibitor and inhibited cells, Fig.~\ref{fig:condinfo_appendix} illustrates communication among the inhibited cells themselves. Unlike the asymmetric flow involving inhibitor cells, the exchange here is symmetric, reflecting the equivalence of inhibited cells.

\end{document}


\title{Supplementary material -- Collective is different: Information exchange and speed-accuracy trade-offs in self-organized patterning}
\newtheorem{theorem}{Theorem}[section]
\newtheorem{definition}{Definition}
\newtheorem{corollary}{Corollary}[theorem]
\newtheorem{lemma}[theorem]{Lemma}
\newcommand{\Hmap}{\mathcal{H}}
\titlespacing*{\subsection}{0pt}{1em}{0.5em}

\author{Ashutosh Tripathi}
\affiliation{Department of Mathematics, Massachusetts Institute of Technology,
77 Massachusetts Avenue, Cambridge, MA 02139, USA}
\affiliation{Center for Computational Biology, Flatiron Institute, 162 5th Ave, New York, NY 10010, USA}
\author{J\"orn Dunkel}
\affiliation{Department of Mathematics, Massachusetts Institute of Technology,
77 Massachusetts Avenue, Cambridge, MA 02139, USA}
\author{Dominic J. Skinner}
\affiliation{Center for Computational Biology, Flatiron Institute, 162 5th Ave, New York, NY 10010, USA}
\maketitle
\setlength{\parindent}{0pt}
\tableofcontents

\section{Model description}
\label{section1cellulardecisionmaking}
 
Consider a multicellular system with cells $\alpha=1,\cdots,M$. The arrangement of these cells is described by the adjacency matrix $(A_{ij})$. Each cell is described by a tuple, $C_\alpha=(u_\alpha,s_\alpha)$, which consists of a discretized internal state variable $u_\alpha \in \{0,\dots,N\}$, and a binary receiver state variable $s_\alpha \in \{0,1\}$, i.e. the signal was received or not.  The total state of the system is described by $X = (C_\alpha)_{\alpha=1}^{M}$. The underlying system is a continuous time Markov chain $(X_t)_{t\geq0}$. At a time-point, only one cell can change its state (internal or signal receiving) while the other cells remain unchanged. For cell $\alpha$, the following stochastic transitions, with the corresponding rates, can occur:
\begin{align}\label{eq:Reaction}
(u_\alpha, s_\alpha)& \xrightarrow{f^+(u_\alpha, s_\alpha)} (u_\alpha+1,s_\alpha) \\
(u_\alpha, s_\alpha)& \xrightarrow{f^-(u_\alpha, s_\alpha)} (u_\alpha-1,s_\alpha) \\
(u_\alpha, 0)& \xrightarrow{k^+(\alpha)} (u_\alpha,1) \\
(u_\alpha, 1)& \xrightarrow{k^-} (u_\alpha,0)
\end{align}
where 
$$k^+(\alpha)=\sum_{\beta\in\text{nbhd}(\alpha) }A_{\alpha,\beta}g(u_\beta),$$
is the rate at which a cell receives a signal and is a function of the neighboring cells' internal states. The transition rates are specified by 
\begin{itemize}
    \item Two $N \times 2$ matrices, $f^+$ and $f^-$ that describe the rates at which cells increase/decrease their internal state, 
    \item A $(N+1) \times 1$ vector $g$ that describes the intensity at which a cell signals to its neighbors,
    \item A scalar $k^{-}$ that specifies the rate at which a signal turns off, and is independent of the cell state.
\end{itemize}
We assume that each rate is individually bounded above by $1$, effectively setting the fastest time-scale in our model. We have further assumed that the transition rates are the same for all the cells in the system, reflecting the homogeneity of the cells.

These rates parametrically determine a cell's behavior, and in total define a strategy denoted by a $n_p\times 1$ vector $\mathbf{p}$, 
$${\mathbf{p}}=[f^{+},f^{-},g,k^-],$$
where the earlier matrices have been flattened down to a vector, and $n_p=5N+2$. 

These rates define a rate matrix, $Q=(q_{ij})$, where $Q_{ij}$ is the rate of transition from state $j$ to state $i$ for $i\neq j$. For $i=j$, we define $Q_{ii}=-\sum_{j\neq i}Q_{ji}$. For brevity, define $q_i=-Q_{ii}=\sum_{j\neq i}Q_{ji}.$  

Throughout, at $t=0$ we start in the initially homogeneous state $\alpha^*=((1,0),(1,0),(1,0)).$

To enforce boundary conditions we set $f^+(0,s)=0$ and $f^-(N,s)=0$, so if  a cell reaches one of the absorbing states ($0$ or $N$), the cell's internal state is stuck there. Thus, the total number of free parameters is $5N-2$. While in an absorbing state, the receiver state, $s_\alpha$, of a cell can still change. We define terminal states as a state where all cells are in an absorbing state, and no more internal-state dynamics can occur. Mathematically, these terminal states $\mathcal{T}$ are defined as
$$\mathcal{T} = \{X=((u_\alpha,s_\alpha))_\alpha \mid \forall \alpha,\,u_\alpha \in \{0,N\}\,, s_\alpha\in\{0,1\} \}.$$ This set consists of terminal states that achieve a target pattern (``good'') and those that do not (``bad''). As a result, $\mathcal{T}=\mathcal{T}^G \bigsqcup \mathcal{T}^B.$ The superscripts $G$ and $B$ indicate the ``good'' and ``bad'' terminal sets. For a system with $M=3$ cells, where all cells neighbor each other (as in the main text), we choose the desirable patterns to be those where only one cell is the inhibitor, and the remaining two cells are inhibited, i.e. 
$$\mathcal{T}^G=\{X=((u_1,s_1),(u_2,s_2),(u_3,s_3))|(u_1,u_2,u_3) \in \{(0,0,N), (0,N,0), (N,0,0)\}, \,s_1,s_2,s_3 \in\{0,1\}\}.$$

\section{Optimal patterning}
In order to define the optimization problem, we first define the average time taken to pattern as well as the error rate.

Let $\tau_j$ be the mean hitting time of the process hitting any state in $\mathcal{T}$ beginning at $j.$ Using convention in \cite{Norris_1997}, note that for $j\notin \mathcal{T}$
\begin{equation}\tau_j=\frac{1}{q_j}+\frac{1}{q_j}\sum_{k\neq j}Q_{kj}\tau_k.
\label{eq_hitting_eq1}
\end{equation}
which can be rearranged to $$0=1+\sum_{k}Q_{kj}\tau_k.$$ More concretely, mean hitting times are defined as a consequence of the following theorem~\cite{Norris_1997}
\begin{theorem}[Expected Hitting Time]
\label{thmexphittime}
Suppose $q_i >0$ for all $i\notin \mathcal{T}$. The vector of mean hitting times $\boldsymbol{\tau} = (\tau_i)_{i\in S}$ is the minimal non-negative solution to the system of linear equations
\begin{equation*}
\begin{cases}
\tau_i = 0 & \text{for } i \in \mathcal{T}, \\
1+\sum_{j} Q_{ji} \tau_j = 0& \text{for } i \notin \mathcal{T}.
\end{cases}
\end{equation*}
\end{theorem}
Specifically, if any other non-negative solution $\tilde{\tau}_i$ exists, then $\tau_i \leq \tilde{\tau}_i$ for all $i$.

To define the error rate of the system, we use that hitting probabilities, $h_i=\mathbb{P}_i[\text{Eventually reach state in } \mathcal{R} | \text{in state } i]$, for any set of states $\mathcal{R}$ satisfy the following theorem~\cite{Norris_1997}:
\begin{theorem}[Hitting Probabilities]
\label{thmexphitprob}
Suppose $q_i >0$ for all $i\notin \mathcal{R}$. The vector of hitting probabilities $\mathbf{h} = (h_i)_{i \in S}$ is the minimal non-negative solution to the system of linear equations
\begin{equation*}
\begin{cases}
h_i = 1 & \text{for } i \in \mathcal{R}, \\
\sum_{j} Q_{ji} h_j = 0& \text{for } i \notin \mathcal{R}.
\end{cases}
\end{equation*}
\end{theorem}
In our system, we want to define the error as the probability of hitting a bad state but not a good state. We adjust the above formulation to solve the probability of never hitting $\mathcal{T}^G$ but hitting $\mathcal{T}^B$. In particular 
\begin{theorem}[Adjusted Hitting Probabilities]
\label{thmadjexphitprob}
Suppose $q_i >0$ for all $i\notin \mathcal{T}$. The vector of hitting probabilities $\boldsymbol{\epsilon} = (\epsilon_i)_{i \in S}$ is the minimal non-negative solution to the system of linear equations
\begin{equation*}
\begin{cases}
\epsilon_i = 0 & \text{for } i \in \mathcal{T}^G, \\
\epsilon_i = 1 & \text{for } i \in \mathcal{T}^B, \\
\sum_{j} Q_{ji} \epsilon_j = 0& \text{for } i \notin \mathcal{T}.
\end{cases}
\end{equation*}
\end{theorem}

To find the strategy which minimizes the average time to reach a terminal state, with error at most $\epsilon_{tol}$, starting from an initial state $\alpha^*$, we solve 
\begin{align*}
    &\inf_{\mathbf{p},\mathbf{\tau},\mathbf{\epsilon}}\,\tau_{\alpha^*}\\
    \text{subject to: }& \nonumber\\
    &\boldsymbol{{\tau}}\geq 0, \\
    &0\leq \mathbf{p},\boldsymbol{\epsilon} \leq 1, \,\\
    &1+\sum_{j\notin \mathcal{T}} Q_{ji} \tau_j = 0 \quad\text{for } i \notin \mathcal{T}\\
    &\sum_{j} Q_{ji} \epsilon_j = 0 \quad\text{for } i \notin \mathcal{T}\\
    & \epsilon_{\alpha^*}\leq\epsilon_{tol},\\
    & {\tau}_i = 0  \quad\text{for } i \in \mathcal{T}\\
    & \epsilon_i = 0  \quad\text{for } i \in \mathcal{T}^G\\
    & \epsilon_i = 1  \quad\text{for } i \in \mathcal{T}^B,
\end{align*}
where we suppress the dependence $Q_{ij}=Q_{ij}(\mathbf{p}).$ We have expanded our domain to minimize over $\boldsymbol{\tau}$ and $\boldsymbol{\epsilon}$ as well:  using the earlier theorems, the constraints encode the information for how $\boldsymbol{\tau}$ and $\boldsymbol{\epsilon}$ depend on the instruction set $\mathbf{p}$.

The formulation above implicitly avoids the cases where $q_i=0$ for a non-terminal state i.e. the cases where theorem  \ref{thmexphittime} and theorem \ref{thmexphitprob} fail. Indeed, if $q_i=0$ for some $i\notin\mathcal{T}$, then 
$$1-q_i\tau_i+\sum_{j\neq i}Q_{ji}\tau_j=1-0\cdot\tau_i+\sum_{j\neq i}0\cdot\tau_j=1\neq0,$$
which does not satisfy the above optimization constraints. Further, any parameter set within $0<\mathbf{p} \leq 1$ is guaranteed to result in a unique and finite $\boldsymbol{\tau}$ and $\boldsymbol{\epsilon}$. Thus, the interior of the optimization only searches over systems where the non-terminal states are transient in nature and there is always a non-zero probability of leaving any such state ($q_i>0$) and a non-zero probability of hitting one of the terminal states from that state $(\tau_{i}<\infty).$

Notice that the rate matrix is, by definition, linear in the parameters. Thus, the problem above is a Quadratically Constrained Quadratic Program (QCQP). These are generically NP-hard to solve~\cite{Nocedal2006, Luo2010}, and we must resort to finding local minima in computationally tractable cases using known optimization techniques.

To solve the optimization problem, we use interior point method~\cite{Nocedal2006}. We perform the minimization across $\sim 100$ initial conditions, including random initial parameters. For random initializations, the solver often converges to an infeasible solution or fails to converge to a solution in $\sim 10,000$ iterates. As a result, we also use the converged solutions from the random initializations, alongside solutions for nearby error values (that have been computed) to provide us with a reasonable initial guess. This method gives us a locally optimal solution, and an upper bound on the global optimal solution. However, repeated runs across multiple initializations and for multiple error rates result in a smooth Pareto front, suggesting that the loss landscape is reasonable and that the global optimum is achievable. We further verify the validity of these minima by using the limited-memory Broyden–Fletcher–Goldfarb–Shanno algorithm (L-BFGS)~\cite{Nocedal2006}.

\subsection{Choice of \emph{M} and \emph{N}}
We focus on a fully interacting triad of cells, because it is the minimal configuration that requires decentralized symmetry breaking with a non-trivial neighborhood. As each cell has $2(N+1)$ states, an $M$ cell system would have $2^M(N+1)^M$ different states. While Gillespie simulations of much larger systems are possible, optimization as we have done for $M=3$, rapidly becomes numerically infeasible. The choice of $N=6$ and $M=3$ (which leads to $14^3=2744$ states) results in a system which is fine-grained enough to capture sharp response thresholds, whilst also allowing the optimal patterning and information computations to be tractable.

\subsection{A sampling-based method for finding solutions}
As noted above, building and inverting the full rate matrix quickly becomes infeasible as $(M,N)$ grows because the number of states is $(2(N+1))^M$. To explore larger systems, we require an approximate solution to our problem, and ideally one that converges to the true solution in some limit. To do so, we will use a Monte-Carlo estimate of the patterning times $\tau_{\alpha^*}$ and patterning errors $\epsilon_{\alpha^*}$ computed using $N_{s}$ independent Gillespie simulations, or stochastic simulation algorithm (SSA), and use this estimate, together with an estimate of the derivative, to optimize the parameters $\mathbf p=[f^+,f^-,g,k^-]$ .

For each SSA simulation, we simulate the system until a horizon $T_{\mathrm{hor}}$. Let $T$ be the hitting time of the terminal set $\mathcal T=\mathcal T^G \sqcup \mathcal T^B$ for a given simulation, assuming that it reaches a terminal state. If the system never enters a terminal state in the simulation, set $T=T_{\mathrm{hor}}$. For this simulation we record 
$$b=\mathbf 1[X_{T}\in\mathcal T^B].$$
Once we have $N_s$ simulations, we take Monte-Carlo approximants for the mean patterning time and error, 
$$\hat\tau_{\alpha^*}=\frac{1}{N_s}\sum_{i=1}^{N_s}T_{_i},\quad \hat\epsilon_{\alpha^*}=\frac{1}{N_s}\sum_{i=1}^{N_s}b_{_i},$$ which follow from the fact that $\tau_{\alpha^*}=\mathbb{E}[T]$ and $\epsilon_{\alpha^*}=\mathbb{E}[b]$ for $T_{hor}\rightarrow \infty$. Since the probability of a trajectory not reaching a terminal state decays exponentially as $T_{hor}\to\infty$, we simply set $T_{hor}$ to be large enough that in practice all of our $N_s$ samples reach a terminal state. Thus, instead of the original problem, we solve the approximate problem 
\begin{align*}
    &\min_{\mathbf{p}}\,\hat\tau_{\alpha^*}\\
    \text{subject to: }& \nonumber\\
    &0\leq \mathbf{p},\boldsymbol{\epsilon} \leq 1, \,\\
    & \hat\epsilon_{\alpha^*}\leq\epsilon_{tol}.\\
\end{align*}

For the original constrained problem
\[
\min_{\mathbf p} \;\tau_{\alpha^*}(\mathbf p)
\quad\text{subject to}\quad
\epsilon_{\alpha^*}(\mathbf p)\le \epsilon_{\mathrm{tol}},
\]
we found the optimal solution to lie on a monotonic Pareto front and hence imposing an equality constraint, $\epsilon_{\alpha^*}=\epsilon_{tol}$ results in the same optimized solutions. Hereafter, we work with this adjusted problem. We solve this problem using gradient descent with an augmented Lagrangian~\cite{Nocedal2006, Deng2025}
\[
\mathcal{L}_\rho(\mathbf p,\lambda)
=\tau_{\alpha^*}(\mathbf p)
+\lambda\big(\epsilon_{\alpha^*}(\mathbf p)-\epsilon_{\mathrm{tol}}\big)
+\tfrac{\rho}{2}\big(\epsilon_{\alpha^*}(\mathbf p)-\epsilon_{\mathrm{tol}}\big)^2,
\quad \lambda\in \mathbb{R},\ \rho>0.
\]
To do gradient descent and dual ascent, we need to  estimate the gradient $g_\mathbf{p}=\nabla_\mathbf{p}\mathcal{L}$, and hence we need a way to estimate the gradients of  $\tau_{\alpha^*}$ and $\epsilon_{\mathrm{tol}}$ with respect to the parameters $\mathbf p$.

Consider a function of the SSA simulation path $\omega$, $f(\omega)$, and suppose we want to compute the gradient of $\mathbb{E}[f(\omega)]$. Using the log-derivative trick~\cite{Mohamed2020},
\begin{align*}
    \nabla_\mathbf{p} \mathbb{E}[f(\omega)] &= \nabla_\mathbf{p} \int d\omega\, f(\omega) p(\omega;\mathbf{p})\\
    &=\int d\omega\, f(\omega) \nabla_\mathbf{p} p(\omega;\mathbf{p})\\
    &=\int d\omega\, f(\omega)p(\omega;\mathbf{p}) \nabla_\mathbf{p} \log p(\omega;\mathbf{p})=\mathbb{E}[f(\omega)\nabla_\mathbf{p} \log p(\omega;\mathbf{p})].
\end{align*}
To evaluate this expectation with a Monte-Carlo average, for each path $\omega$, we need to calculate the score function $S(\omega;\mathbf{p})=\nabla_\mathbf{p} \log p(\omega;\mathbf{p})$. For a continuous time Markov process (CTMC), we can determine the form of the score function exactly. Consider an arbitrary CTMC with the transition rates $\{a_r(\mathbf{x};\mathbf{p})\}$ and $\lambda(\mathbf{x};\mathbf{p})=\sum_{r}a_r(\mathbf{x};\mathbf{p})$. Then, for a path $\omega=\{(t_{k},\mathbf{x}_{t_k}, a_{r_k})\}_{k=1}^{K}$ which contains $K$ transitions at time $t_k$ to states $\mathbf{x}_{t_k}$ due to reaction $a_{r_k}$, the log probability density of the path measure $p(\omega;\mathbf{p})$ is
$$\log p(\omega;\mathbf{p}) = \sum_{k=1}^K\log a_{r_k}(\mathbf{x}_{t_k^-};\mathbf{p})-\int_{0}^{T_{hor}}\lambda(\mathbf{X}_t;\mathbf{p})\,dt.$$ Since $\lambda$ is constant outside of transitions, the log path measure can be simplified as  
$$\log p(\omega;\mathbf{p}) = \sum_{k=1}^K\log a_{r_k}(\mathbf{x}_{t_k^-};\mathbf{p})-\sum_{k=1}^{K}\Delta t_{k}\lambda(\mathbf{x}_{t_k^-};\mathbf{p}),$$
where $\Delta t_k=t_{k+1}-t_k$ with $t_{K+1}=T_{hor}$. Thus, the score function is 
$$S(\omega;\mathbf{p})=\sum_{k=1}^K\nabla_\mathbf{p}\log a_{r_k}(\mathbf{x}_{t_k^-};\mathbf{p})-\sum_{k=1}^{K}\Delta t_{k}\nabla_\mathbf{p}\lambda(\mathbf{x}_{t_k^-};\mathbf{p}) = \sum_{k=1}^K\nabla_\mathbf{p}\log a_{r_k}(\mathbf{x}_{t_k^-};\mathbf{p})-\sum_{k=1}^{K}\Delta t_{k}\left(\sum_{r}\nabla_\mathbf{p}a_r(\mathbf{x}_{t^-_k};\mathbf{p})\right).$$

Then, using that $\tau_{\alpha^*}=\mathbb{E}[T]$ and $\epsilon_{\alpha^*}=\mathbb{E}[b]$ (in the $T_{hor}\to\infty$ limit), we can deduce that $g_\tau=\nabla_{\mathbf{p}}\tau_{\alpha^*} = \mathbb{E}[TS(\omega;\mathbf{p})]$ and $g_\epsilon=\nabla_{\mathbf{p}}\epsilon_{\alpha^*} = \mathbb{E}[bS(\omega;\mathbf{p})]$. Then $g_\mathbf{p}=\mathbb{E}[TS(\omega;\mathbf{p})]+\lambda\mathbb{E}[bS(\omega;\mathbf{p})] +\rho(\mathbf{\epsilon}_{\alpha^*}-\epsilon_{tol})\mathbb{E}[bS(\omega;\mathbf{p})].$

However, these Monte-Carlo estimators for the gradients can have large variance and hence we use control variates to reduce the variance~\cite{Mohamed2020}. Consider the score function $S\in \mathbb{R}^d$. As earlier, our aim is to calculate $\nabla_{\mathbf{p}} \mathbb{E}[f(\omega)]=\mathbb{E}[f(\omega)S(\omega;\mathbf{p})].$ One way to introduce control variates is to use constant baseline $c$ so that we compute $\mathbb{E}[h]$ where $ h = (f(x)-c)S(\omega;\mathbf{p})$ instead. Due to $\mathbb{E}[S]=0,$ we know that this modified approximant is still an unbiased estimator. We aim to find the constant $c$ that minimizes the total variance, $\phi(c)= \mathrm{tr}(\mathrm{Cov}(h)) = \mathbb{E}[\|h\|^2]-\|\mathbb{E}[fS]\|^2=\mathbb{E}[(f-c)^2\|S\|^2]-\|\mathbb{E}[fS]\|^2.$ Expanding, we obtain $\phi(c)=-2c\mathbb{E}[f\|S\|^2]+c^2\mathbb{E}[\|S\|^2] + const.$, from which we see that the \textit{optimal baseline}, $c^*$, is $$c^*=\frac{\mathbb{E}[f\|S\|^2]}{\mathbb{E}[\|S\|^2]}.$$ We approximate the expectations in the numerator and denominator, with a plug-in estimator $$\hat{c}^* = \frac{\frac{1}{N_s}\sum_{i=1}^{N_s}f_i\|S_i\|^2}{\frac{1}{N_s}\sum_{i=1}^{N_s}\|S_i\|^2}.$$ Due to the strong law of large numbers and the continuous mapping theorem, the plug-in estimator asymptotically converges to the optimal baseline $\hat{c}^*\xrightarrow{\mathrm{a.s}}c^*$.

As a result, we estimate $g_\tau$ and $g_\epsilon$ using an optimal constant baseline (as above). Given $N_s$ independent SSA paths, their score vectors $S_i\in\mathbb R^{5N-2}$, terminal times $T_i$ and bad-state indicators $b_i$, we set
\[
c_T = \frac{\sum_{i=1}^{N_s} T_i \|S_i\|^2}{\sum_{i=1}^{N_s} \|S_i\|^2},
\qquad
c_b = \frac{\sum_{i=1}^{N_s} b_i \|S_i\|^2}{\sum_{i=1}^{N_s} \|S_i\|^2},
\]
and use
\[
\hat{g}_\tau
=\frac1{N_s}\sum_{i=1}^{N_s} (T_i-c_T)\,S_i,
\qquad
\hat{g}_\epsilon
=\frac1{N_s}\sum_{i=1}^{N_s} (b_i-c_b)\,S_i.
\]
The stochastic gradient of the augmented Lagrangian then reads
\[
\hat{g}_{\mathbf{p}}
= \hat{g}_\tau
+\bigl(\lambda+\rho(\hat\epsilon_{\alpha^*}-\epsilon_{\mathrm{tol}})\bigr)
\hat{g}_\epsilon.
\]

\begin{algorithm}[t]
\caption{Projected AMSGrad to solve the approximate patterning problem}
\label{algo:SGD}

\KwIn{Initial parameters $\mathbf p_0\in[0,1]^{5N-2}$, dual $\lambda_0$ (default $\lambda_0=0$), penalty $\rho>0$, error tolerance $\epsilon_{\mathrm{tol}}$, SSA simulation horizon $T_{\mathrm{hor}}$, number of independent SSA simulations per iteration $N_s$, maximum iterations $N_{\mathrm{SGD}}$,
AMSGrad hyperparameters $(\eta_{\max},\eta_{\min},T_0,T_{\mathrm{mult}},\beta_1,\beta_2)$}
\KwOut{Locally optimal parameters $\mathbf{p}^*$}

\For{$t\gets1$ \KwTo $N_\mathrm{SGD}$}{
    \tcp{A single SGD iteration}
    Run $N_s$ SSA trajectories under $\mathbf p_t$ up to $T_{\mathrm{hor}}$; collect $\{T_i,b_i,S_i\}_{i=1}^{N_s}$.\;
    \tcp{Monte-Carlo estimate}
    $\hat\tau \gets \frac{1}{N_s}\sum_i T_i$,\quad $\hat\epsilon \leftarrow \frac{1}{N_s}\sum_i b_i$.\;
    \tcp{Control variate optimal baseline estimates}
    $c_T \gets \dfrac{\sum_i T_i\|S_i\|^2}{\sum_i \|S_i\|^2}$,\quad
    $c_b \gets \dfrac{\sum_i b_i\|S_i\|^2}{\sum_i \|S_i\|^2}$.\;
    \tcp{Gradients of time and error}
    $\hat{g}_\tau \gets \dfrac{1}{N_s}\sum_i (T_i-c_T)S_i$,\quad
    $\hat{g}_\epsilon \gets \dfrac{1}{N_s}\sum_i (b_i-c_b)S_i$.\;

    \tcp{Augmented-Lagrangian gradient}
    $\gamma_t \gets \lambda_t + \rho(\hat\epsilon-\epsilon_{\mathrm{tol}})$\;
    $\hat g_\mathbf{p} \gets \hat{g}_\tau + \gamma_t\,\hat{g}_\epsilon$.\;

    \tcp{AMSGrad moment updates}
    $m_t \gets \beta_1 m_{t-1} + (1-\beta_1)\hat g_\mathbf{p}$\;
    $v_t \gets \beta_2 v_{t-1} + (1-\beta_2)\hat g_\mathbf{p}^2$\;
    $\hat v_t \gets \max(\hat v_{t-1}, v_t)$\;
    $\hat m_t \gets m_t/(1-\beta_1^t)$,\quad $\hat v_t \leftarrow \hat v_t/(1-\beta_2^t)$\;
    
    \tcp{Cosine-annealed learning rate, as described in~\cite{Loshchilov2017}}
    $\eta_t \gets \text{CosineLR}(t;\eta_{\max},\eta_{\min},T_0,T_{\mathrm{mult}})$\;

    \tcp{Projected primal update, projecting into the box $\mathcal{B}=[0,1]^{5N-2}$}
    \tcp{need a small parameter $\varepsilon$ to regularize possible division by zero}
    $\varepsilon \gets 10^{-12}$
    $\mathbf p_{t+1}\leftarrow \Pi_{\mathcal{B}}\bigl(\mathbf p_t - \eta_t\,\hat m_t/(\sqrt{\hat v_t}+\varepsilon)\bigr)$\;
    
    \tcp{Dual ascent}
  $\lambda_{t+1}\leftarrow  \lambda_t + \rho\,(\hat\epsilon-\epsilon_{\mathrm{tol}})$.\;
}

\KwRet{$\mathbf{p}^*=\mathbf{p}_{t+1}$}\;
\end{algorithm}

For our problem, for a given $(M,N)$ and cell topology adjacency matrix $(A_{ij})$, we have $5N-2$ parameters due to the absorbing fates, i.e. $\mathbf{p}\in \mathcal{B}=[0,1]^{5N-2}$. We use standard constrained projected stochastic gradient descent (SGD), where in our case the stochasticity arises from the Monte-Carlo sampling, and at each step we project the parameters $\mathbf{p}$ back into the box. This method, as with any gradient descent, is sensitive to the initial parameter guess. In all results shown, the primal variable $\mathbf p$ is updated using an AMSGrad optimizer~\cite{Reddi2019} with a cosine–annealed learning-rate schedule~\cite{Loshchilov2017}. We maintain first and second moment estimates $(m_t,v_t)$ of the stochastic gradients, use the element-wise maximum of the second moments as in AMSGrad, and with the current learning rate $\eta_t$ (obtained using the suggested routine in~\cite{Loshchilov2017}):
\[
\mathbf p_{t+1}
=\Pi_\mathcal{B}
\Bigl(
\mathbf p_t - \eta_t\,\frac{m_t}{\sqrt{\widehat v_t}+\varepsilon}
\Bigr),
\]
where $\Pi_\mathcal{B}$ denotes projection onto the relevant hypercube.
The dual variable is updated by
$\lambda_{t+1} =\lambda_t + \rho (\hat\epsilon_{\alpha^*}^{(t)}-\epsilon_{\mathrm{tol}}).$ The overall algorithm is summarized in Algorithm~\ref{algo:SGD}. 

We use $N_s=500$–$1000$ SSA simulations per iteration, with each simulation being run for $T_{\mathrm{hor}}=300$ time units. We perform a short warm-start phase followed by a longer run which runs for a maximum of $N_{SGD}=100,000$ iterations, with early stopping based on mean feasibility of the error constraint and approximate stabilization of the mean hitting time in a sliding window of recent iterates. 

\begin{figure*}[t]
    \centering
    \includegraphics{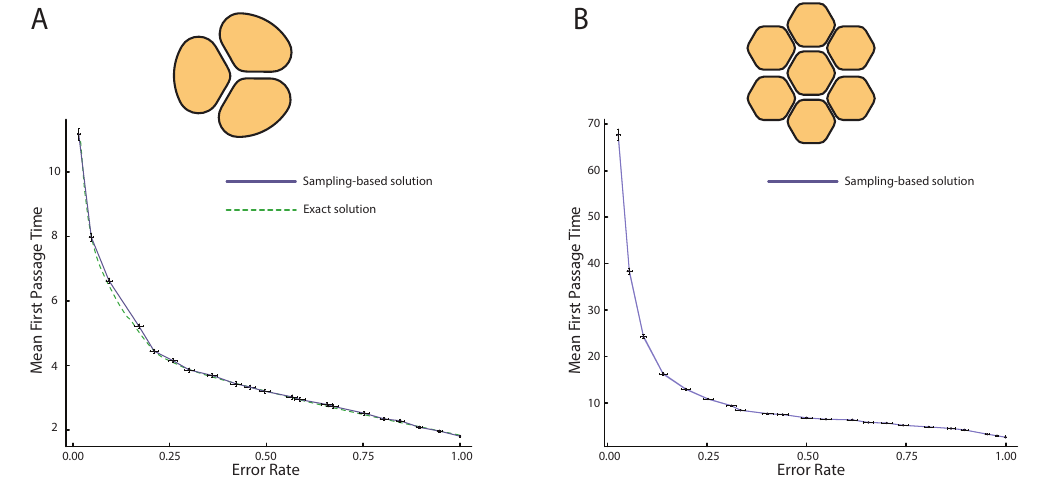}
    \caption{Sampling-based optimization method yields the same qualitative Pareto front. (A) For the three-cell system, as in the main text, the approximate Pareto front (purple, solid) recovers the same Pareto front obtained exactly from the interior point method (green, dashed). (B) The approximate solver can find the Pareto front for larger systems, such as the seven-cell system demonstrated here, resulting in a similarly shaped Pareto front. For the approximate solutions, $n=10,000$ simulations were used to find the mean and variance for the figures, and the error bars are $1.96\times$ the standard deviation.}
    \label{fig:SGD}
\end{figure*}

Fig~\ref{fig:SGD}(A) shows the result for $(M,N)=(3,6)$, the model considered in the main text, obtained from SGD and compares it to the exact results obtained earlier through the interior point method. As expected, the sampling based method does a good job at approximating the true Pareto front.

\subsection{A seven-cell model of patterning}
\begin{figure*}[t]
    \centering
    \includegraphics{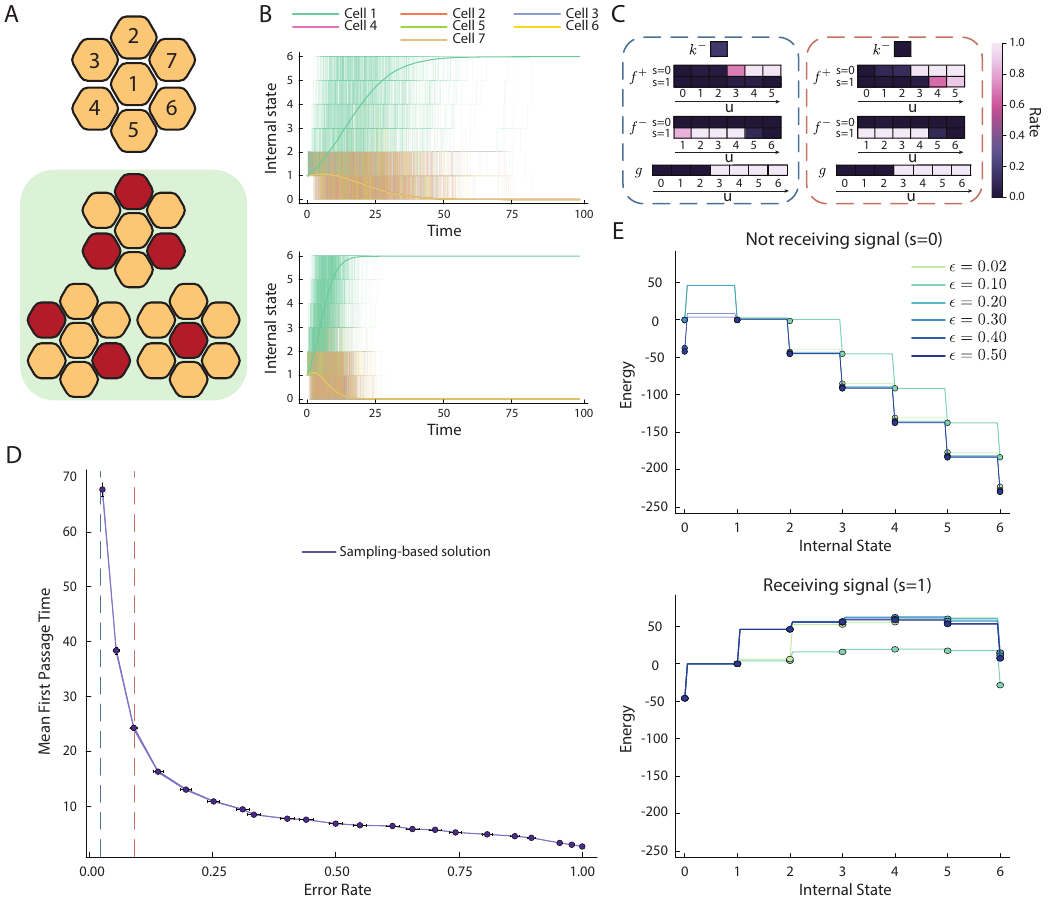}
    \caption{Comparison of optimal strategies for the seven-cell case. (A) The seven cells are arranged to form a hexagonal tile, and the cell numbers are assigned as in the schematic (left). We classify target patterns into three classes: (i) a single central inhibitor, (ii) three alternating inhibitors on the outer ring, and (iii) two opposite inhibitors on the outer ring. Accounting for rotational symmetry, this yields six distinct “good” terminal states. (B) Average internal state for each cell conditioned on a successful trajectory where the inhibitor state was reached by cell 1 (the central cell), for $\epsilon=0.02$ (left) and $\epsilon=0.10$ (right). 1000 stochastic realizations are also shown. (C) Optimized model parameters for $\epsilon=0.02$ (top) and $\epsilon=0.10$ (bottom). (D) The approximate Pareto front obtained using the sampling-based optimization approach was computed using $10,000$ Gillespie, with error bars showing $1.96\times$standard error. (E) Energy landscapes for the approximate optimal strategies show similar qualitative trends as the optimal parameters for the three-cell model in the main text.}
    \label{fig:SevenCell}
\end{figure*}

With this new sampling-based method, we can extend our analysis to larger systems. Going beyond the three-cell system studied elsewhere in the text, here we consider a model of seven cells, representing an asymmetric two–dimensional epithelial region with a non-trivial neighborhood structure. In contrast to the fully connected three cell motif in the main text, each outer cell now interacts with the central cell and its two ring neighbors, while the central cell interacts with all six outer cells (Fig.~\ref{fig:SevenCell}A). As the number of states is $14^7= 105,413,504,$ it is infeasible to directly solve the patterning problem with standard interior-point implementations. For this geometry we therefore rely exclusively on the sampling–based SGD scheme described in the previous subsection. Fig~\ref{fig:SGD}(B) shows the approximate Pareto front found for this larger $(M,N)=(7,6)$ system. 

We define the target set $\mathcal{T}^G$ by grouping the terminal configurations into the three symmetry classes shown in Fig.~\ref{fig:SevenCell}A: (i) a single central inhibitor, (ii) three alternating inhibitors on the outer ring, and (iii) two opposite outer inhibitors. Accounting for rotations, these classes comprise six distinct ``good'' terminal states. This choice mirrors the three–cell case, where good patterns are those with a single inhibitor and two inhibited cells, but now allows multiple distinct patterns in which the inhibitor cells can be either the central cell or cells on the ring and no two inhibitor cells are in contact.
Panels~\ref{fig:SevenCell}B–C illustrate optimal strategies at two representative points on the approximate Pareto front, for $\epsilon = 0.02$ (left/top) and $\epsilon = 0.10$ (right/bottom). As in the three–cell system, low error requires a slower, more graded separation between inhibitor and neighbors, whereas relaxing the error constraint allows the system to commit much more rapidly. The corresponding rate functions in Fig.~\ref{fig:SevenCell}C exhibit similar qualitative structure as for $M=3$.

\subsection{Landscape interpretation of transition rates}
Given the optimal rates $f^\pm, g,k^-$, we would like an intuitive way to interpret this particular patterning strategy. We also want to make a connection to phenomenological models of cell fate patterning which invoke a Waddington landscape-like metaphor, made mathematically precise~\cite{Corson2012, Corson2017}.
To do so, for each cell and given the cell's signal receiving state is $s$, we will visualize and interpret the ``landscape'' that this cell is navigating.

First note that for a CTMC with rate matrix $\mathbf{W}$, that is time reversible ($W_{ij} > 0 \implies W_{ji} > 0$), the rate matrix can be parametrized as 
\begin{equation}
    W_{ij}=e^{-\left(B_{ij}-E_j-\frac{F_{ij}}{2}\right)},
\end{equation}
where $E_j$ is a vertex parameter, $B_{ij}=B_{ji}$ is a symmetric edge parameter and $F_{ij}=-F_{ij}$ is an asymmetric edge parameter~\cite{Owen2020}. This decomposition can be interpreted as describing a system evolving in an energy landscape with wells of depth $E_j$, energy barriers of height $B_{ij}$ driven by forces $F_{ij}$, where the transition rates follow an Arrhenius-like expression.

Our system is not constrained to be time reversible, and certain transition rates can be formally zero. We can replace $W_{ij}=0$ with $W_{ij}=\varepsilon$  where $0<\varepsilon \ll 1$, for any rate that is formally zero. We check that this does not impact the error rate or patterning time in any $O(1)$ way, by recomputing these quantities. We have not constrained our system thermodynamically, but any such constraint would enforce reversibility. Having done this, the symmetric rates, $B_{ij}$, are $$B^s_{u,u+1}=\frac{1}{2}[(E^s_u-\log(f^+(u,s)))+(E^s_{u+1}-\log(f^-(u+1,s)))],$$ and the asymmetric rates are 
$$F^s_{u,u+1}=-[(E^s_u-\log(f^+(u,s)))-(E^s_{u+1}-\log(f^-(u+1,s)))],$$
where we are taking $f^+(N,s)=0$ and $f^-(0,s)=0$ and have replaced all other zero transition rates with $\varepsilon$.

This decomposition is not unique as we have freedom to choose the vertex energies $E_i$. One choice is to demand that $F_{ij}=0$. Setting $E^s_1=0$, this implies
$$E^s_{u+1}=E^s_u+\log(f^-(u+1,s))-\log(f^+(u,s)),$$ and thus determines all the vertex energies $E_{i}$ and consequently the barrier heights $B_{ij}$. This choice removes the non-equilibrium forces $F_{ij}$ and allows us to interpret the dynamics as equilibrium transitions on an energy landscape with Arrhenius-like rates. To be clear, we are not demanding that the system is equilibrium, the full system is non-equilibrium and cannot be decomposed without non-symmetric transition rates. Rather, this construction should be viewed as a mathematically precise description of a single cell's internal state dynamics given its signal receiving state.

\section{Information theory}

We are interested in studying communication between cells, and to do so, we use information theory. Since our model is a continuous time system, we briefly discuss the information theoretic background required for analysis of these system.

\subsection{\label{SIsec: InfoTheoryBackground}Information theory for continuous time processes: background}

Following conventions in~\cite{Spinney2016, Spinney2017}, we can define the concept of mutual information and transfer entropy for paths $x_{t_0}^t$ and $y_{t_0}^t$.
\begin{definition}
For the (continuous time) stochastic processes $x_{t_0}^T$ and $y_{t_0}^T$, the \textbf{mutual information} in time $[t_0,t]$ is given by
\begin{align*}
I(x_{t_0}^T ; y_{t_0}^T)
=\mathbb{E}\left[\log\frac{d\mathbb{P}^{XY}}{d(\mathbb{P}^X\times\mathbb{P}^Y)}\right].
\end{align*}
\end{definition}
Where here $\mathbb{P}^{XY}$ is the joint path measure associated with the combined trajectory $\{x_{t_0}^T,y_{t_0}^T\}$and $\mathbb{P}^{X}$ and $\mathbb{P}^{Y}$ are the marginal path measures corresponding to $x_{t_0}^T$ and $y_{t_0}^T$, respectively. The term inside the logarithm denotes the Radon-Nikodym derivative between $\mathbb{P}^{XY}$ and $\mathbb{P}^{X}\times\mathbb{P}^{Y}.$

We can also define the transfer entropy rate:
\begin{definition}
For the (continuous time) stochastic processes $x_{0}^T$ and $y_{0}^T$, the \textbf{transfer entropy rate} is defined as
\begin{align*}
\dot{\mathcal{T}}_{{Y}\to{X}}(t)
=&\lim_{dt\to 0}\frac{1}{dt}\mathbb{E}\left[\log{\frac{d\mathbb{P}_{t+dt}[{x}_{t+dt}|{{x}}^{t}_{0},{{y}}^{t}_{0}]}{d\mathbb{P}_{t+dt}[{x}_{t+dt}|{{x}}^{t}_{0}]}}\right],
\end{align*}
\end{definition}

where, $x^t_{0}$ denotes the full trajectory from $0$ to $t$ (i.e $\{x(\tau)|0\leq\tau<t\}$). Throughout we will take trajectories as starting from time $0$, although elsewhere the transfer entropy rate is often defined starting from $t-r$ for some fixed offset $r$. We can take the integral of this to define the transfer entropy:

\begin{definition}
For the (continuous time) stochastic processes $x_{0}^T$ and $y_{0}^T$, the (accumulated) \textbf{transfer entropy} is defined as
\begin{equation}
\mathcal{T}_{Y\to X}\big(t)=\int_{0}^{t}\dot{\mathcal{T}}_{{Y}\to{X}}(t')dt'.
\end{equation}
\end{definition}
Note that our definitions here may differ from elsewhere in the literature~\cite{Spinney2017}, since we always begin the stochastic process at time $0$.

As we are working with jump processes on a discrete state space, any particular realization $\{x_{0}^{T},y_{0}^T\}$ can be described by a set of states and transition times. Suppose for $x$ that there are $N_x$ transitions through states $x_i$ at times $t_i$ for $0\leq i \leq N_x$ with $x_i = x(t)$ for $t_i\leq t < t_{i+1}$ and $t_0=0$ by convention. It can be shown that~\cite{Spinney2017, Moor2023}
\begin{equation}
\mathcal{T}_{{Y}\to{X}}(T)=\mathbb{E}\left[ \sum_{i=1}^{N_x}\log{\frac{W[x_{i}|x^{t_i^-}_{0},y^{t_i^-}_{0}]}{W[x_{i}|x^{t_i^-}_{0}]}}
+\int_{0}^{T}\left(\lambda_{x}[{{x}}_{0}^{t'}]-\lambda_{x|y}[{{x}}_{0}^{t'},{{y}}_{0}^{t'}]\right)dt'\right],
\end{equation}
where the transition rates $W$ and escape rates $\lambda$ are defined as 
\begin{align*}
W[{x}'|{{x}}_{0}^{t},{{y}}_{0}^{t}]
&=\lim_{dt\to 0} \frac{1}{dt}\mathbb{P}[x(t+dt) = {x}'|{{x}}_{0}^{t},{{y}}_{0}^{t}],\nonumber\\
W[{x}'|{{x}}_{0}^{t}]&=\lim_{dt\to 0} \frac{1}{dt}\mathbb{P}[x(t+dt)={x}'|{{x}}_{0}^{t}],\\
\lambda_{x}[{{x}}_{0}^{t}]&=\sum_{{x}'\neq {x}^{-}_t}W[{x}'|{{x}}_{0}^{t}],\\
\lambda_{x|y}[{{x}}_{0}^{t},{{y}}_{0}^{t}]&=\sum_{{x}'\neq {x}^{-}_t}W[{x}'|{{x}}_{0}^{t},{{y}}_{0}^{t}],
\end{align*}
and $W[{x}'|{{x}}_{0}^{t_i^-},{{y}}_{0}^{t_i^-}]
=\lim_{t\nearrow t_i} W[{x}'|{{x}}_{0}^{t},{{y}}_{0}^{t}] $. 
The expectation is taken over different realizations of the stochastic process, each of which may have differing $N_x$ and differing transition times. Note that, $\mathbb{E}[\lambda_x[x^t_{0}]]=\mathbb{E}[\lambda_{x|y}[x_{0}^t,y_{0}^t]]$ as both expressions are simply the average transition rate out of the state $x$ at time $t$ averaged over all possible path
histories. We can therefore equivalently write this as
\begin{equation}\label{eq:TE}
    \mathcal{T}_{X\rightarrow Y}(T) = \mathbb{E}\left[\sum_{i=1}^{N_x}\log\frac{W[x_i|x_{0}^{t_i^-},y_{0}^{t_i^-}]}{W[x_i|x_{0}^{t_i^-}]}\right].
\end{equation}

We can also derive the following expression for the transfer entropy rate, since
\begin{align}
   \mathbb{E}\left[\log{\frac{d\mathbb{P}_{t+dt}[{x}_{t+dt}|{{x}}^{t}_{0},{{y}}^{t}_{0}]}{d\mathbb{P}_{t+dt}[{x}_{t+dt}|{{x}}^{t}_{0}]}}\right] &= \mathbb{E}\left[ (N_x(t+dt)-N_x(t))\log{\frac{W[x_{t+dt}|x^{t}_{0},y^{t}_{0}]}{W[x_{t+dt}|x^{t}_{0}]}}
+\int_{t}^{t+dt}\left(\lambda_{x}[{{x}}_{0}^{t'}]-\lambda_{x|y}[{{x}}_{0}^{t'} 
{{y}}_{0}^{t'}]\right)dt' \right] \\ \notag
&= \mathbb{E}\left[ \mathbb{I}_{x(t+dt)\neq x(t)}\log{\frac{W[x_{t+dt}|x^{t}_{0},y^{t}_{0}]}{W[x_{t+dt}|x^{t}_{0}]}}\right]
\\ \notag
&= \mathbb{E}_{\{x_0^t,y_0^t\}}\left[ \sum_{x'\neq x(t)} \mathbb{P}[x(t+dt) = {x}'|{{x}}_{0}^{t},{{y}}_{0}^{t}]\log{\frac{W[x_{t+dt}|x^{t}_{0},y^{t}_{0}]}{W[x_{t+dt}|x^{t}_{0}]}}\right],
\end{align}
where in the $dt\to0$ limit,
\begin{equation}\label{eq:TER}
\dot{\mathcal{T}}_{{Y}\to{X}}(t)
=\mathbb{E}\left[\sum_{x'\neq x(t)}\left(\log{\frac{W[x'|x^{t}_{0},y^{t}_{0}]}{W[x'|x^{t}_{0}]}}\right)W[x'|x_0^t,y_0^t]\right].
\end{equation}

Equations~\eqref{eq:TE} and~\eqref{eq:TER} provide a way of computing $\mathcal{T}_{Y\to X}$  and $\dot{\mathcal{T}}_{Y\to X}$ by approximating the expectation as a Monte-Carlo sum, provided we can compute conditional rates like $W[x'|x_0^t, y_0^t]$ and $W[x'|x_0^t]$, which we will detail later. 

It turns out that a lower variance estimate of equation~\eqref{eq:TER} can be computed at almost no additional cost by effectively including a control variate. Generically, one can improve the variance of an unbiased estimator $\hat{\theta}$ by adding a zero mean variable $Z$, to create a new estimator $\hat\theta_c = \hat\theta + cZ$ with the optimal choice of $c$ being $c = -\text{Cov}(\hat\theta,Z)/\text{Var}(Z)$. With $\hat\theta$ being the Monte-Carlo estimate of equation~\ref{eq:TER}, an obvious choice for $Z$ would be to use those same samples to take a Monte-Carlo estimate of $\mathbb{E}[\lambda_{x}[{{x}}_{0}^{t}]-\lambda_{x|y}[{{x}}_{0}^{t}, 
{{y}}_{0}^{t}]]$, which is indeed expectation zero. The covariance cannot be computed exactly, but, supposing $\delta[x'] = W[x'|x_0^t]-W[x'|x_0^t,y_0^t]$ is small,

\begin{align*}
    c &= -\frac{\mathbb{E}\left[ \sum_{x'\neq x(t)} \left(\log{\frac{W[x'|x^{t}_{0},y^{t}_{0}]}{W[x'|x^{t}_{0}]}}W[x'|x_0^t,y_0^t]\right)(\lambda_{x}[{{x}}_{0}^{t}]-\lambda_{x|y}[{{x}}_{0}^{t}, 
{{y}}_{0}^{t}]]) \right]}{\mathbb{E}\left[ (\lambda_{x}[{{x}}_{0}^{t}]-\lambda_{x|y}[{{x}}_{0}^{t}, 
{{y}}_{0}^{t}]])^2 \right]}\\
    &\approx -\frac{\mathbb{E}\left[ \sum_{x'\neq x(t)} \left( -\delta[x'] + O\left( \frac{\delta[x']^2}{W[x'|x^{t}_{0},y^{t}_{0}]} \right)\right)(\lambda_{x}[{{x}}_{0}^{t}]-\lambda_{x|y}[{{x}}_{0}^{t}, 
{{y}}_{0}^{t}]]) \right]}{\mathbb{E}\left[ (\lambda_{x}[{{x}}_{0}^{t}]-\lambda_{x|y}[{{x}}_{0}^{t}, 
{{y}}_{0}^{t}]])^2 \right]}\\
    &\approx 1,
\end{align*}
where we have used that $\sum_{x'} \delta[x] = \lambda_{x}[{{x}}_{0}^{t}]-\lambda_{x|y}[{{x}}_{0}^{t}]$.

As a result, we shall use the following for the computation of the rates, 
\begin{equation}
\dot{\mathcal{T}}_{{Y}\to{X}}(t)
=\mathbb{E}\left[\sum_{x'\neq x(t)}\left(\log{\frac{W[x'|x^{t}_{0},y^{t}_{0}]}{W[x'|x^{t}_{0}]}}\right)W[x'|x_0^t,y_0^t] -\sum_{x'\neq x(t)}(W[x'|x_0^t,y_0^t]-W[x'|x_0^t])\right].
\label{TERate}
\end{equation}
This inclusion of the rates as control variates is also discussed in Ref.~\cite{das2024}.

For the continuous time stochastic jump processes  that we are interested in, where $x$ and $y$ do not simultaneously change, the mutual information between $x$ and $y$ is simply the sum of the transfer entropies,

\begin{equation}\label{eq:TEDecomp}
    I^{XY}(t)=\mathcal{T}_{X\rightarrow Y}(t)+\mathcal{T}_{Y\rightarrow X}(t).
\end{equation}
This can be seen by decomposing the mutual information $I^{XY}(t+dt)$ as
\begin{align*}
     \mathbb{E}\left[\log \frac{d\mathbb{P}(x_0^{t+dt},y_0^{t+dt})}{d(\mathbb{P}(x_0^{t+dt})\times\mathbb{P}(y_0^{t+dt}))} \right]
    &=\mathbb{E}\left[\log\frac{d\mathbb{P}(x_0^{t},y_0^{t})}{d(\mathbb{P}(x_0^{t})\times\mathbb{P}(y_0^{t}))}\right]+\mathbb{E}\left[\log\frac{d\mathbb{P}(x_{t+dt}|x_0^{t},y_0^{t})}{d\mathbb{P}(x_{t+dt}|x_0^{t})}\right]+\\&\mathbb{E}\left[\log\frac{d\mathbb{P}(y_{t+dt}|x_0^{t},y_0^{t})}{d\mathbb{P}(y_{t+dt}|y_0^{t})}\right]+\mathbb{E}\left[\log\frac{d\mathbb{P}(x_{t+dt},y_{t+dt}|x_0^{t},y_0^{t})}{d(\mathbb{P}(x_{t+dt}|x_0^{t},y_0^{t})\times\mathbb{P}(y_{t+dt}|x_0^{t},y_0^{t}))}\right] \\\notag
    &= I^{XY}(t) + dt \dot{\mathcal{T}}_{Y\to X}(t) + dt \dot{\mathcal{T}}_{X\to Y}(t) + o(dt),
\end{align*}
where the final expectation vanishes as $dt\to0$ essentially because the probability of $x$ and $y$ both jumping is $O(dt^2)$ and hence at the $O(dt)$ level the behavior of $x$ and $y$ appears independent, which can be proved by Taylor expansion. Alternatively, Ref.~\cite{Moor2023} show the transfer entropy decomposition in equation~\ref{eq:TEDecomp} by starting from Jacod's formula for the Radon-Nikodym derivative  
for a counting process and separating this directly into terms that are recognizably the transfer entropies.

\subsection{Computation of rates for CTMC on finite discrete state space with hidden states}
The remaining difficulty in calculating transfer entropy rates and their integrals is in computing quantities of the form $ W[x'|x_0^t,y_0^t]$. These are challenging due to the presence of latent variables, $z$, which must be marginalized over. Returning to the definition of $W$, if only one of $x$, $y$, or $z$ can change in each transition, and if $x'\neq x(t)$, we have that  
\begin{align}
W[{x}'|{{x}}_{0}^{t},{{y}}_{0}^{t}]
&=\lim_{dt\to 0} \frac{1}{dt}\mathbb{P}[x(t+dt) = {x}'|{{x}}_{0}^{t},{{y}}_{0}^{t}], \nonumber\\
&= \lim_{dt\rightarrow0}\frac{1}{dt}\sum_{y',z',z}\mathbb{P}[(x,y,z)_{t+dt}=(x',y',z')|x_0^t,y_0^t,z(t)]\mathbb{P}[z(t)|x_0^t,y_0^t] \nonumber\\
&= \sum_{z} Q_{(x',y,z)(x,y,z)}\mathbb{P}[z(t)|x_0^t,y_0^t],
\label{Eq:defW_XY}
\end{align}
and similarly
\begin{align}
    W[x'|x_0^t] 
    &=\sum_{y,z}Q_{(x',y,z)(x,y,z)}\mathbb{P}[y(t),z(t)|x_0^t],
    \label{Eq:defW_X}
\end{align}
where $Q$ is the full transition rate matrix.

To aid our efforts, further define 
\begin{align*}
    \pi^A(z,t) &= \mathbb{P}[\bar{a}(t)=z|a_0^t],
\end{align*}
where $A$ represents some set of variables, for instance $a(t) = [u_1(t),u_2(t)]$, $\bar{a}$ is a vector of the remaining variables that are not in $A$, and calculating $\pi^A$ is our current challenge.

As discussed in the main text, for stochastic reaction networks, terms like $\pi^A$, obey a filtering equation~\cite{Duso2018, Moor2023}.
As in the main text, consider a stochastic reaction network with $K$ reaction channels, $n$ chemical species $Z_1,\dots,Z_n$, with vector $Z(t)$ recording the copy number of each species at time $t$,  and the $k^{th}$ reaction occurring at a rate $\lambda_k(Z(t))$. Take $R_A$ as the set of reactions that involve a change in $A$, $dN_k(t)$ is the increment of $N_k(t)$ which counts the number of times reaction $k$ has occurred, and $\lambda_k^A = \mathbb{E}[\lambda_k(Z(t))| a_0^T]$ is the  expected rate of the $k^{th}$ reaction given $a_0^T$. Then
\begin{align}
    d\pi^A(\bar{a},t) = &\sum_{k\in S_{\bar{A
    }}} \left[\lambda_k(\bar{a}-\nu_k^{\bar{A}},a(t))\pi^A(\bar{a}-\nu_k^{\bar{A}},t) - \lambda_k(\bar{a},a(t))\pi^A(\bar{a},t)  \right]dt \\\notag 
    & - \sum_{k\in R_A} [\lambda_k(\bar{a},a(t)) - \lambda_k^A(t) ]\pi^A(\bar{a},t)dt  + \sum_{k\in R_A} \left[ \frac{\lambda_k(\bar{a}-\nu_k^{\bar{A}},a(t))}{\lambda_k^a(t)}\pi^A(\bar{a}-\nu_k^{\bar{A}},t) - \pi^A(\bar{a},t)\right]dN_k(t),
\end{align}
where  $\nu_k^{\bar{A}}$ is a vector that represents the stoichiometric change in $\bar{A}$ after the $k^{th}$ reaction, and $S_{\bar{A}}$ is the set of reactions that only involve species in $\bar{A}$. The first summation represents a master-like equation representing the evolution of the latent variables under reactions that are exclusive to them, the second summation represents an updated belief about the latent variables given that no reaction has occurred and the third summation represents an updated belief given that reaction $k$ occurred. 

Quantities of the form $\mathbb{P}(x(t) | y_0^T)$, can be computed by marginalizing $\pi^Y$ over all latent variables except $x$. Crucially, our model, as specified by equation \eqref{eq:Reaction} can be interpreted as a specifying a stochastic reaction network for the species $(u_1,s_1,u_2,s_2,u_3,s_3)$. Moreover, any change of these species can only be due to exactly one reaction, and each reaction only changes one species. Therefore, for $A = [x(t),y(t)]$,

\begin{align}
        d\pi^A(z,t)&= \,dt\sum_{z^*\neq z}\pi(z^*,t)Q_{(x,y,z),(x,y,z^*)}-\pi(z,t)Q_{(x,y,z^*),(x,y,z)}\nonumber\\
        &-dt\sum_{(x^*,y^*)\neq (x,y)}\pi(z,t)\left[Q_{(x^*,y^*,z),(x,y,z)}-\sum_{\tilde{z}}\pi(\tilde{z},t)Q_{(x^*,y^*,\tilde{z}),(x,y,\tilde{z})}\right] \nonumber\\
        &+\sum_{(x^*,y^*)\neq (x,y)}\pi(z,t)\left(\frac{Q_{(x^*,y^*,z),(x,y,z)}}{\sum_{\tilde{z}}\pi(\tilde{z},t) Q_{(x^*,y^*,\tilde{z}),(x,y,\tilde{z})}} - 1\right)dN_{(x,y)\rightarrow (x^*,y^*)}(t) ,
        \label{Eq:FilteringEqPi_XY}
\end{align}
where $N_{(x,y)\rightarrow (x^*,y^*)}(t)$ is the counting processes for the transition $(x,y)\rightarrow (x^*,y^*)$ at time $t$. Since only one reaction, say the $k^{th}$, could be implicated in the transition $(x,y,z)\rightarrow (x^*,y^*,z)$, we have that $Q_{(x^*,y^*,z)(x,y,z)} = \lambda_k(x,y,z)$. Analogous SDEs govern the evolution of similar quantities, such as when $A = [x(t)]$. 

\subsection{Dynamic correlational information}
Recently, a related quantity to mutual information called correlational information (CI) has been proposed~\cite{Bruckner2024}.  For random variables $X_1,X_2,\dots,X_K$, it is defined as
\begin{equation}
    CI(t) = \frac{1}{K}\mathbb{E}\left[\log \frac{P(X_1(t),\dots,X_K(t))}{\prod_{i=1}^K P_i(X_i(t))}\right],
    \label{Eq:CIStatic}
\end{equation}
where $P_i$ is the marginal probability for variable $X_i$. A related measure of correlation, $K\times CI(t)$, has been previously defined in information theoretic literature and is often referred to as ``multi-information'' or ``total-correlation''~\cite{Watanabe1960}. 

For continuous-time jump processes, we extend \eqref{Eq:CIStatic} to trajectories. Specializing to three coordinates $(x,y,z)$, as earlier, we define \emph{dynamic} correlation information as
\begin{equation*}
    CI(x_0^t,y_0^t,z_0^t) = \frac{1}{3}\mathbb{E}\left[\log\frac{d\mathbb{P}[x_0^t,y_0^t,z_0^t]}{d(\mathbb{P}[x_0^t]\times\mathbb{P}[y_0^t]\times\mathbb{P}[z_0^t])}\right].
    \label{eq:CIpath}
\end{equation*}
We assume, as throughout, that $x,y,z$ are components of a CTMC on a finite state space and that simultaneous jumps do not occur. A standard $dt-$ expansion analogous to the mutual-information decomposition in Section~\ref{SIsec: InfoTheoryBackground} yields

\begin{align*}
&CI(x_0^{t+dt},y_0^{t+dt},z_0^{t+dt})
\\&=CI(x_0^t,y_0^t,z_0^t)+\frac{1}{3}\mathbb{E}\left[\log\frac{d\mathbb{P}[x_{t+dt}|x_0^t,y_0^t,z_0^t]}{d\mathbb{P}[x_{t+dt}|x_0^t]} +\log\frac{d\mathbb{P}[y_{t+dt}|x_0^t,y_0^t,z_0^t]}{d\mathbb{P}[y_{t+dt}|y_0^t]}+\log\frac{d\mathbb{P}[z_{t+dt}|x_0^t,y_0^t,z_0^t]}{d\mathbb{P}[z_{t+dt}|z_0^t]}\right]+\frac{1}{3}\mathbb{E}[R],
\end{align*}
where the remainder term $R$ collects terms involving joint increments of two or more coordinates, 
$$\mathbb{E}[R]=\mathbb{E}\left[\log\frac{d\mathbb{P}[x_{t+dt},y_{t+dt},z_{t+dt}|x_0^t,y_0^t,z_0^t]}{d\mathbb{P}[x_{t+dt}|x_0^t,y_0^t,z_0^t]d\mathbb{P}[y_{t+dt}|x_0^t,y_0^t,z_0^t]d\mathbb{P}[z_{t+dt}|x_0^t,y_0^t,z_0^t]}\right].$$
As with mutual information, the remainder term reflects higher-order dependencies and $\mathbb{E}[R]=O(dt^2)$. Indeed, in case no jump occurs the numerator and denominator in $R$ agree up to $O(dt)$; and in case of a single jump, $R=O(dt)$ but it is multiplied by a probability $O(dt)$. Therefore, $\mathbb{E}[R]$ does not contribute to the differential equation for the dynamic correlation information. Integrating the differential equation yields
\begin{equation}
    CI(x_0^t,y_0^t,z_0^t)=\frac{1}{3}\left(\mathcal{T}_{(x,y)\rightarrow z}+\mathcal{T}_{(y,z)\rightarrow x}+\mathcal{T}_{(z,x)\rightarrow y}\right).
\end{equation}
For $M=3$ cells with $(x,y,z)=(u_1,u_2,u_3)$, the symmetry of our model implies that all three transfer entropies are equal. As a result, the dynamic correlation information reduces to
$$CI_{(u_1,u_2,u_3)}(t)=\mathcal{T}_{(u_1,u_2)\rightarrow u_3}.$$ 

\subsection{Algorithm for computing the cumulative transfer entropy and mutual information}

We now describe the algorithm in more detail. For a CTMC, let $\xi$ denote the full state, and divide this full state into a set of latent variables and observed variables.

We denote the set of observed variables $A$, and let $\bar{A}$ denote the remaining latent variables, so that at any time we may write $\xi(t) = (a(t),\bar{a}(t))$, partitioning into observed variables $a(t)$ and latent variables $\bar{a}(t)$. To compute information metrics, we require
\begin{itemize}
    \item The full CTMC's transition rate matrix, $Q_{ij}$, 
    \item A coarse-graining function that takes a full trajectory $\xi_0^T$ and outputs (i) the observed trajectory $a_0^T$, (ii) the observed jump times, (iii) the complete set of values $A$ can take, $\mathcal{A}_{poss},$ and the set of values that $\bar{A}$ can take, $\bar{\mathcal{A}}_{poss}.$
\end{itemize} 
We then perform the following steps:
\\

\paragraph{Step 1: Draw SSA trajectories.}
First, we draw  $N_{\mathrm{sim}}$
independent realizations of the CTMC with the Gillespie (SSA) algorithm. For our model parameters we found $N_{\mathrm{sim}}\sim10^3$–$10^4$
provides a satisfactory estimate. To determine the appropriate $N_{\mathrm{sim}}$, the running sample variance of $\widehat{\mathcal T}$ and $\widehat I$ can be used to decide when additional trajectories are unnecessary.

\begin{algorithm}[t]
\caption{Monte-Carlo estimation of transfer entropy, transfer entropy rate, mutual information and mutual information rate}
\label{algo:estimate_TE_MI}

\KwIn{Transition–rate matrix $Q$, number of simulated trajectories $N_\mathrm{sim}$, time steps $\mathbb{T}_q=\{t_i\}_{i=1}^{T_q}$}
\KwOut{Estimates $\widehat{\mathcal T}_{X\!\to Y}, \widehat{\mathcal T}_{Y\!\to X}$ and $\widehat{I}^{XY}$ for different time-points $t_1,\cdots,t_{T_q}$}

$\forall q:\; \bigl[\mathcal T^{XY}_{\mathrm{sum}},\mathcal T^{YX}_{\mathrm{sum}}\bigr](t_q)\gets0$ \tcp*{Initialize accumulators for integrals (arrays of length $T_q$)}

$\forall q:\; \bigl[\dot{\mathcal{T}}^{XY}_{\mathrm{sum}},\dot{\mathcal{T}}^{YX}_{\mathrm{sum}}\bigr](t_q)\gets0$ \tcp*{Initialize accumulators for rates (arrays of length $T_q$)}

\For{$i\gets1$ \KwTo $N_\mathrm{sim}$}{
    \tcp{Monte-Carlo loop through the trajectories}
    Simulate a full path $\chi^{(i)}=\bigl(x^{(i)}_{0:T},\,y^{(i)}_{0:T},\,z^{(i)}_{0:T}\bigr)$ with Gillespie's algorithm, this has jumps at $t\in \mathbb{T}_{\mathrm{jump}}$\;
    
    Compute $\{\mathbb Q^{XY}(t)\},\{\mathbb Q^{X}(t)\}\text{ and } \{\mathbb Q^{Y}(t)\}$ for all $t\in \mathbb{T}_{\mathrm{jump}}$;
    
    Compute $\{\mathbf{b}^{XY}(t)\},\{\mathbf b^{X}(t)\}\text{ and } \{\mathbf b^{Y}(t)\}$, for all $t\in  \mathbb{T}_{\mathrm{jump}}$;

    Compute $\{\mathbf{g}^{XY}(t)\},\{\mathbf g^{X}(t)\}\text{ and } \{\mathbf g^{Y}(t)\}$, for all $t\in \mathbb{T}_{\mathrm{jump}}$; 
    
    Solve the filtering equation to obtain $\boldsymbol{\pi}_{Z|X,Y}, \boldsymbol{\pi}_{Z|X}, \text{ and } \boldsymbol{\pi}_{Z|Y}$ for all $t\in \mathbb{T}_q\cup\mathbb{T}_{\mathrm{jump}}$; 

    Evaluate $W^{(i)}_{X|X,Y}(t)$ and $W^{(i)}_{X|X}(t)$ for all time-point $t\in \mathbb{T}_q \cup \mathbb{T}_{\mathrm{jump}}$;
    
    Evaluate $W^{(i)}_{Y|X,Y}(t)$ and $W^{(i)}_{Y|Y}(t)$ for all time-points $t\in \mathbb{T}_q \cup \mathbb{T}_{\mathrm{jump}}$;
    
    Compute the \emph{path-wise} transfer entropy $\mathcal \{\mathcal T^{(i)}_{X\!\to Y}[t]\}_{t\in \mathcal{T}_q}$ and $\{\mathcal T^{(i)}_{Y\!\to X}[t]\}_{t\in \mathcal{T}_q}$\;

    Compute the \emph{path-wise} transfer entropy rate  $\{\dot{\mathcal{T}}^{(i)}_{X\!\to Y}[t]\}_{t\in \mathcal{T}_q}$ and $\{\dot{\mathcal{T}}^{(i)}_{Y\!\to X}[t]\}_{t\in \mathcal{T}_q}$\;
    
    Addition to $\mathcal{T}^{XY}_{\mathrm{sum}},\, \mathcal{T}^{YX}_{\mathrm{sum}},\,\dot{\mathcal{T}}^{XY}_{\mathrm{sum}},\mathrm{and}\, \dot{\mathcal{T}}^{YX}_{\mathrm{sum}}$ with the path-specific values at each time-step $t_q$\;
}

\ForEach{$t_q\in \mathbb{T}_q$}{
    \tcp{Averaging the information metrics}
    $\widehat{\mathcal T}_{X\!\to Y}(t_q), \widehat{\mathcal T}_{Y\!\to X}(t_q)\gets \dfrac{\mathcal T^{XY}_{\text{sum}}(t_q)}{N_\mathrm{sim}} ,\dfrac{\mathcal T^{YX}_{\text{sum}}(t_q)}{N_\mathrm{sim}}$\tcp*{for the transfer entropies}

    $\widehat{\dot{\mathcal{T}}}_{X\!\to Y}(t_q), \widehat{\dot{\mathcal{T}}}_{Y\!\to X}(t_q)\gets \dfrac{\dot{\mathcal{T}}^{XY}_{\text{sum}}(t_q)}{N_\mathrm{sim}} ,\dfrac{\dot{\mathcal{T}}^{YX}_{\text{sum}}(t_q)}{N_\mathrm{sim}}$\tcp*{for the transfer entropy rates}

    $\widehat{I}_{XY}(t_q)\gets \widehat{{\mathcal{T}}}_{X\!\to Y}(t_q)+\widehat{{\mathcal{T}}}_{Y\!\to X}(t_q)$\tcp*{for the mutual information}

    $\widehat{\dot{I}}_{XY}(t_q)\gets \widehat{\dot{\mathcal{T}}}_{X\!\to Y}(t_q)+\widehat{\dot{\mathcal{T}}}_{Y\!\to X}(t_q)$\tcp*{for the mutual information rate}
}

\KwRet{$\bigl\{\widehat{\mathcal T}_{X\!\to Y}(t_q)\bigr\},\bigl\{\widehat{\dot{\mathcal{T}}}_{X\!\to Y}(t_q)\bigr\}, \bigl\{\widehat{\mathcal T}_{Y\!\to X}(t_q)\bigr\}, \bigl\{\widehat{\dot{\mathcal{T}}}_{Y\!\to X}(t_q)\bigr\}, \bigl\{\widehat{I}_{XY}(t_q)\bigr\}, \bigl\{\widehat{\dot{I}}_{XY}(t_q)\bigr\}$}\;

\end{algorithm}

\paragraph{Step 2: Build the latent-variable rate matrix $\mathbb{Q}^A(t)$.}
For each trajectory, we need to calculate the local information metrics, as defined earlier. To do so, we need to compute the latent probabilities $\pi^A(\bar{A},t)$.  At any given time, this distribution is  a vector $\boldsymbol{\pi_{\bar{A}|A}}$, which is of length $|\mathcal{A}_{\mathrm{poss}}|$, and evolves under the filtering equation.  To solve this equation, we first compute $\mathbb{Q}^A(t)$, a matrix of dimensions $|\bar{\mathcal{A}}_{\mathrm{poss}}|\times |\bar{\mathcal{A}}_{\mathrm{poss}}|$, representing the transition matrix restricted to only the latent variables, so that $\mathbb{Q}^A_{\bar{a} \bar{a}'}(t) = Q_{\xi \xi'}$ where $\xi = (a,\bar{a})$, $\xi' = (a,\bar{a}')$, and $a(t)$ is the observed variable at time $t$.

\paragraph{Step 3: Compute Bayesian update vectors  $\mathbf{b}^{A}(t)$.}
One term in the filtering equation  accounts for change in probability if no observable transition occurs. This involves calculating $\mathbf{b}^{A}(t)$, a vector of dimensions $|\bar{\mathcal{A}}_{\mathrm{poss}}|$, for the entire trajectory, which we define as $$\mathbf{b}^{A}_{\bar{a}}(t)=\sum _{a'\neq a(t)}Q_{(a',\bar{a}),(a(t),\bar{a})}.$$ 

\paragraph{Step 4: Compute jump transition vectors  $\mathbf{g}^{A}(t)$.}
One term in the filtering equation accounts for changes in $\pi^A$ after $A$ is observed jumping. This involves calculating $\mathbf{g}^{A}(t)$, a vector of dimensions $|\bar{\mathcal{A}}_{\mathrm{poss}}|$, to be evaluated at the jump times of $A$, which we define as 
$$\mathbf{g}^{A}_{\bar{a}}(t)=Q_{(a(t),\bar{a})(a(t^-),\bar{a})}
$$

\paragraph{Step 5: Calculate conditional probabilities $\boldsymbol{\pi}_{\bar{A}|A}$(t) for each trajectory.}
Once we have calculated the above quantities for the entire trajectory, we are in a good shape to calculate the probabilities $\pi_{\bar{A}|A}(t)$ for the entire trajectory. Indeed, using the SDE (\ref{Eq:FilteringEqPi_XY}), we note that for $t$ between jump times of $A$ on the specific trajectory, $\tau_i \leq t <\tau_{i+1}$, the evolution of $\pi_{\bar{A}|A}$ is governed by the following non-linear ODE 
$$\frac{d}{dt}\boldsymbol{\pi_{\bar{A}|A}}(t)=\mathbb{Q}^A(\tau_{i})\boldsymbol{\pi_{\bar{A}|A}}-\boldsymbol{\pi_{\bar{A}|A}}\odot\left(\mathbf{b}^{A}(\tau_i)-\mathbf{b}^{A}(\tau_i)\cdot\boldsymbol{\pi_{\bar{A}|A}}\mathbf{1}\right),$$
where $\odot$ represents the Hadamard product and $\mathbf{1}$ is a vector of $1$s. 

On the other hand, at the jump times, $t=\tau_{i+1},$ we must have that 
$$\pi_{\bar{A}|A}(\tau_{i})=\frac{\pi_{\bar{A}|A}(\tau_{i}^{-})\odot \mathbf{g}^{A}(\tau_i)}{\pi_{\bar{A}|A}(\tau_{i}^{-})\cdot \mathbf{g}^{A}(\tau_i)}.$$
\paragraph{Step 6: Build the coarse-grained rate matrices $W[\cdot|A]$.}
Having computed the relevant probabilities $\boldsymbol{\pi}_{\bar{A}|A}(t)$ for relevant $A$ for one trajectory, we can now evaluate the coarse-grained rate matrices $W[\cdot|A]$ for this by using equation~\eqref{Eq:defW_XY}, or similar. 
\paragraph{Step 7: Monte Carlo estimation and averaging}
Repeat the steps above for each of the Gillespie realizations to calculate the full coarse-grained rates for each trajectory realization. Once done, we can use equations~\eqref{eq:TE} and \eqref{TERate} to calculate the path-wise transfer entropy, path-wise transfer entropy rate. These are averaged over all realizations to obtain Monte-Carlo estimates of the transfer entropy and transfer entropy rate. We also use \eqref{eq:TEDecomp} to calculate the estimates of the mutual information and mutual information rate for the process. All steps are summarized in Algorithm \ref{algo:estimate_TE_MI}.

\section{Optimization of final state mutual information}
In this section, we consider the final state of a three-cell system, and ask which probability distributions minimize or maximize the final state mutual information between cells. Calling the three cell's final states  $x,y,z\in\{0,N\}$, we define $\mathbb{P}(x,y,z)$ as $\mathbb{P}(0,0,0) = a$, $\mathbb{P}(0,0,N) = b$, $\mathbb{P}(0,N,N) = c$, $\mathbb{P}(N,N,N) = d$ with the remaining probabilities constrained by the fact that each cell is identical. The joint probability distribution of one cell and the remaining cells, $(x,(y,z))$ is given in Table \ref{tab:TableJointDistrthreecell3}, whereas the joint distribution of a pair of cells is given in Table~\ref{tab:TableJointDistrthreecell2}.
\begin{table}[h]
    \centering
    \begin{tabular}{|l|l|l|l|l|}
    \hline
        \diagbox{ $x$ }{$(y,z)$}     & $(0,0)$ & $(0,N)$ & $(N,0)$ & $(N,N)$ \\ \hline
    $0$ & $a$     & $b$  & $b$     & $c$    \\ \hline
    $N$ & $b$     & $c$ & $c$     & $d$     \\ \hline
    \end{tabular}
    \caption{Joint distribution of the final states in the three-cell system, $(x,(y,z))$. Here, $a,b,c,d\geq0$ and $a+3b+3c+d=1.$}
    \label{tab:TableJointDistrthreecell3}
\end{table}
\begin{table}[h]
    \centering
    \begin{tabular}{|l|l|l|}
    \hline
       \diagbox{ $x$ }{$y$}      & $0$ & $N$ \\ \hline
    $0$ & $a+b$     & $b+c$     \\ \hline
    $N$ & $b+c$     & $c+d$     \\ \hline
    \end{tabular}
    \caption{Joint distribution of the final states of two of the cells in the three-cell system, $(x,y)$.}
    \label{tab:TableJointDistrthreecell2}
\end{table}

The parameters must sum to one, so $a+3b+3c+d =1$, and since the error rate is fixed, we have $3b = 1 - \epsilon$. Further, since $d = \epsilon - a - 3c$, we can consider the  distribution to be a function of just two parameters, $a$ and $c$. All parameters are non-negative, including $d$, and hence  the feasible region is defined by 
\begin{align}
    0\leq a &\leq \epsilon \\ \notag
    0\leq c &\leq \epsilon/3 \\ \notag
    a + 3c &\leq \epsilon.
\end{align}

Overall to find maxima and minima of both $I(x:y)$, $I(x:(y,z))$, we take the following approach:
\begin{enumerate}[label=(\roman*)]
    \item Identify stationary points in the interior and compute $I$.
    \item Restricting to one of the boundaries, find stationary points along that boundary and compute $I$.
    \item Evaluate at the three corner points of the feasible region.
\end{enumerate}
We may find stationary points analytically or through numerical root finding. From this finite set of possibilities, we then identify the global minimum and maximum, by numerically evaluating $I$ if necessary

\subsection{Pairwise mutual information}
To compute $I(x;y) = f(a,c)$, let $q=\mathbb{P}(x=0)=a+2b+c$, then
\begin{align*}
    I(x;y) &= (a+b)\log\left(\frac{a+b}{q^2}\right)+2(b+c)\log\left(\frac{b+c}{q(1-q)}\right)+(c+d)\log\left(\frac{c+d}{(1-q)^2}\right)\\
    &=(a+b)\log(a+b)+2(b+c)\log(b+c)+(\epsilon-2c-a)\log(\epsilon-2c-a)\\&-2(a+2b+c)\log(a+2b+c)-2(1-(a+2b+c))\log(1-(a+2b+c)).
\end{align*}
Taking the gradients of $f$ with respect to $a$ and $c$ gives,
\begin{align*}
    \frac{\partial f}{\partial a} &=\log(a+b)-\log(\epsilon-2c-a)-2\log(a+2b+c)+2\log(1-(a+2b+c)),\\
    \frac{\partial f}{\partial c} &=2\log(b+c)-2\log(\epsilon-2c-a)-2\log(a+2b+c)+2\log(1-(a+2b+c)).\\
\end{align*}

At stationary point, $\frac{\partial f}{\partial a}=0=\frac{\partial f}{\partial c}$,  and hence
\begin{align}
    (a+b)(1-(a+2b+c))^2&=(\epsilon-2c-a)(a+2b+c)^2\\
    (b+c)(1-(a+2b+c))&=(\epsilon-2c-a)(a+2b+c).
\end{align}
Together, these imply that $(b+c)^2=(\epsilon-2c-a)(a+b),$ from which one can deduce that
$$(a+b)(1-(a+2b+c))=((a+2b+c)-(a+b))(a+2b+c) \implies (a+2b+c)^2=(a+b).$$
Taking the square root of both sides (only the positive square root is possible), we find $c+b=\sqrt{a+b}-(a+b)$. Since $\sqrt{x}-x \leq 1/4$ for $x\geq 0$ we can conclude a feasible stationary point exists only if $b\leq \frac{1}{4} \implies \epsilon\geq \frac{1}{4}.$ In the case of $\epsilon \geq \frac{1}{4},$ one stationary solution is $a = c = \epsilon/3 - 1/12$. For these stationary points, $a+b = (\sqrt{a+b})^2$, $c+b = \sqrt{a+b}(1 - \sqrt{a+b})$, and $c+d = (1 - \sqrt{a+b})^2$, at which point the joint distribution in Table~\ref{tab:TableJointDistrthreecell2} factorizes, and hence $I$ is $0$. Therefore, we conclude $\min I(x;y)=0$ for $\epsilon\geq \frac{1}{4}$. 

To find the minima for $\epsilon\leq \frac{1}{4}$, or for to find the maxima for any $\epsilon$, we need to search the boundary of the feasible set. The boundaries are at $a=0$, at $c=0$ and at $\epsilon -3c-a=0$. Let's analyze these boundaries in order:
\begin{itemize}
    \item $\boldsymbol{c=0}:$ For this boundary, we define $$g(a)=(a+b)\log(a+b)+2b\log(b)+(\epsilon-a)\log(\epsilon-a)-2(a+2b)\log(a+2b)-2(1-(a+2b))\log(1-(a+2b)).$$ 
    Note that $g(a)=f(a,0).$ Then, $g'(a)=0$ gives us that the stationary points must satisfy $(a+b)(1-(a+2b))^2=(\epsilon-a)(a+2b)^2$, i.e. 
    $$2(a+2b)^3-3(a+2b)^2+(1+2b)(a+2b)-b=0.$$
    Noting that two of the solutions of this equation are from $(a+2b)^2=(a+b),$ we can factorize
    $$((a+2b)^2-(a+b))(2(a+2b)-1)=0.$$
    Thus, on this boundary, there are two local minima at $(a+2b)^2=(a+b)$ and a local maxima at $a+2b=\frac{1}{2}$, for $\epsilon \geq \frac{1}{4}$. The minima satisfy $b = \sqrt{a+b} - (a+b)$, where since $\sqrt{x} - x$ is a unimodal function in $0\leq x \leq 1$, and  since $\sqrt{1/4} - (1/4) \geq b$ for $b\leq 1/4$, we note that all three extremal points lie within the interval.   For $\epsilon<\frac{1}{4},$ we note that the the derivative $g'(a)>0$ for $0\leq a\leq \epsilon,$ and so there are no stationary points within the interval.

    \item $\boldsymbol{a=0}:$ For this boundary, we define $$g(c)=b\log(b)+2(b+c)\log(b+c)+(\epsilon-2c)\log(\epsilon-2c)-2(c+2b)\log(c+2b)-2(1-(c+2b))\log(1-(c+2b)).$$ 
    Note that $g(c)=f(0,c).$ Again, $g'(c)=0$ gives us that the stationary points must satisfy $(b+c)(1-(2b+c))=(\epsilon-2c)(2b+c),$ i.e.
    $$(2b+c)^2-b=0.$$
    This implies that $c=\sqrt{b}-2b$, with only the positive root being possible.
    
    \item $\boldsymbol{\epsilon-3c-a=0}:$
    In this case $a=\epsilon-3c$ and $\epsilon-2c-a=c.$ Along this line, the mutual information becomes
    \begin{align*}
    g(c)=&(b+\epsilon-3c)\log(b+\epsilon-3c)+2(b+c)\log(b+c)+c\log(c)
    \\\notag &-2(\epsilon-2c+2b)\log(\epsilon-2c+2b)-2(1-(\epsilon-2c+2b))\log(1-(\epsilon-2c+2b)).
    \end{align*}
    The points where the derivative $g'(c)=0$ are the solutions to $$c(b+c)^2(1-(2c+b))^4=(2c+b)^4(1-2b-3c)^3.$$
    We find these through points by numerically solving for roots to the above equation, and then evaluating the function $g$ at these stationary points.
\end{itemize}
All that remains is to evaluate the mutual information at the corner points. The values are 
\begin{align*}
    f(0,0) &=-b\log(b)+\epsilon\log(\epsilon)-4b\log(2)-2(1-2b)\log(1-2b) \\\notag
f(\epsilon,0)&=(\epsilon+b)\log(\epsilon+b)+2b\log(b)-2(\epsilon+2b)\log(\epsilon+2b)-2(1-(\epsilon+2b))\log(1-(\epsilon+2b)) \\\notag
f(0,\epsilon/3)&=b\log(b)+2\left(b+\frac{\epsilon}{3}\right)\log(b+\frac{\epsilon}{3})+\frac{\epsilon}{3}\log\frac{\epsilon}{3}-2\left(\frac{\epsilon}{3}+2b\right)\log\left(\frac{\epsilon}{3}+2b\right)-2\left(1-(\frac{\epsilon}{3}+2b)\right)\log\left(1-(\frac{\epsilon}{3}+2b)\right).
\end{align*}

\begin{figure}[t]%
\centering
\includegraphics[width=0.5\linewidth]{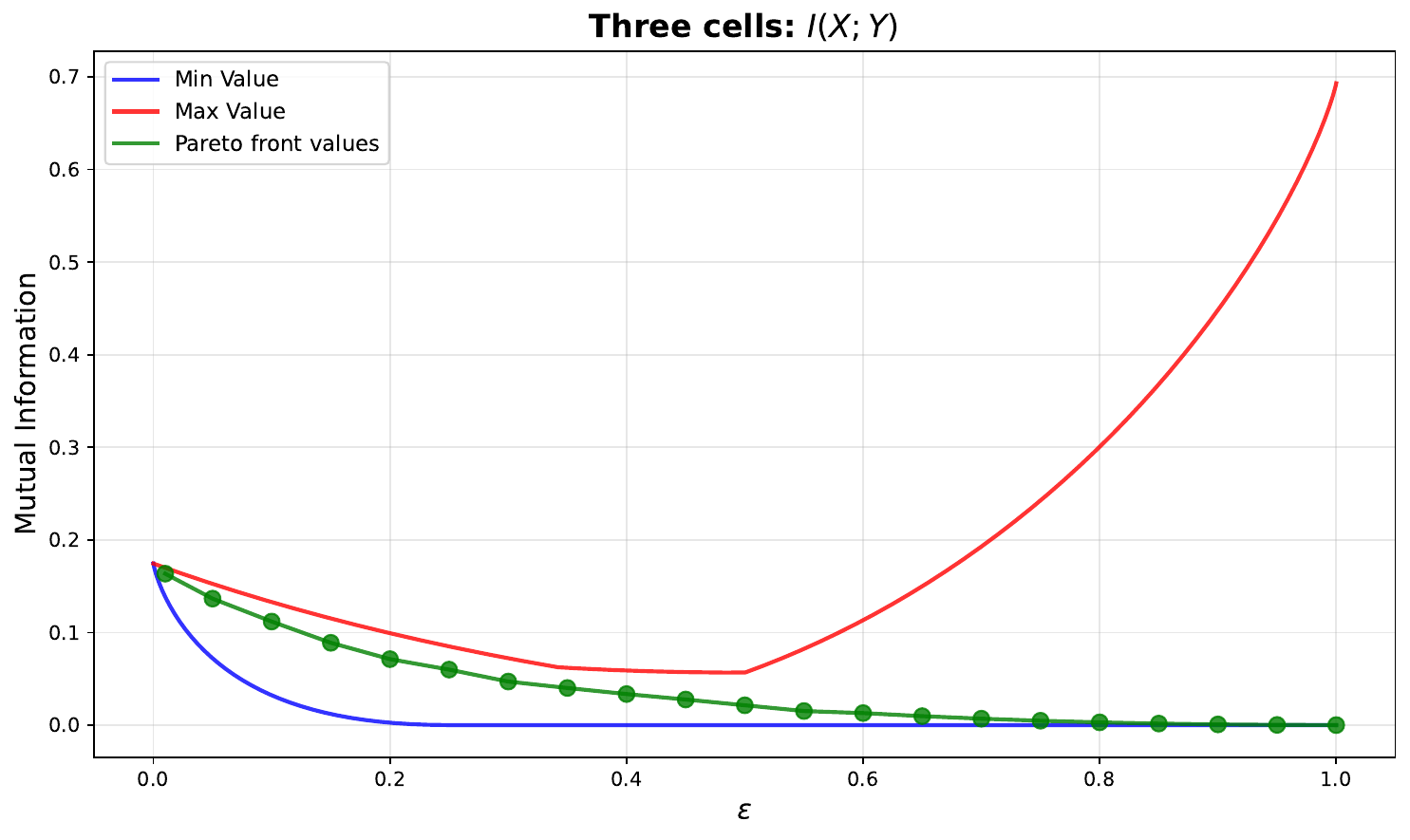}
\caption{Maximum and minimum values of final state mutual information between a pair of cells in a three-cell system. Values are plotted against the error rate $\epsilon$, and are shown along with the values of numerically optimized strategies which lie on the Pareto front.}
\label{Fig:threecellmutinfplot1}
\end{figure}

Overall, the results are shown in Fig.~\ref{Fig:threecellmutinfplot1}.

\subsection{Mutual information between one cell and its neighbors}
For the final state mutual information between a cell and its two neighbors, we will proceed similar to before. Once again, the mutual information is a function of $a,c$ only (given $\epsilon$), and the feasible region is also the same as before. The mutual information can thus be written as 
\begin{align*}
    I(x;(y,z)) &=a\log(a)+3b\log(b)+3c\log(c)+(\epsilon-3c-a)\log(\epsilon-3c-a)\\
    &-(a+b)\log(a+b)-2(b+c)\log(b+c)-(\epsilon-2c-a)\log(\epsilon-2c-a)\\&-(a+2b+c)\log(a+2b+c)-(1-(a+2b+c))\log(1-(a+2b+c))\equiv f(a,c).
\end{align*}

Similar to before, we first identify the stationary points of $I$ in the interior. For this, we compute the gradient,
\begin{align*}
    \frac{\partial f}{\partial a} &=\log(a)-\log(\epsilon-3c-a)-\log(a+b)+\log(\epsilon-2c-a)-\log(a+2b+c)+\log(1-(a+2b+c)),\\
    \frac{\partial f}{\partial c} &=3\log(c)-3\log(\epsilon-3c-a)-\log(b+c)+2\log(\epsilon-2c-a)-\log(a+2b+c)+\log(1-(a+2b+c)).
\end{align*}
Immediately we note that, due to the $\log a$ term that $\partial f/\partial a \to -\infty$ as $a \to 0$ for fixed $c$, and similar for $c\to0$. Similarly, the gradient of $f$ in the direction $(1,3)$ tends to $+\infty$ as you approach $a+3c=\epsilon$. Hence the global minimum is an interior point, whereas the global maxima could be at the boundaries or an interior point.

At stationary point, $\frac{\partial f}{\partial a}=0=\frac{\partial f}{\partial c}$ we have the following system of equations
\begin{align}
    a(1-(a+2b+c))(\epsilon-2c-a)&=(a+b)(\epsilon-3c-a)(a+2b+c),\\
    c^3(1-(a+2b+c))(\epsilon-2c-a)^2&=(b+c)^2(\epsilon-3c-a)^3(a+2b+c).
\end{align}

To find the stationary points, we use a symbolic algebra package to solve the above equations. Upon finding solutions within the feasible region, we compute the value of $I$ numerically.

Next, we turn to finding the stationary points of the functions constrained to the boundary.

\begin{itemize}
    \item $\boldsymbol{c=0}:$ For this boundary, we define
    \begin{align*}
        g(a) &=a\log(a)+b\log(b)+(\epsilon-a)\log(\epsilon-a)\\
        &-(a+b)\log(a+b)-(\epsilon-a)\log(\epsilon-a)\\&-(a+2b)\log(a+2b)-(1-(a+2b))\log(1-(a+2b)).
    \end{align*}
    Again, $g'(a)=0$ tells us that the stationary points satisfy $(a+b)(1-(a+2b))^2=(\epsilon-a)(a+2b)^2$, or 
    $$(\epsilon-a)(2a^2+(5b-1)a+2b^2)=0.$$
    Thus, on this boundary, the stationary points are at $a=\epsilon$ (a corner point), and $a=\frac{1-5b\pm \sqrt{(1-b)(1-9b)}}{4}$. The latter stationary points exist only for $b\leq \frac{1}{9}$, i.e. $\epsilon\geq \frac{2}{3}.$

    \item $\boldsymbol{a=0}:$ For this boundary, we define 
    \begin{align*}
        g(c) &=2b\log(b)+3c\log(c)+(\epsilon-3c)\log(\epsilon-3c)\\
        &-2(b+c)\log(b+c)-(\epsilon-2c)\log(\epsilon-2c)\\&-(2b+c)\log(2b+c)-(1-(2b+c))\log(1-(2b+c)).
    \end{align*}
    Note that $g(c)=f(0,c).$ Again, $g'(c)=0$ gives us that the stationary points must satisfy 
    $$c^3(1-(2b+c))(\epsilon-2c)^2=(b+c)^2(\epsilon-3c)^3(2b+c).$$
   We find the roots of this equation numerically, finding that a single solution exists and is a local minima along the boundary.

    \item $\boldsymbol{\epsilon-3c-a=0}:$
    In this case $a=\epsilon-3c$ and $\epsilon-2c-a=c.$ Using this, we can simplify the mutual information constrained to this line as
    \begin{align*}
        g(c) &=(\epsilon-3c)\log(\epsilon-3c)+3b\log(b)+2c\log(c)\\
        &-(\epsilon-3c+b)\log(\epsilon-3c+b)-2(b+c)\log(b+c)-\\&-(\epsilon-2c+2b)\log(\epsilon-2c+2b)-(1-(\epsilon-2c+2b))\log(1-(\epsilon-2c+2b)).
    \end{align*}
    As earlier, we need to find the values of $c$ where derivative $g'(c)=0.$ These turn out to be the solutions to $$c^2(\epsilon-3c+b)^3(\epsilon-2c+2b)^2=(\epsilon-3c)^3(b+c)^2(1-(\epsilon-2c+2b))^2.$$
       As with $a=0$, we find the roots of this equation numerically and evaluate the mutual information at these points.
\end{itemize}

All that remains is to evaluate the mutual information at the corner points. The values are 
\begin{align*}
    f(0,0) &= -2b\log(2b)-(1-2b)\log(1-2b) \\\notag
f(\epsilon,0)&=\epsilon\log(\epsilon)+b\log(b)-(b+\epsilon)\log(b+\epsilon)-(\epsilon+2b)\log(\epsilon+2b)-(1-(\epsilon+2b))\log(1-(\epsilon+2b)) \\\notag
f(0,\epsilon/3)&=2b\log(b)+\frac{2}{3}\epsilon\log(\epsilon/3) +2/3\log(3) -\left(\frac{2}{3}- \frac{\epsilon}
{3}\right)\log\left(\frac{2}{3}- \frac{\epsilon}
{3}\right)-\left(\frac{1}{3} + \frac{\epsilon}
{3}\right)\log\left(\frac{1}{3} + \frac{\epsilon}
{3}\right).
\end{align*}

\begin{figure}[t]
\centering
\includegraphics[width=0.5\linewidth]{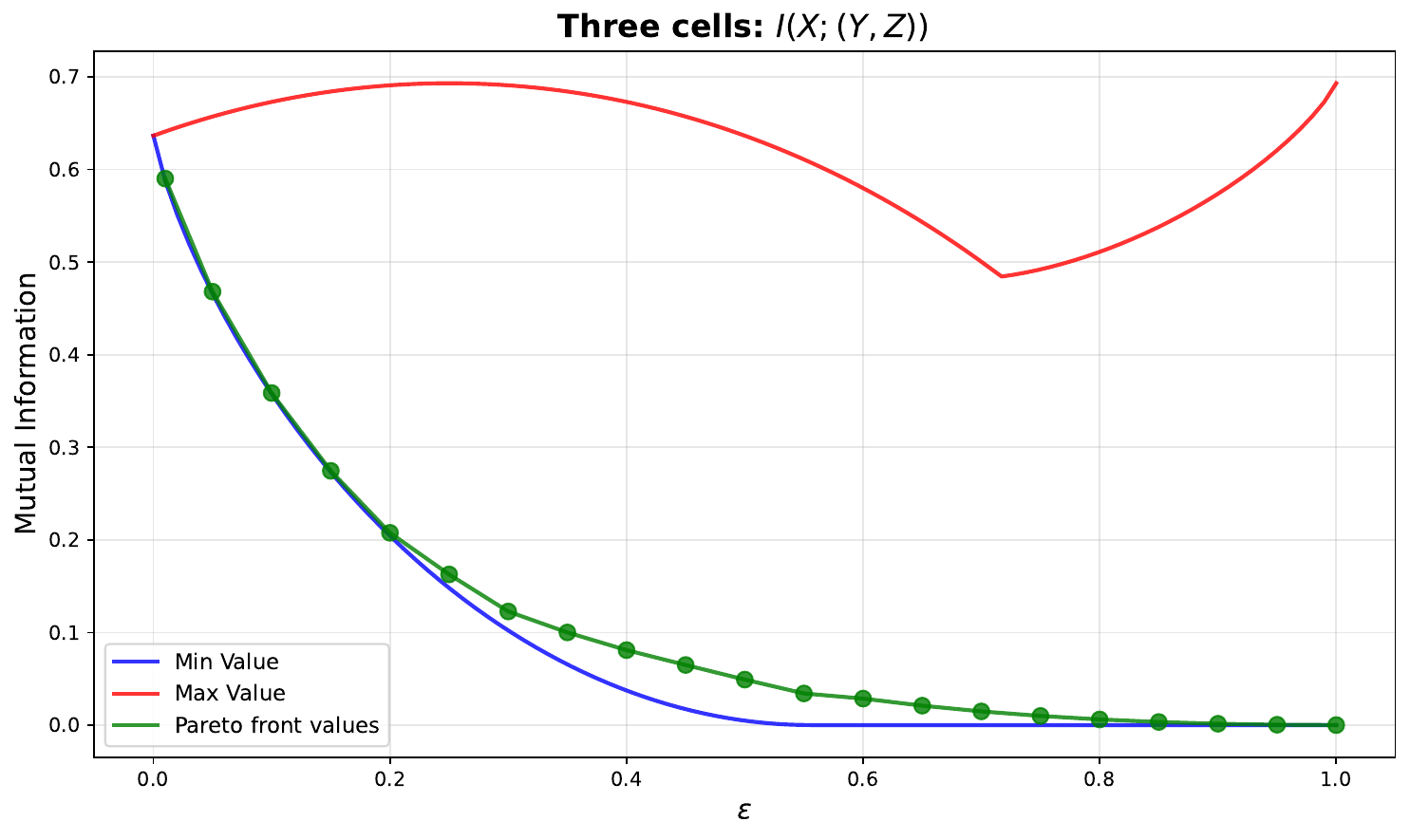}
\caption{Maximum and minimum values of final state mutual information between one cell and the remaining cells in a three-cell system. Values are plotted against the error rate $\epsilon$, and are shown along with the values of numerically optimized strategies which lie on the Pareto front.}
\label{Fig:threecellmutinfplot2}
\end{figure}

Overall, the results are shown in Fig.~\ref{Fig:threecellmutinfplot2}.
\subsubsection{Tightness of bounds}
These results give us a bound on the final state mutual information for any system with three symmetric binary variables. 
It is worth noting that this does not necessarily mean that there exists a set of parameters $\mathbf{p}$ such that the model achieves these extremal values. The maximum (minimum) found above is an upper (lower) bound for the actual maximum (minimum) final state mutual information that is achievable in our model.

\subsection{Calculation of information metrics from data}
For calculation of mutual information between two random variables, $X$ and $Y$, from data, we use the Kraskov– Stögbauer–Grassberger (KSG) estimator~\cite{Kraskov2004}. Suppose the data set has $N$ points, $\{(X_i,Y_i)\}_{i=1}^{N}$ and that $X$ and $Y$ belong to some metric space with norms $\|\cdot\|_X$ and $\|\cdot\|_Y$ respectively. Consider the metric space defined by $Z=(X,Y)$, with the sup-norm $\|z-z'||=\max\{\|x-x'\|_X,\|y-y'\|_Y\}.$ Let $\epsilon_k(i)/2$ be the distance from a data point $z_i$ to its $k^{\mathrm{th}}$ nearest-neighbor, and let $\epsilon^X_k(i)/2$ and $\epsilon^Y_k(i)/2$ be between the same points, projected in the $X$ and $Y$ subspaces. Let $n^X(i)$ be the number of points in $x_j$ that are a distance less than $\epsilon_k(i)/2$ from $x_i$ (and analogously define $n^Y(i)$). Then, the KSG estimator is 
\begin{equation}
    \hat{I}(X;Y) = I((X_i,Y_i)_{i=1}^{N})= \psi(k)+\psi(N)-\frac{1}{N}\sum_{i=1}^{N}(\psi(n^X(i))+\psi(n^Y(i))),
    \label{eq:KSGEstimator}
\end{equation}
where $\psi(z)$ is the digamma function, $\psi(z)=\frac{d}{dz}(\log\Gamma(z))$.

For computational purposes, as noted in~\cite{Kraskov2004}, a value of $k>1$ should be chosen; however, if $k$ is taken too large, the resulting increase in systematic error can outweigh the increase in statistical accuracy. In our calculations we therefore fix $k=4$ as a representative choice, while also verifying Sec.~\ref{subsec:analysisexperimental data} that our qualitative findings are robust across different values of $k$. More importantly, we check the error of estimator by performing permutation tests.

To test the error of the estimator we create a null data, preserving the marginals of the data but removing the statistical dependence. To do this, we draw $\sigma_1,\cdots,\sigma_K$ i.i.d. random permutations of $\{1,\cdots,N\}.$ For each random permutation, we then estimate the mutual information, 
$$s^*=I((X_i,Y_i)_{i=1}^{N}),\quad s_j=I((X_{\pi_j(i)},Y_i)_{i=1}^{N}).$$ 
Next, we sort all $\{s_j\}$ to obtain the order statistics $s_{(1)}\leq \cdots\leq s_{(K)}$. The mean under permutation test is $\bar{s}=\frac{1}{K}\sum_{j}s_j.$ The $(1-\alpha) \%$ confidence interval, on the other hand, is $[s_{(l)},s_{(u)}],$ where $$l=\left\lfloor \frac{\alpha}{2} K\right\rfloor,\quad  u=\left\lceil \left(1-\frac{\alpha}{2}\right)K\right\rceil.$$

The standard deviation for this estimator can be quantified using the Jackknife variance estimator, 
$$\widehat{\mathrm{se}}=\sqrt{\frac{N-1}{N}\sum_{i=1}^{N}(s^*_{-(i)}-\hat{s^*})},$$
where $s^*_{-(i)}=I((X_j,Y_j)_{j=1, j\neq i}^{N}),\quad \hat{s^*}=\frac{1}{N}\sum_{i=1}^{N}s^*_{-(i)}.$ This tends to be biased upwards~\cite{Efron1981} and as a result is a conservative estimate of the variance.

\section{Local versus global optimum}
Consider the situation where the group of cells are working with a cooperatively optimal common strategy, $\mathbf{p}^*$. Now, suppose a cell, say cell $1$, is tweaked to \textit{locally optimize} its strategy $\mathbf{p}_1$, finding a strategy that optimizes the collective optimum while keeping the strategies of all other cells fixed. We say that such a cell $3$ is acting ``greedily''.

Without loss of generality, say we do this with cell $3$. Then, the optimization problem for this tweaked cell is 
\begin{align*}
    &\inf_{\mathbf{p}_3}\,\tau_{\alpha^*}(\mathbf{p}^*, \mathbf{p}^*, \mathbf{p}_3)\\
    \text{subject to: }& \nonumber\\
    &\boldsymbol{{\tau}}\geq 0, \\
    &0\leq \mathbf{p}_3,\boldsymbol{\epsilon} \leq 1, \,\\
    &1+\sum_{j\notin \mathcal{T}} Q_{ji} \tau_j = 0 \quad\text{for } i \notin \mathcal{T}\\
    &\sum_{j} Q_{ji} \epsilon_j = 0 \quad\text{for } i \notin \mathcal{T}\\
    & \epsilon_{\alpha^*}\leq\epsilon_{tol},\\
    & {\tau}_i = 0  \quad\text{for } i \in \mathcal{T}\\
    & \epsilon_i = 0  \quad\text{for } i \in \mathcal{T}^G\\
    & \epsilon_i = 1  \quad\text{for } i \in \mathcal{T}^B,
\end{align*}
where we have naturally amended our notations to account for the heterogeneity in the cell strategies.

\begin{figure*}[t]
    \centering
    \includegraphics{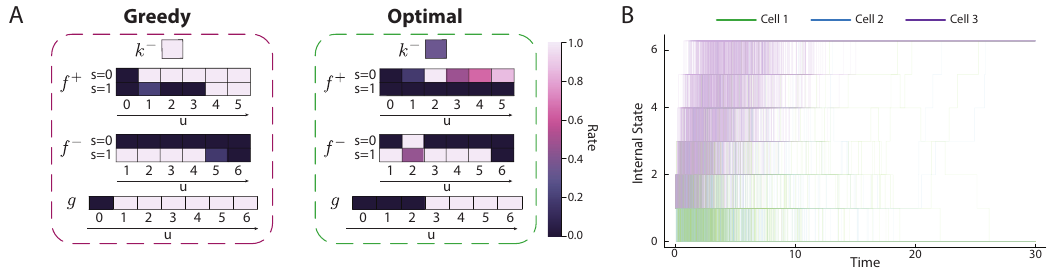}
    \caption{Local optimization over a single cell leads to that cell acting greedily. (A) Parameters for the optimally greedy cell (left), and collectively optimal system at $\epsilon=0.02$, as shown in main text Fig~2C. (B) Gillespie simulations of the model with one greedy cell, where in $n=1000$ simulations, the greedy cell (cell $3$) ends ups as an inhibitor $945$ times. }
    \label{fig:LocalParameters}
\end{figure*}

It is easy to notice that after solving this modified optimization problem,
$$\tau_{\alpha^*}(\mathbf{p}^*, \mathbf{p}^*, \mathbf{p}^*_3)\leq\tau_{\alpha^*}(\mathbf{p}^*, \mathbf{p}^*,\mathbf{p}^*).$$
Thus, a single greedy cell in a tissue where the remaining cells are acting collectively optimally (and have fixed strategies) forms patterns quicker for the same accuracy. The strategy that the greedy cell adopts is more assertive than the collectively optimal solution,  Fig. \ref{fig:LocalParameters}, leading this cell to end up as an inhibitor more often than other cells. 

If we let all cells adopt the single-cell assertive strategy, the entire system performs sub-optimally.

\section{Speed-Error optimization does not optimize for information flow}

\begin{figure*}[t]
    \centering
    \includegraphics{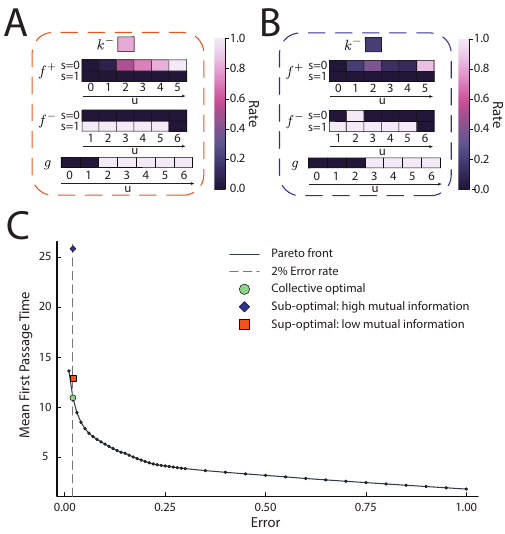}
    \caption{The speed-error optimal patterning model does not optimize for mutual information flow between cells. (A) Low mutual information and (B) high mutual information sub-optimal solutions used in the main text. (C) The two suboptimal solutions, alongside the collectively optimal solution at $\epsilon=0.02$ and the Pareto frontier.}
    \label{fig:NotOptimalInfo}
\end{figure*}

As established in the main text, at a fixed error, the optimal solution need not be one that maximizes nor minimizes the mutual information shared with the other cells. For the computations in the main text, we manually find two strategies at error rate $\epsilon=0.02$. The parameters corresponding to these higher and lower error rates are illustrated in Fig \ref{fig:NotOptimalInfo} along with the speed of patterning on the Pareto front.
\section{Experimental data}
\label{sec:experimental data}
\subsection{Description of the data}

\begin{figure*}[t]
    \centering
    \includegraphics{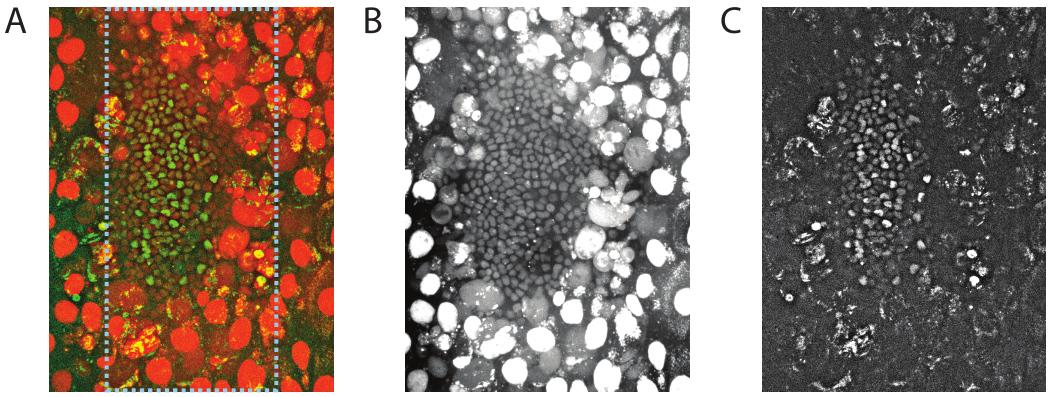}
    \caption{(A) Max projection of raw image, showing a nuclear marker (red) and Scute intensity (green). Blue dashed box indicates the region of interest. (B)  Max projections of the nuclear marker channel. (C) Max projection of the GFP-Sc channel. All projections show the same frame.}
    \label{fig:rawimages}
\end{figure*}
Experimental data was obtained from Ref.~\cite{Phan2024}, and consists of time-lapse confocal microscopy videos of the anterior dorsal histoblast nest in Drosophila pupal abdomen development. Imaging is performed during the window where on sensory organ precursor (SOP) cells are patterning, from approximately $14$ to $24$ hours after puparium formation (APF). The dataset captures the dynamics of Scute protein expression using a functional GFP-tagged knock-in version of Scute (GFP-Sc) expressed from the endogenous locus. Figure~\ref{fig:rawimages} shows examples frames from some of these experimental videos.

Three movies were obtained from individual pupae, each spanning $12-16$ hours of continuous imaging. Any time point of each video consists of a z-stack of  $660 \times 900 \times 24$ pixels, with each voxel size being $0.39\times0.39\times1.33 \,\mu m^3$. There are a total of $310$ frames in each video, and the time-lapse between each frame is $2.5$ minutes. At around frame $42$ in each experiment, the microscope was adjusted, resulting z-stack shifting by about $5-6$ slices. A minor adjustment is also observed around frame $94-95$ of $2-3$ slices.

\subsection{Segmenting and filtering cells}

We start from the $3$D segmentation of nuclei from Ref~\cite{Phan2024}, which in turn applies the methods in Ref.~\cite{Corson2017}, followed by additional filtering. We further filter the cells by manually defining a region of interest (ROI), drawn to include only the patterning cells. This region remains the same across all frames. For each time point, we then ignore cells outside of this ROI. An example ROI is shown in Fig. \ref{fig:rawimages}.

For Fig.~8 of the main text,  we tracked an example SOP cell using Ultrack~\cite{bragantini2024ultrack, bragantini2024ucmtracking}. 

\subsection{Extracting fluorescence intensity}

\begin{figure*}[t]
    \centering
    \includegraphics{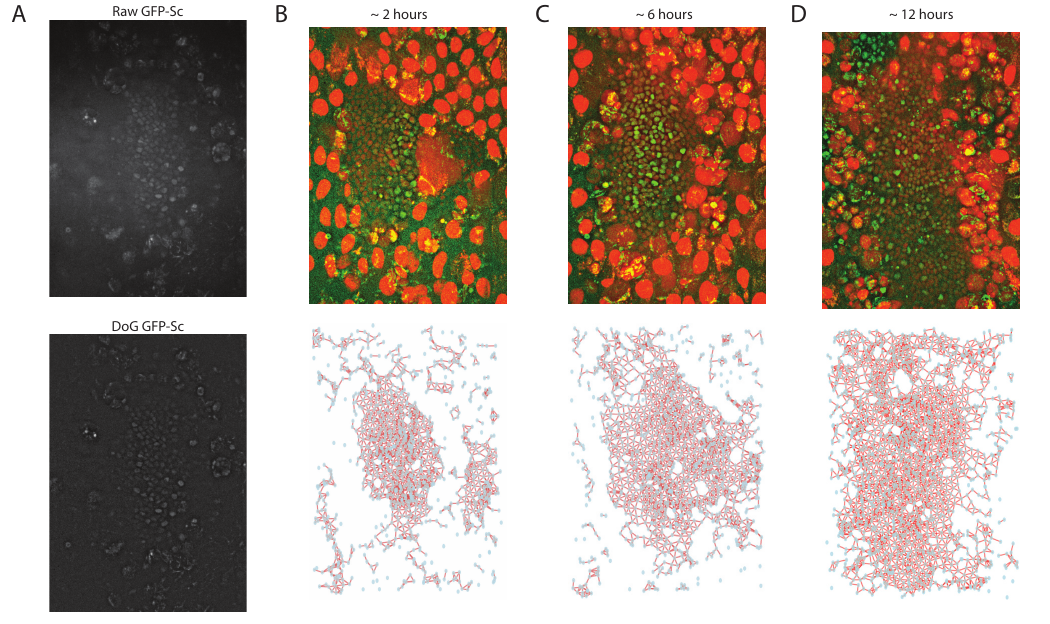}
    \caption{(A) Example raw GFP-Sc snapshot of a z-slice from a video (top) alongside the same image under DoG (bottom). DoG enhances the local effects, and removes the broad image trends. (B-D) Three snapshots from one of the videos at 3 different time points: $2$ hours (B), $6$ hours (C) and $12$ hours since the start of the experiment. Max z-projection of the videos are shown (top) alongside the z-projection of the graph reconstructed from cell segmentation (bottom). Red lines are the edges, while the blue circles are the centroids of the cell segments.}
    \label{fig:TracesResidualandCellGraph}
\end{figure*}
To measure the Scute intensity of each cell, and to compare a cell's intensity to its neighbor, we first need to remove regional background intensity shifts (such as those introduced by auto-fluorescence). To do so, we perform a two-dimensional Gaussian blur on the videos at each time-point and for each z-slice. We use scikit-image library in python with standard deviation of $10$ pixels, which roughly translates to $3.9\, \mu m$, and gives a full-width-at-half-maximum of $9-10\,\mu m$---roughly the diameter of the largest cell of interest. This value is large enough to remove the background noise effectively, and reduce the artificial correlation between neighboring cells added due to background noise, while being small enough to not induce further spurious interactions between neighboring cells. Then we take the residual of the original z-stack image and the gaussian blurred image to get a residual image. This is essentially the standard difference-of-Gaussians algorithm, with $\sigma_1\rightarrow0$ and $\sigma_2=\sigma$. An example is shown in Fig.~\ref{fig:TracesResidualandCellGraph}.

For each segmented cell, say cell $\alpha$, we measure the median of the channel intensity in the segmented voxels, $u^*_\alpha(t)$. We also measure the mean intensity and standard-deviation of intensities across the all the voxels in the z-stack at each time $t$, $\mu(t)$ and $\sigma(t)$, we then normalize the raw intensities by performing a population wide z-scoring, $$u_\alpha(t)=\frac{u^*_\alpha(t)-\mu(t)}{\sigma(t)}.$$

\subsection{Finding cellular neighborhoods}

To identify which cells are neighbors at any given time, we take the following sequence of steps:
\begin{itemize}
    \item Take the Delaunay triangulation in $3$-dimensions using the centroids of the filtered and segmented cells.
    \item Filter the edges by length, removing all edges that are larger than $d_{\mathrm{max}}.$ We use $d_{\mathrm{max}}=12.0\,\mu m$, which is taken to be slightly larger than the mode of the edge lengths of the Delaunay triangulation.
\end{itemize}
This results in a neighborhood adjacency matrix $A_{ij}(t)$ where $A_{ij}(t)=1$ if two cells are identified as neighbors, and $A_{ij}=0$ otherwise. We compute this matrix for each frame, and it changes across frames due to cell rearrangement, division, and death. Example adjacency graphs for the same experiment at multiple time-points are shown in Fig.~\ref{fig:TracesResidualandCellGraph}.

\subsection{Information calculation on experimental data}
\label{subsec:analysisexperimental data}
At a given time point, we now have a set of normalized scute intensities $\{u_i(t)\}$, as well as an adjacency matrix $A_{ij}(t)$. From this, we define the average intensity of the neighboring cells as 
\begin{equation}
    v_i(t) = \frac{\sum_j A_{ij}(t)u_j(t)}{\sum_j A_{ij}(t)}.
\end{equation}
We now have a set of pairs, $\{(u_i(t),v_i(t)\}$ of a cell's intensity and its neighbors average intensity. We apply the KSG estimator to this set of pairs, as described earlier, to estimate the mutual information $I(u(t);v(t))$ at some time point.
\begin{figure*}[t]
    \centering
    \includegraphics{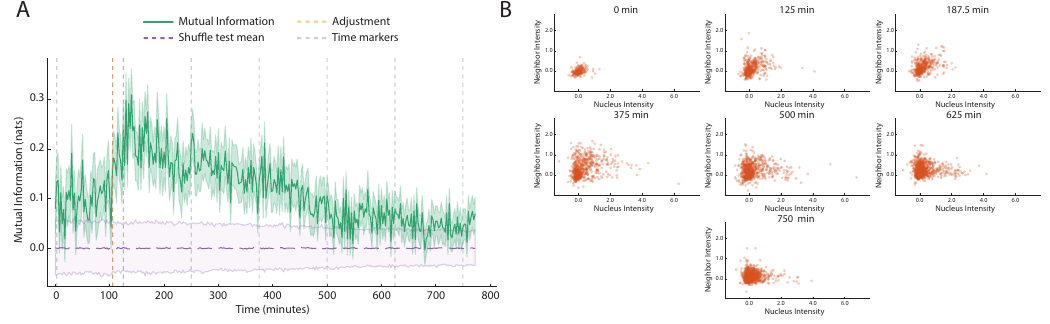}
    \caption{Detailed look at mutual information calculation for Experiment $2$. (A) Static mutual information calculation for the experiment (green) with ribbon of $\pm$ std. deviation. Also shown are $95\%$ confidence interval and mean from permutation tests (purple), the time-point of z-stack adjustment (orange dashed line). (B) Scatter plot of nucleus intensity, $u(t)$, and neighbor mean intensity, $v(t)$, at different time-points, marked by gray dashed line in (A).}
    \label{fig:MIdetails}
\end{figure*}
The information calculation for one specific experiment (experiment $2$) is illustrated in Fig~\ref{fig:MIdetails}, which also shows the raw data points at various stages of patterning. 

The non-monotonic mutual information trend can be described by the following series of steps:
\begin{itemize}
    \item Time $0-125$ min (pre-selection): During this early phase, GFP-Sc intensities are low and fairly uniform across the pro-neural cluster, with only a few cells expressing Scute. As a result, cell-neighbor pairs carry little information: mutual information is low but slightly above the shuffled baseline.
    \item Time $125-375$ min (onset of lateral inhibition): As some cells increase in Sc levels while the neighboring cells decrease, strong cell-neighbor dependence emerges. At the same time, the single-cell intensity distribution broadens (marginal variance increases). These changes, higher marginal entropy and lower conditional entropy, drive a marked increase in mutual information, which peaks during this period. 
    \item Time $375-750$ min (fate resolution): Lateral inhibition enters its final phase, and the system begins resolving cell fates. Although GFP-Sc intensity in SOP cells declines later in this window~\cite{Phan2024}, the drop in mutual information begins earlier. This suggests the decline is primarily due to fate resolution rather than changes in signal intensities.
    
    The final low level of mutual information is due to a combination of:
    \begin{enumerate}[label=(\roman*)]
        \item \textbf{Reduced marginal variance}: Intermediate states disappear, lowering marginal entropy.
        \item \textbf{Fewer informative pairs}: SOP cells and their neighbors commit to a fate and reduce their Scute expression. This leads to both the central cell and its neighbors having low Scute levels, producing numerous low-low pairs.
    \end{enumerate}
    Together, these effects reduce mutual information. While the first factor (i) also appears in our minimal model of $M=3$ cells, the second (ii) does not.
\end{itemize}
These trends are observed in  all three experimental videos. 

\begin{figure}[!h]
    \centering
    \includegraphics{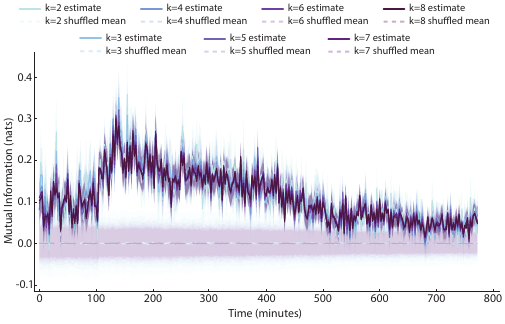}
    \caption{The non-monotonic trend remains robust across different choices of nearest neighbors for the estimator. For experiment E$2$, mutual information was calculated using the KSG estimator with $k=2,\dots,8$, along with the corresponding standard deviations from jackknife resampling, and the mean and $95\%$ confidence intervals from permutation tests. Although the absolute values vary with $k$, the non-monotonic trend with a transient peak is consistent.}
    \label{fig:MIdiffkdetails}
\end{figure}

We further examine the robustness of the trend with respect to different choices of $k$, as shown in Fig.~\ref{fig:MIdiffkdetails}. Specifically, we compute the mutual information for $k \in \{2,3,4,5,6,7,8\}$ and observe that, although the absolute estimator values vary with $k$, the overall trend and the order of magnitude of the mutual information values consistently persist across all tested values.

The mutual information time series in the main text was smoothed using a one-dimensional Gaussian filter with standard deviation $\sigma=3$ frames, corresponding to roughly $7.5$ minutes.

\bibliographystyle{abbrv}